\pdfoutput=1
\documentclass[a4paper,12pt]{article}
\usepackage{jheppub,esint,psfrag}
\usepackage[utf8]{inputenc}
\usepackage{bm}
\usepackage{booktabs}
\usepackage{makecell}
\usepackage{amsfonts}
\usepackage{amssymb}
\usepackage{amsmath}
\usepackage{hyperref}
\usepackage{mathrsfs}

\newcommand{\ep}{\varepsilon}

\def\I{{\rm i}}
\def\D{{\rm d}}
\newcommand{\z}{&&\hspace*{-1cm}}

\title{Transcendental structure \\ of multi-loop massless two-point correlators \\ and RG functions}
\author[a]{N. Volchanskiy,}
\author[a]{A.~V. Kotikov,}
\author[b]{I.~A. Kotikov,}
\author[a,c]{R.~N.~Lee}

\affiliation[a]{Joint Institute for Nuclear Research, 141980 Dubna, Russia}

\affiliation[b]{Moscow State University, Faculty of Physics, 119991 Moscow, Russia}

\affiliation[c]{ Budker Institute of Nuclear Physics, 630090 Novosibirsk, Russia}

\emailAdd{kotikov@theor.jinr.ru}
\emailAdd{kotikov.ia22@physics.msu.ru}
\emailAdd{r.n.lee@inp.nsk.su}
\emailAdd{nikolay.volchanskiy@gmail.com}

\abstract{We present recent results on the $\pi$-dependent contributions to massless Euclidean two-point correlators, as well as to generic anomalous dimensions and $\beta$-functions. We review the derivation of a complete all-order expression for hatted single zeta values based on the Landau--Khalatnikov--Fradkin transformation \cite{Kotikov:2019bqo}. A new compact integral representation for this expression is also provided. By performing a deep $\varepsilon$-expansion of four-loop master integrals, we extend the results of refs.~\cite{Baikov:2018wgs,Baikov:2018gap,Baikov:2019zmy} to multiple zeta values (MZVs) of weights up to 16. Specifically, we find that, up to this transcendental weight, the $\varepsilon$-dependent hatted MZVs absorb all Riemann zeta values of even argument from the master integrals, except for $\zeta_{12}$, $\zeta_{3} \zeta_{12}$, and $\zeta_{14}$ (whereas $\zeta_{16}$ is completely absorbed). Combining this property with 't~Hooft's pole equations, we derive constraints on the transcendental structure of renormalization group (RG) functions up to fifteen loops, which even enables us to infer the explicit closed form of certain constraints as functions of the perturbative order. Using these identities, we predict the $\pi$-dependent terms in the RG functions at eight and nine loops for $\text{O}(n)$ $\phi^4$ theory, at six loops for a general non-Abelian gauge theory coupled to fermions, and at seven loops for QED with $N_f$ fermion flavors. All existing multi-loop calculations of the RG functions in these models agree with the constraints on the $\pi$-dependent terms. Furthermore, all-order large-$N_f$ computations corroborate our predictions at the next-to-be-calculated perturbative orders in the gauge theories.}

\begin{document}
\maketitle
\section{Introduction}

Consider the multi-loop structure of Euclidean propagator-type functions (in refs.~\cite{Baikov:2018wgs,Baikov:2018gap,Baikov:2019zmy} they are called p-functions). Nearly four decades ago, it was observed that all contributions proportional to $\zeta_4 = \pi^4/90$ cancel out in the Adler function at three loops~\cite{Gorishnii:1990vf}. Two decades later, it was shown that the four-loop contribution is also $\pi$-free and that a similar fact holds for the coefficient function of the Bjorken sum rule~\cite{Baikov:2010je}. There is now mounting evidence, see, e.g.\ refs.~\cite{Baikov:2016tgj,Chetyrkin:2017bjc,Chetyrkin:2017bjcAddendum,Herzog:2017ohr,Ruijl:2018poj,Davies:2017hyl,Moch:2018wjh,Vogt:2018miu,Herzog:2018kwj}, that various massless Euclidean physical quantities exhibit striking regularities in terms proportional to even $\zeta$-function values, $\zeta_{2n}$, e.g.\ $\pi^{2n}$ with $n$ being a positive integer\footnote{Note also that, within a Schwinger-Dyson equation approach in fixed dimension, renormalized Euclidean massless correlators have been shown to be expressed only in terms of odd zeta-values \cite{Kreimer:2006ua,Bellon:2008de}.}. Such puzzling facts have recently led to the formulation of the ``no-$\pi$ theorem'' \cite{Baikov:2018wgs,Baikov:2018gap,Baikov:2019zmy}. The latter is based on the observation~\cite{Broadhurst:1999xk,Baikov:2010hf} that the $\ep$-dependent transformation of the $\zeta$-values, $\zeta_{2n} \to 0$ and $\zeta_{2n+1} \to \hat\zeta_{2n+1}(\ep)$,
\begin{equation}
\hat{\zeta}_3 \equiv \zeta_3 + \frac{3\ep}{2} \zeta_4 - \frac{5\ep^3}{2} \zeta_6,~~ \hat{\zeta}_5 \equiv \zeta_5 + \frac{5\ep}{2} \zeta_6,~~
\hat{\zeta}_7 \equiv \zeta_7\, ,
\label{hatZe}
\end{equation}
eliminates even zetas from the expansion of four-loop p-integrals. Generalizations of (\ref{hatZe}) to 5, 6, and 7 loops are available in refs.~\cite{Baikov:2018wgs,Baikov:2018gap,Baikov:2019zmy,Georgoudis:2018olj,Georgoudis:2021onj}. The expansions~(\ref{hatZe}) and their extensions provide a way to predict the terms $\sim \pi^{2n}$ in higher orders of perturbation theory (see refs.~\cite{Baikov:2018wgs,Baikov:2018gap,Baikov:2019zmy} and sections~\ref{sec:PiConstraints}--\ref{sec:PredictionsYM} below).

The generalization of (\ref{hatZe}) to the case of hatted single zeta values $\hat{\zeta}_{2n+1}$ with arbitrary $n$ has been done in closed form in \cite{Kotikov:2019bqo} with the help of the Landau-Khalatnikov-Fradkin (LKF) transformation. The LKF transformation~\cite{Landau:1955zz,Fradkin:1955jr} elegantly relates the fermion propagator in two different $\xi$-gauges (and similarly for the fermion-photon vertex) in QED and related theories. Its most important applications~(see \cite{Kotikov:2019bqo} and references therein) are related to the study of the gauge covariance of Schwinger-Dyson equations and their solutions. Other applications~\cite{Bashir:2002sp,Jia:2016wyu} focus on estimating large orders of perturbation theory. Indeed, and this will play a crucial role in what follows, the non-perturbative nature of the LKF transformation allows one to fix some of the coefficients of the all-order expansion of the fermion propagator. Starting with a perturbative propagator in some fixed gauge, say $\eta$, all the coefficients depending on the difference between the gauge fixing parameters of the two propagators, $\xi - \eta$, are fixed by a weak coupling expansion of the LKF-transformed initial one. Such estimations have been carried out for QED in various dimensions~\cite{Bashir:2002sp,Jia:2016wyu,Gusynin:2020cra,Pikelner:2020mga}, for generalizations to brane worlds~\cite{Ahmad:2016dsb,James:2019ctc}, and for more general SU(N) gauge theories~\cite{DeMeerleer:2018txc,DeMeerleer:2019kmh}.

Note that the results in \cite{Kotikov:2019bqo} contain only single zeta values, whereas the results in \cite{Baikov:2018wgs,Baikov:2018gap,Baikov:2019zmy} also contain multiple zeta values (MZV)
\begin{equation}\label{MZVdef}
\zeta_{m_1,\dots,m_k}=\sum\limits_{i_1=1}^\infty\sum\limits_{i_2=1}^{i_1-1}
\dots\sum\limits_{i_k=1}^{i_{k-1}-1}\prod\limits_{j=1}^k\frac{\mbox{sgn}(m_j)^{i_j}}{i_j^{|m_j|}}\,,
\end{equation}
the consideration of which is the focus of the present study.

In this paper we continue a systematic investigation \cite{Lee:2010cga,Lee:2010ug,Lee:2011jt} of the four-loop massless propagator diagrams and present results for the deep $\ep$-expansion of {\em all} the master integrals up to transcendentality weight sixteen. Studying the structure of the $\varepsilon$-expansion of the master integrals provides information on the form of the hatted MZVs, following directly the Baikov and Chetyrkin approach \cite{Baikov:2018wgs,Baikov:2018gap,Baikov:2019zmy}. 

In order to obtain deep $\ep$-expansions, we use the exact in $d$ results of ref.~\cite{Lee:2011jt} obtained using DRA method. We calculate the emerging multiple sums using the \texttt{SummerTime} package \cite{Lee:2015eva}. The high-precision numerical coefficients of $\ep$-expansion (about 3000 decimal digit) are then recognized using PSLQ algorithm \cite{PSLQ} as $\mathbb{Q}$-linear combinations of multiple zeta values. Such a deep $\ep$-expansion allowed us to construct the linear combinations of master integrals with the property of homogeneous transcendentality\footnote{In particular, this expansion provides some information on the structure of the results of the higher-loop integrals. Such a study of the two-point and three-point two-loop diagrams with masses has been done earlier in   \cite{Fleischer:1998nb,Fleischer:1999hp,Kotikov:2007vr}.} (see discussions below).

High-loop master integrals and renormalization-group (RG) functions reveal deep algebraic and number-theoretic structures. This is notably illustrated by the no-$\pi$ theorem of refs.~\cite{Baikov:2018wgs,Baikov:2018gap,Baikov:2019zmy}. The possibility of absorbing Riemann zeta values of even arguments, $\zeta_{2n}$, into the $\varepsilon$-expansions of hatted MZVs results in numerous constraints on the transcendental content of RG functions. Specifically, the rational prefactors of powers of $\pi^2$ in an RG function coefficient at a given loop order can be determined from the rational prefactors of MZVs in lower-loop coefficients. The $\varepsilon$-expansions of hatted MZVs up to weight 16 allow us to derive the complete set of relations for RG functions up to the nine-loop order. We can even go beyond that order---in fact, we venture to formally construct relations up to 15 loops---to infer and verify the general closed forms of these relations as functions of the perturbative order. These general relations can then be used to formulate conjectures about $\pi^2$-dependent terms at yet-to-be-calculated loop orders of RG functions in specific quantum field theory models.

The rest of the paper is organized as follows. In section~\ref{sec:hat-zetas}, we present the exact all-order expression for hatted single zeta values together with the main steps of its derivation based on the LKF transformation. The section ends with a new integral representation for the hatted MZVs, which is proven in appendix~\ref{app:LeeIntegral}. In section~\ref{sec:hatMZVs}, we consider the $\ep$-expansions of the four-loop two-point master integrals and the hatted MZVs. These results for the hatted zeta values lead to relations among the rational coefficients in $\beta$-functions and anomalous dimensions (ADs), to which section~\ref{sec:PiConstraints} is dedicated. Section~\ref{sec:PredictionsPhi4} contains our predictions for the $\pi$-dependent parts of the eight- and nine-loop corrections to the $\beta$-function and ADs in the $\text{O}(n)$ $\varphi^4$ model. Section~\ref{sec:PredictionsYM} presents analogous predictions for a general non-Abelian gauge theory at six loops and QED at seven loops. A concise summary of our main results is given in section~\ref{sec:summary}, while the most cumbersome expressions from sections~\ref{sec:hat-zetas}--\ref{sec:PiConstraints} are deferred to the appendices.

\section{\label{sec:hat-zetas}Hatted Riemann zeta function values}
\subsection{LKF transformation}
\label{sec:LKF:x-space}

In this section we consider QED\footnote{The results in the case of scalar QED are very similar and can be found in ref. \cite{Kotikov:2019bqo}.} in a Euclidean space of dimension $d$ ($d=4-2\ep$). The general forms of the fermion propagator in the momentum and $x$-space representations $S_F(p,\xi)$ and $S_F(x,\xi)$, in a given gauge $\xi$, read
\begin{equation}
S_F(p,\xi) = \frac{i}{\hat{p}} \, P(p,\xi) \, ,~~ S_F(x,\xi) =  \hat{x} \, X(x,\xi) \, ,
\label{SFp}
\end{equation}
where the matrix-tensor structures, namely, the factors $\hat{p}$ and $\hat{x}$ containing Dirac $\gamma$-matrices, have been explicitly isolated from the scalar parts. The two representations, $S_F(x,\xi)$ and $S_F(p,\xi)$, are related by the Fourier transform which is defined as
\begin{equation}
S_F(p,\xi) = \int \frac{\D^dx}{(2\pi)^{d/2} } \, e^{\I px} \, S_F(x,\xi) \, , \qquad
S_F(x,\xi) = \int \frac{\D^dp}{(2\pi)^{d/2} } \, e^{-\I px} \, S_F(p,\xi) \, .
\label{SFp2x}
\end{equation}

The celebrated LKF transformation connects in a very simple way the fermion propagator in two different gauges, e.g.\ $\xi$ and $\eta$. In dimensional regularization, it reads \cite{Kotikov:2019bqo}:
\begin{equation}
S_F(x,\xi) = S_F(x,\eta)\, e^{\I D(x)},~~D(x)=-i\Delta e^2\mu^{2d}  \int \frac{\D^dp}{(2\pi)^{d/2} } \, \frac{e^{-\I px}}{p^4},~~\Delta=\xi-\nu \, .
\label{LKFN}
\end{equation}

We may now proceed to calculate $D(x)$. In order to do so, it is possible to use the following simple formulas
for the Fourier transform of massless propagators (see, e.g.\ \cite{Kotikov:2018wxe,Kotikov:2019gqp,Kotikov:2021kbu}):
\begin{equation}
\int \D^dx \, \frac{e^{\I px}}{x^{2\alpha}} \, = \frac{2^{2\tilde{\alpha}} \pi^{d/2} a(\alpha)}{p^{2\tilde{\alpha}}},~~
\int \D^dp \, \frac{e^{-\I px}}{p^{2\alpha}} \, = \frac{2^{2\tilde{\alpha}} \pi^{d/2} a(\alpha)}{x^{2\tilde{\alpha}}}, ~~
a(\alpha)=\frac{\Gamma(\tilde{\alpha})}{\Gamma(\alpha)},~~\tilde{\alpha}=\frac{d}{2}-\alpha \, ,
\label{SFx2pN}
\end{equation}
This makes the $\ep$-dependence of $D(x)$ explicit:
\begin{equation}
D(x)=\frac{\I\,\Delta\,A}{\ep}\,\Gamma(1-\ep)\, (\pi\mu^2 x^2)^{\ep} , \qquad A=\frac{\alpha_{\rm em}}{4\pi}=\frac{e^2}{(4\pi)^2}\, .
\label{DxN}
\end{equation}
From eq.~(\ref{DxN}), we see that $D(x)$ contains a common factor $\Delta A$ accompanied by the singularity $\ep^{-1}$.

\subsection{LKF transformation in momentum space}
\label{sec:LKF:p-space:F}

Let us assume that, for a given gauge-fixing parameter $\eta$, the fermion propagator $S_F(p,\eta)$ with external momentum $p$ takes the form (\ref{SFp}), where  $P(p,\eta)$ is given by
\begin{equation}
P(p,\eta) = \sum_{m=0}^{\infty} a_m(\eta)\, A^m \,{\left(\frac{\tilde{\mu}^2}{p^2}\right)}^{m\ep} \, ,~~ \tilde{\mu}^2= 4\pi \mu^2 \, ,
\label{Peta}
\end{equation}
where $a_m(\eta)$ are the coefficients of the loop expansion of the propagator and $\tilde{\mu}$ is the renormalization scale, which lies between the MS-scale $\mu$ and the $\overline{{\rm MS}}$-scale $\overline{\mu}$. The LKF transformation then shows that, for another gauge parameter $\xi$, the fermion propagator can be expressed as
\begin{equation}
\hspace{-5pt}P(p,\xi) = \sum_{m=0}^{\infty} a_m(\xi)\, A^m\, {\left(\frac{\tilde{\mu}^2}{p^2}\right)}^{m\ep} \, ,
\label{Pxi} 
\end{equation}
where
\begin{equation}
a_m(\xi) = a_m(\eta) \, \frac{\Gamma(2-(m+1)\ep)}{\Gamma(1+m\ep)}\,
\sum_{l=0}^{\infty} 
\frac{\Gamma(1+(m+l)\ep)\,\Gamma^l(1-\ep)}{l!\,\Gamma(2-(m+l+1)\ep)} \, \frac{(\Delta \, A)^l}{(-\ep)^l} \, {\left(\frac{\tilde{\mu}^2}{p^2}\right)}^{l\ep} \, .
\label{axi}
\end{equation}
In order to derive (\ref{axi}), we used the fermion propagator $S_F(p,\eta)$ with  $P(p,\eta)$ given by (\ref{Peta}), performed the Fourier transform to
$S_F(x,\eta)$ and applied the LKF transformation (\ref{LKFN}). As a final step, we took the inverse Fourier transform and obtained 
$S_F(p,\xi)$ with  $P(p,\xi)$ given by (\ref{Pxi}).

\subsubsection{Scale fixing}

Following ref. \cite{Kotikov:2019bqo}, we consider only the case of the so-called MS-like schemes.
In these schemes, specific terms arising from the application of dimensional regularization must be fixed.
This procedure will be called {\it scale fixing} and will play a crucial role in our analysis.
	
Let us first recall that the $\overline{{\rm MS}}$-scale $\overline{\mu}$ is related to the previously defined scale $\tilde{\mu}$ via $\overline{\mu}^2=\tilde{\mu}^2\, e^{-\gamma}$, where $\gamma$ is the Euler constant. An advantage of the $\overline{{\rm MS}}$-scale is that it removes the Euler constant $\gamma$ from the $\ep$-expansion. Moreover, it is well known that, the final results for two-point massless diagrams do not display any $\zeta_2$\footnote{Strictly speaking, $\zeta_2$ can appear in some formulas such as sum rules in deep-inelastic scattering. They originate from an analytic continuation \cite{Kotikov:2005gr,Kazakov:1987jk,Kotikov:1994mi} of certain special forms of p-integrals. We will not consider this case in the present study.}. Therefore, it is convenient to choose a scale which also eliminates $\zeta_2$ from the intermediate steps of the calculation. For this purpose, we considered two different scales in ref.~\cite{Kotikov:2019bqo}. 
	
The first is the popular $G$-scale \cite{Chetyrkin:1980pr}. Following ref. \cite{Broadhurst:1999xk}, we used a slight modification of this scale in ref. \cite{Kotikov:2019bqo}. We refer to this variant as the $g$-scale, which removes an additional factor of $1/(1-2\ep)$ from the one-loop result.

Moreover, in ref.~\cite{Kotikov:2019bqo} we introduced a new scale which is based on calculations of massless diagrams performed by Vladimirov who appended \cite{Vladimirov:1979zm} an additional factor $\Gamma(1-\ep)$ to each loop contribution. The latter corresponds multiplying the scale by a factor of $\Gamma^{-1}(1-\ep)$. Following ref. \cite{Kotikov:2019bqo}, we refer to this scale as the minimal Vladimirov-scale, or MV-scale, which is defined as
\begin{equation}
\mu_{{\rm MV}}^{2\ep}=\frac{\tilde{\mu}^{2\ep}}{\Gamma(1-\ep)} \, .
\label{Vla}
\end{equation}
The use of the MV-scale leads to simpler results compared with the $g$-scale. Hence, the MV-scale is more appropriate for our analysis and all our subsequent results are will be given in this scale. Differences arising from the use of the $g$-scale can be found in ref.~\cite{Kotikov:2019bqo}. 

In the MV-scale, we can rewrite the result (\ref{axi}) in the following general form:\footnote{In the case of scalar QED the factor  $(1-(m+1)\ep)/(1-(m+l+1)\ep)$ is absent (see \cite{Kotikov:2019bqo}).}
\begin{equation}
a_m(\xi) = a_m(\eta)
\sum_{l=0}^{\infty} \, \frac{1-(m+1)\ep}{1-(m+l+1)\ep} \,
\Phi_{\rm MV}(m,l,\ep) 
\, \frac{(\Delta \, A)^l}{(-\ep)^l l!} \, {\left(\frac{\mu_{\rm MV}^2}{p^2}\right)}^{l\ep} \, ,
\label{axi.1}
\end{equation}
where
\begin{equation}
\Phi_{{\rm MV}}(m,l,\ep)=\frac{\Gamma(1-(m+1)\ep)\Gamma(1+(m+l)\ep)\Gamma^{2l}(1-\ep)}{\Gamma(1+m\ep)\Gamma(1-(m+l+1)\ep)} \, .
\label{Phi:V:def}
\end{equation}
In eq.~(\ref{axi.1}), the factor $(1-(m+1)\ep)/(1-(m+l+1)\ep)$ has been explicitly isolated from $\Phi_{\rm MV}(m,l,\ep)$ in order to ensure uniform transcendental level, i.e.\ the same value of $s$ for $\zeta_s$ at every order of the $\ep$-expansion of $\Phi_{\rm MV}(m,l,\ep)$ (see below).

\subsubsection{MV-scale}

The $\Gamma$-function $\Gamma(1+\beta\ep)$ has the following expansion:

\begin{equation}
\Gamma(1+\beta\ep) = \exp \Big[ -\gamma \beta \ep + \sum_{s=2}^{\infty}\, (-1)^s \, \eta_s \beta^s \ep^s \Bigr],~~ 
\eta_s = \frac{\zeta_s}{s} \, .
\label{Gamma:exp}
\end{equation}
Substituting eq.~(\ref{Gamma:exp}) into eq.~(\ref{Phi:V:def}) yields the following form for the factor $\Phi_{{\rm MV}}(m,l,\ep)$:
\begin{equation}
\Phi_{{\rm MV}}(m,l,\ep)= \exp \Big[ \sum_{s=2}^{\infty}\,\eta_s \, p_s(m,l) \, \ep^s \Bigr]\, ,
\label{Phi:V}
\end{equation}
where
\begin{equation}
p_s(m,l)=  (m+1)^s-(m+l+1)^s + 2l + 
(-1)^s \Bigl\{(m+l)^s-m^s\Bigr\}\, ,
\label{ps:V}
\end{equation}
and, as expected from the MV-scale, we have:
\begin{equation}
p_1(m,l)=0,~~ p_2(m,l)=0 \, .
\label{ps:V:1-2}
\end{equation}

As can be seen from eq.~(\ref{Phi:V}), $\Phi_{{\rm MV}}(m,l,\ep)$ contains $\zeta_s$ values of a given weight (or transcendental level) $s$ as a prefactor of $\ep^s$. This property strongly constrains the coefficients of the $\ep$-series thereby simplifying our analysis. It is reminiscent of the property previously found in ref.~\cite{Kotikov:2000pm}. When judiciously used, it sometimes allows one to derive results without any calculations (as demonstrated in ref.~\cite{Kotikov:2002ab,Kotikov:2004er,Bianchi:2013sta}). In other cases, it simplifies the structure of the results, which can then be predicted as an ansatz in a highly straightforward manner (see refs.~\cite{Fleischer:1998nb,Kotikov:2007cy,Bajnok:2008qj,Lukowski:2009ce,Marboe:2014sya,Marboe:2016igj} and the review in \cite{Kotikov:2010gf}).

\subsection{Solution to the recurrence relations}
\label{sec:LKF:proof}

The polynomial \(p_s(m,l)\) from eq.~(\ref{ps:V}) is separated into contributions from even and odd $s$ values, leading to the following recurrence relations:
\begin{equation}
	p_{2k} = p_{2k-1} + L p_{2k-2} + p_{3}, ~~ p_{2k-1} = p_{2k-2} + L p_{2k-3} + p_{3}, ~~  L=l(l+1) \, . 
\label{p2k-1:V}
\end{equation}
Specific to the MV-scheme, these relations  depend only on $L$, which leads to significant simplifications.

Indeed, by representing $p_{2k} $ as
\begin{equation}
p_{2k}=\sum_{s=2}^{k} p_{2s-1} \, C_{2k,2s-1} \, =  \sum_{m=1}^{k-1} p_{2k-2m+1} \, C_{2k,2k-2m+1} \, ,
\label{p2k:V:general}
\end{equation}
it is possible to determine the exact $k$-dependence of the coefficients $C_{2k,2s-1}$, which take the following form:
\begin{equation}
C_{2k,2k-2m+1} = b_{2m-1}
\, \frac{(2k)!}{(2m-1)! \, (2k-2m+1)!} \, ,
\label{C:V:structure}
\end{equation}
where the coefficients $b_{2m-1}$ are given by
\begin{equation}
b_{2m-1} = \frac{(2^{2m} - 1)}{m} \, B_{2m} \, .
\label{b:V:expressionB}
\end{equation}
Here, $B_{2m}$ are the Bernoulli numbers.

\subsection{All-order $\epsilon$-expansion of hatted $\zeta$-values}

At this stage, it is convenient to express the argument of the exponential on the r.h.s.\ of eq. (\ref{Phi:V}) as follows:
\begin{equation}
\sum_{s=3}^{\infty}\,\eta_s \, p_s \, \ep^s = \sum_{k=2}^{\infty}\,\eta_{2k} \, p_{2k} \, \ep^{2k} +
\sum_{k=2}^{\infty}\,\eta_{2k-1} \, p_{2k-1} \, \ep^{2k-1} \, .
\label{Phi:V:exp}
\end{equation}
Using eq.~(\ref{p2k:V:general}), the first term on the r.h.s.\ of eq.~(\ref{Phi:V:exp}) can be expressed as:
\begin{equation}
\sum_{k=2}^{\infty}\,\eta_{2k} \, p_{2k} \, \ep^{2k} = \sum_{k=2}^{\infty}\,\eta_{2k}  \, \ep^{2k} \, \sum_{s=2}^k p_{2s-1} \,
C_{2k,2s-1} =
\sum_{s=2}^{\infty} p_{2s-1} \, \sum_{k=s}^{\infty}\,\eta_{2k} \, C_{2k,2s-1} \, \ep^{2k} \, .
\label{Phi:V:exp1}
\end{equation}
Consequently, eq.~(\ref{Phi:V:exp}) can be written as $\sum_{s=2}^{\infty}\,\hat{\eta}_{2s-1} \, p_{2s-1} \, \ep^{2s-1}$, where
\begin{equation}
\hat{\eta}_{2s-1} = \eta_{2s-1} +  \sum_{k=s}^{\infty}\,\eta_{2k} \, C_{2k,2s-1} \, \ep^{2(k-s)+1},~~
C_{2k,2s-1} = b_{2k-2s+1} \, \frac{(2k)!}{(2s-1)! \, (2k-2s+1)!} \, .
\label{hEta}
\end{equation}
Thus, eq.~(\ref{Phi:V}) can be represented as:
\begin{equation}
\Phi_{{\rm MV}}(m,l,\ep) = \exp \Big[ \sum_{s=2}^{\infty}\,\hat{\eta}_{2s-1} \, p_{2s-1} \, \ep^{2s-1} \Bigr]
= \exp \Big[ \sum_{s=2}^{\infty}\,\frac{\hat{\zeta}_{2s-1}}{2s-1} \, p_{2s-1} \, \ep^{2s-1} \Bigr] \, ,
\label{Phi:V2}
\end{equation}
where
\begin{equation}
\hat{\zeta}_{2s-1} = \zeta_{2s-1} +  \sum_{k=s}^{\infty}\,\zeta_{2k} \, \hat{C}_{2k,2s-1} \, \ep^{2(k-s)+1} 
\label{hZeta}
\end{equation}
with
\begin{equation}
\hat{C}_{2k,2s-1} = \frac{2s-1}{2k} \, C_{2k,2s-1}  
= b_{2k-2s+1} \, \frac{(2k-1)!}{(2s-2)! \, (2k-2s+1)!} \, .
\label{hC}
\end{equation}
Together with eqs. (\ref{hC}) and (\ref{b:V:expressionB}), eq.~(\ref{hZeta}) provides an exact expression for the hatted $\zeta$-values in terms of the standard ones, valid to all orders in $\ep$. Moreover, eq.~(\ref{hZeta}) leads to a very compact integral representation for $\hat{\zeta}_{2m+1}$
\begin{equation}
\hat{\zeta}_{2m+1} =\frac{1}{(2m)!} \, \int_0^1 \, \frac{dt }{1-t}\,\frac{2}{t^{\ep}+1} \, \ln^{2m}t\,,  \label{hzeta2m1}
\end{equation}
i.e.\ the transform $\zeta_{2m+1} \to \hat{\zeta}_{2m+1}$ corresponds to the simple substitution $1 \to 2/(t^{\epsilon}+1)$ within the respective integrands (see the integral representation for $\zeta_{2m+1}$ in eq.~\eqref{zeta2m1}). The result (\ref{hzeta2m1}) is proven in appendix~\ref{app:LeeIntegral}.

\section{\label{sec:hatMZVs}${\varepsilon}$-dependence of hatted multiple zeta values}

\subsection{\label{sec:MZVnotation}Notation for MZVs, their products and $\varepsilon$-expansions}

The $\ep$-expansion of MZV \eqref{MZVdef} is postulated in the form
\begin{align}\label{eq:zetaexpansatz}
	\hat\zeta_{\boldsymbol{\sigma}} = \sum_{n \geqslant 0} \hat\zeta_{\boldsymbol{\sigma}}^{(n)} \ep^n,
	\qquad
	\hat\zeta_{\boldsymbol{\sigma}}^{(n)} = \sum_{2m + \lvert \boldsymbol{\Lambda} \rvert = \lvert \boldsymbol \sigma \rvert + n} \hat\zeta_{\boldsymbol{\sigma}(n)}^{2m;\boldsymbol{\Lambda}} \zeta_{2m;\boldsymbol{\Lambda}}.
\end{align} 
Here, the coefficients $\hat\zeta_{\boldsymbol{\sigma}(n)}^{2m;\boldsymbol{\Lambda}}$ are rational numbers; $m$ is a non-negative integer, $m \in \mathbb{N}$, bold lowercase Greek letters denote multi-indices, e.g., $\boldsymbol{\sigma}=n_1,n_2,\dots n_{d(\boldsymbol{\sigma})}$, $1 \leqslant n_i \in \mathbb{N}$ for $i$ running from 1 to the depth of the multi-index $d(\boldsymbol{\sigma})$ (the depth of MZV is the same as the depth of the corresponding multi-index), and bold uppercase Greek letters denote multi-multi-indices (or multi$^2$-indices) $\boldsymbol{\Lambda} = \boldsymbol{\lambda}_1; \dots ; \boldsymbol{\lambda}_L$. MZV with a multi$^2$-index is defined as a product of MZVs with its constituent multi-indices, e.g.,
\begin{align}\label{eq:zetaexpansatz2}
	\zeta_{2m;\boldsymbol{\Lambda}}
	=
	\zeta_{2m} \zeta_{\boldsymbol{\Lambda}}
	=
	\zeta_{2m} \prod_{a=1}^{L} \zeta_{\boldsymbol{\lambda_a}}.
\end{align} 
The sums $\lvert \boldsymbol{\lambda}_a \rvert = \sum_{i=1}^{d(\boldsymbol{\lambda}_a)} n_{ai}$ and $\lvert \boldsymbol{\Lambda} \rvert = \sum_{a=1}^L \lvert \boldsymbol{\lambda}_a \rvert$ are the weights of the multi-index $\boldsymbol{\lambda}_a=n_{a1},n_{a2},\dots n_{ad(\boldsymbol{\lambda}_a)}$ and the multi$^2$-index $\boldsymbol{\Lambda}$, respectively. Once again, we consider the weight $\lvert \boldsymbol{\lambda} \rvert $  of the multi-index $\boldsymbol{\lambda}$ and weight of MZV $\zeta_{\boldsymbol{\lambda}}$ to be equivalent. Furthermore, we will use the notion of transcendental weight of an MZV product $\zeta_{\boldsymbol{\Lambda}}$ defined as the weight $\lvert \boldsymbol{\Lambda} \rvert$ of the multi$^2$-index $\boldsymbol{\Lambda}$.

In what follows, if an MZV product is written in the form \eqref{eq:zetaexpansatz}, $\zeta_{2m;\boldsymbol{\Lambda}}$ with an explicitly isolated even index $2m$, it is implied that all the Riemann zeta functions of even arguments in the product have been combined into $\zeta_{2m}$. Put in other words, no sub-multi-index $\boldsymbol{\lambda}_a$, $a=1, \dots, L$, of the multi$^2$-index $\boldsymbol{\Lambda}$ is a single even integer.  

In eq.~\eqref{eq:zetaexpansatz}, only those $\zeta_{2m;\boldsymbol{\Lambda}}$ are allowed that belong to the subset of $\mathbb{Q}$-linearly independent products of MZVs of weight not exceeding $ \lvert \boldsymbol \sigma \rvert + n$. We use the sets of algebraically independent MZVs compiled in \cite{BBV,Kuipers:2011vqa}, which list them explicitly up to weight 28.

The specific values of the rational coefficients $\hat\zeta_{\boldsymbol{\sigma}(n)}^{2m;\boldsymbol{\Lambda}}$ in the expansion \eqref{eq:zetaexpansatz} are determined by the transcendental structure of the two-point master integrals, which can also be expanded in terms of the minimal-depth basis of MZVs~\cite{BBV,Kuipers:2011vqa}. We turn to this topic in the next subsection.

%==================================================
%==================================================
%==================================================

\subsection{\label{sec:masters}Four-loop master-integrals for massless two-point functions up to transcendental weight 16}

\begin{table}
\begin{center}
\begin{tabular}{ >{\centering\arraybackslash}p{0.08\textwidth} >{\centering\arraybackslash}p{0.08\textwidth}  >{\centering\arraybackslash}p{0.08\textwidth}  >{\centering\arraybackslash}p{0.08\textwidth} >{\centering\arraybackslash}p{0.08\textwidth} >{\centering\arraybackslash}p{0.08\textwidth} >{\centering\arraybackslash}p{0.08\textwidth} >{\centering\arraybackslash}p{0.08\textwidth} >{\centering\arraybackslash}p{0.08\textwidth}  }\hline\hline
\multicolumn{9}{c}{complexity level 0}
 \\\hline
$M_{0,1}$ & $M_{1,1}$  & $M_{1,2}$ &  $M_{1,3}$  &  $M_{1,4}$  & $M_{2,3}$  &  $M_{2,4}$  &  $M_{2,5}$ &  $M_{3,1}$
 \\
\parbox[c][0.08\textwidth][c]{0.08\textwidth}{\includegraphics[width=0.08\textwidth]{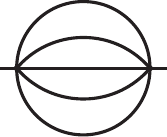}}  
&  \parbox[c][0.08\textwidth][c]{0.08\textwidth}{\includegraphics[width=0.08\textwidth]{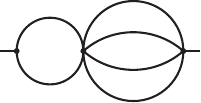}}   
&  \parbox[c][0.08\textwidth][c]{0.08\textwidth}{\includegraphics[width=0.08\textwidth]{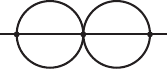}} 
& \parbox[c][0.08\textwidth][c]{0.08\textwidth}{\includegraphics[width=0.08\textwidth]{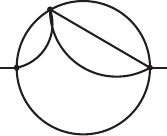}}  
& \parbox[c][0.08\textwidth][c]{0.08\textwidth}{\includegraphics[width=0.08\textwidth]{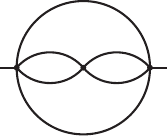}}
& \parbox[c][0.08\textwidth][c]{0.08\textwidth}{\includegraphics[width=0.08\textwidth]{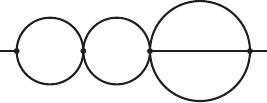}}
& \parbox[c][0.08\textwidth][c]{0.08\textwidth}{\includegraphics[width=0.08\textwidth]{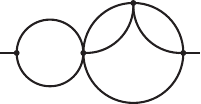}}
& \parbox[c][0.08\textwidth][c]{0.08\textwidth}{\includegraphics[width=0.08\textwidth]{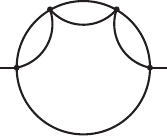}}
& \parbox[c][0.08\textwidth][c]{0.08\textwidth}{\includegraphics[width=0.08\textwidth]{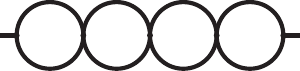}}
\\
\hline\hline
\multicolumn{9}{c}{complexity level 1}
 \\\hline
 \multicolumn{9}{c}{
\begin{tabular}{c c c c c c c c}
$M_{2,1}$ & $M_{2,2}$  & $M_{2,6}$ &  $M_{2,7}$  &  $M_{3,2}$  & $M_{3,3}$  &  $M_{3,4}$  &  $M_{3,6}$ 
 \\
\parbox[c][0.08\textwidth][c]{0.08\textwidth}{\includegraphics[width=0.08\textwidth]{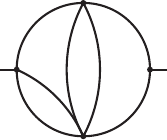}}  
&  \parbox[c][0.08\textwidth][c]{0.08\textwidth}{\includegraphics[width=0.08\textwidth]{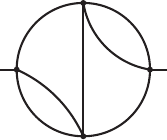}}   
&  \parbox[c][0.08\textwidth][c]{0.08\textwidth}{\includegraphics[width=0.08\textwidth]{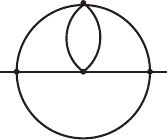}} 
& \parbox[c][0.08\textwidth][c]{0.08\textwidth}{\includegraphics[width=0.08\textwidth]{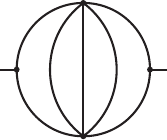}}  
& \parbox[c][0.08\textwidth][c]{0.08\textwidth}{\includegraphics[width=0.08\textwidth]{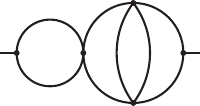}}
& \parbox[c][0.08\textwidth][c]{0.08\textwidth}{\includegraphics[width=0.08\textwidth]{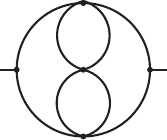}}
& \parbox[c][0.08\textwidth][c]{0.08\textwidth}{\includegraphics[width=0.08\textwidth]{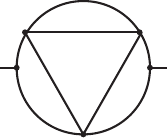}}
& \parbox[c][0.08\textwidth][c]{0.08\textwidth}{\includegraphics[width=0.08\textwidth]{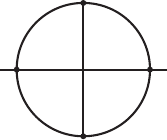}}
\end{tabular}
 }
\\
\hline\hline
\multicolumn{9}{c}{complexity level 2}
 \\\hline
 \multicolumn{9}{c}{
\begin{tabular}{c c c c c c}
$M_{3,5}$ & $M_{4,1}$  & $M_{4,2}$ &  $M_{4,3}$  &  $M_{4,4}$  & $M_{5,2}$  
 \\
\parbox[c][0.08\textwidth][c]{0.08\textwidth}{\includegraphics[width=0.08\textwidth]{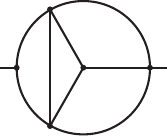}}  
&  \parbox[c][0.08\textwidth][c]{0.08\textwidth}{\includegraphics[width=0.08\textwidth]{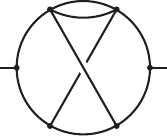}}   
&  \parbox[c][0.08\textwidth][c]{0.08\textwidth}{\includegraphics[width=0.08\textwidth]{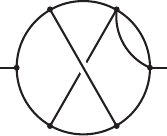}} 
& \parbox[c][0.08\textwidth][c]{0.08\textwidth}{\includegraphics[width=0.08\textwidth]{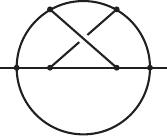}}  
& \parbox[c][0.08\textwidth][c]{0.08\textwidth}{\includegraphics[width=0.08\textwidth]{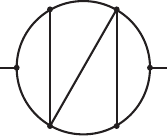}}
& \parbox[c][0.08\textwidth][c]{0.08\textwidth}{\includegraphics[width=0.08\textwidth]{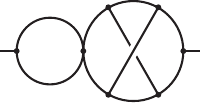}}
\end{tabular}
 }
\\
\hline\hline
\multicolumn{9}{c}{complexity level 3}
 \\\hline
 \multicolumn{9}{c}{
\begin{tabular}{c c c }
$M_{4,5}$ & $M_{5,1}$  & $M_{6,1}$ 
 \\
\parbox[c][0.08\textwidth][c]{0.08\textwidth}{\includegraphics[width=0.08\textwidth]{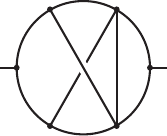}}  
&  \parbox[c][0.08\textwidth][c]{0.08\textwidth}{\includegraphics[width=0.08\textwidth]{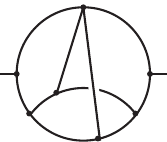}}   
&  \parbox[c][0.08\textwidth][c]{0.08\textwidth}{\includegraphics[width=0.08\textwidth]{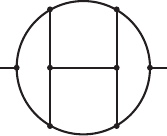}} 
\end{tabular} }
 \\
\hline\hline
\multicolumn{9}{c}{complexity level 4}
 \\\hline
\multicolumn{9}{c}{
\begin{tabular}{c c}
$M_{6,2}$ & $M_{6,3}$ 
 \\
\parbox[c][0.08\textwidth][c]{0.08\textwidth}{\includegraphics[width=0.08\textwidth]{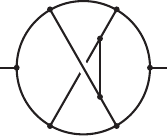}}  
&  \parbox[c][0.08\textwidth][c]{0.08\textwidth}{\includegraphics[width=0.08\textwidth]{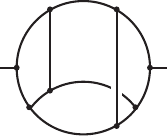}}   
\end{tabular}
}
\\\hline\hline
\end{tabular}
\end{center}
\caption{Four-loop master-integral topologies for massless two-point functions according to \cite{Baikov:2005nv}. The notion of complexity level was introduced in \cite{Lee:2010cga,Lee:2011jt}}
\end{table}

There are 28 master-integrals for massless four-loop two-point functions $M_{i,k}$, where $i=0,\dots 6$ is 5 less than the number of propagators in the corresponding master integral \cite{Baikov:2005nv}. The master integrals can be classified by their complexity level\footnote{The notion of complexity level was first introduced for  three-loop master integrals in \cite{Lee:2010cga}}, which ranges from 0 to 4. An integral of complexity level $n$ is expressed in terms of $n$-fold sums at most. 
In appendix~\ref{app:homrepMI} we construct a basis 
\begin{equation}
	M^\prime_{i,k} = C_{i,k}(\ep)M_{i,k}+\sum_p C_{i,k}^{p}(\ep)M_p,
\label{eq:Relatiion}
\end{equation}
of functions possesing the property of uniform transcendentality (UT). Here $C_{i,k}(\ep)$ and $\ C_{i,k}^{p}(\ep)$ are rational functions of $\ep$ and the sum goes over a subset of master integrals obtained from $M_{i,k}$ by contracting some lines. For $i=1,\ldots,5$ these functions have been presented already in ref.~\cite{Lee:2011jt}, while for the last three integrals, $M_{6,k}$ with $k=1,\ldots,3$ we construct these combinations by guessing the coefficient from the deep $\ep$ expansion that we have at hand. After this construction was completed, we have learned about paper \cite{Bianchi:2025sjc}, where the UT basis was expressed via integrals with squared propagators and/or numerators. We have checked that after IBP reduction the basis of ref. \cite{Bianchi:2025sjc} is expressed via linear combinations of $M^\prime_{i,k}$ with constant coefficients.
For each $M^\prime_{i,k}$ every coefficient of the expansion in $\varepsilon$ is a $\mathbb{Q}$-linear combination of MZV products of the same weight, with the weight increasing by one at each subsequent order in $\varepsilon$ \cite{Lee:2011jt}. This property makes the UT basis convenient for constructing the $\varepsilon$-expansions \eqref{eq:zetaexpansatz} of MZVs. 

We have obtained the $\varepsilon$-expansions of all 28 master integrals $M'_{i,k}$ up to the 16th power of the regularization parameter, i.e.\ up to transcendental weight 16. At this order, the master integrals are found not to contain any transcendental constants other than MZVs. The explicit expressions in terms of the minimal-depth basis of refs.~\cite{BBV,Kuipers:2011vqa} are listed in the auxiliary files of the arXiv submission.

\subsection{The Riemann zeta functions of even arguments in the master-integral transcendental structure}

All nine master integrals of complexity level 0 are expressed in terms of the Euler gamma functions and, therefore, contain only Riemann zeta functions $\zeta_n$, $n =2$, $3, \dots$ in their $\varepsilon$ expansions \cite{Lee:2011jt}. These master integrals can be expressed through $\hat \zeta_n $ in eqs.~\eqref{hZeta} and \eqref{hC} entirely:
\begin{gather}\label{eq:Mhatz}
M_{i,k}\left( \left\{ \zeta_{n} \right\}; \varepsilon\right) = M_{i,k}\left( \left\{ \hat\zeta_{n} \right\} ; \varepsilon \right),
\\  \label{eq:KTformula}
\hat\zeta_{2s} = 0,
\qquad
\hat\zeta_{2s-1} = \zeta_{2s-1} + \sum_{k\geqslant s} \zeta_{2k} \frac{2(2k-1)!\left(2^{2(k-s+1)}-1\right)}{(2s-2)!(2k-2s+2)!}  B_{2(k-s+1)} \varepsilon^{2(k-s)+1},
\end{gather}
where $B_n$ are the Bernoulli numbers.

For the remaining 19 integrals, their $\varepsilon$ expansions contain MZVs of depths greater than 1. If we temporarily set aside all terms proportional to $\zeta_{12}$, $\zeta_{14}$, and $\zeta_{16}$, we can introduce circumflexed MZVs \eqref{eq:zetaexpansatz} so that the master integrals contain no ``even'' Riemann zetas up to weight 10, i.e., up to weight 16, the master integrals can be written as
\begin{multline}\label{eq:Mrem}
M'_{i,k}\left( \left\{\zeta_{\boldsymbol{\sigma}}\right\}; \varepsilon \right) = M'_{i,k}\left( \left\{\hat\zeta_{\boldsymbol{\sigma}}\right\}; \varepsilon \right) 
\\
+ c{}^{(12)}_{i,k} \zeta_{12} \varepsilon^{12}+ c{}^{(14)}_{i,k} \zeta_{14} \varepsilon^{14} + c{}^{(15)}_{i,k} \zeta_3 \zeta_{12} \varepsilon^{15} + c{}^{(16)}_{i,k} \zeta_{16} \varepsilon^{16} + O(\varepsilon^{17}) .
\end{multline}
The corresponding set of circumflexed MZVs contains 86 arbitrary parameters in general. However, 40 of these 86 parameters cancel out in the master integrals, meaning that the coefficients $c{}^{(a)}_{i,k}$ for $a=12$, 14, 15, 16 do not depend on them. These 40 parameters can be set to zero, which significantly simplifies the circumflexed MZVs. The corresponding expressions for the MZVs are given below and in an auxiliary file, while the general expressions are provided in an auxiliary file only. The depth-1 MZVs are given by eq.~\eqref{eq:KTformula}, and expressions for higher depths are as follows.

\textbf{Weight 8:}
\begin{align} \label{eq:zeta53}
\hat\zeta_{5,3} &{}= \left(\zeta _{5,3}-\frac{29}{12} \zeta _8\right)-\frac{15}{2} \zeta _4 \zeta _5 \varepsilon -\frac{2905}{376} \zeta _{10} \varepsilon ^2+\frac{25}{2} \zeta _5 \zeta _6 \varepsilon
   ^3 +\hat\zeta_{5,3(4)}^{12} \zeta _{12} \varepsilon ^4 -\frac{105}{2} \zeta _5 \zeta _8 \varepsilon ^5
\notag\\
&\hphantom{={}}+\hat\zeta_{5,3(6)}^{14} \zeta _{14} \varepsilon^6
  + \left( \hat{\zeta }_{5,3{(7)}}^{3;12} \zeta _3 \zeta _{12} +\frac{765}{2} \zeta _5 \zeta _{10}\right)  \varepsilon ^7
  + \hat{\zeta }_{5,3{(8)}}^{16}\zeta _{16} \varepsilon ^8
   +O\left(\varepsilon ^9 \right).
\end{align}

\textbf{Weight 10:}
\begin{align} \label{eq:zeta73}
 \hat\zeta_{7,3} &{}= \left( \zeta _{7,3}-\frac{793}{94} \zeta _{10} \right) - \left(15 \zeta _5 \zeta _6+21 \zeta _4 \zeta _7\right) \varepsilon + \hat\zeta_{7,3(2)}^{12} \zeta _{12} \varepsilon ^2
\notag\\
&\hphantom{={}}
+\left(35 \zeta _6 \zeta _7+\frac{105}{2} \zeta _5 \zeta _8\right) \varepsilon ^3
+\hat\zeta_{7,3(4)}^{14} \zeta _{14} \varepsilon ^4
\notag\\
&\hphantom{={}}
+ \varepsilon ^5 \left( \hat{\zeta }_{7,3{(5)}}^{3;12}\zeta _3 \zeta _{12} -147 \zeta _7 \zeta _8-378 \zeta _5 \zeta _{10} \right)+ \hat{\zeta}_{7,3{(6)}}^{16} \zeta _{16} \varepsilon ^6 
   +O\left(\varepsilon ^7\right).
\end{align}

\textbf{Weight 11:}
\begin{align} \label{eq:zeta533}
 \hat\zeta_{5,3,3} &{}= \left(\zeta _{5,3,3}-\frac{5}{2} \zeta _5 \zeta _6+3 \zeta _4 \zeta _7+45 \zeta _2 \zeta _9\right)+\left( \hat\zeta_{5,3,3(1)}^{12} \zeta _{12}+\frac{3}{2} \zeta _4 \zeta _{5,3}\right)
   \varepsilon -\frac{105}{16} \zeta _5 \zeta _8 \varepsilon ^2
\notag\\
&\hphantom{={}} +\left( \hat\zeta_{5,3,3(3)}^{14} \zeta _{14}-\frac{5}{2} \zeta _6 \zeta _{5,3}\right)  \varepsilon ^3
+ \left(\hat{\zeta }_{5,3,3{(4)}}^{3;12} \zeta _3 \zeta _{12} +\frac{165}{8} \zeta _5 \zeta _{10} \right) \varepsilon ^4
\notag\\
&\hphantom{={}}
+ \left( \hat{\zeta}_{5,3,3{(5)}}^{16} \zeta _{16} +\frac{21}{2} \zeta _8 \zeta _{5,3} \right) \varepsilon ^5 
+O\left(\varepsilon ^6\right).
\end{align}

\textbf{Weight 12:}
\begin{align} \label{eq:zeta93}
 \hat\zeta_{9,3} &{}= \left(\zeta _{9,3} +\hat\zeta_{9,3(0)}^{12} \zeta _{12}\right)+\left(-\frac{75}{2} \zeta _6 \zeta _7-21 \zeta _5 \zeta _8-\frac{81}{2} \zeta _4 \zeta _9\right) \varepsilon
+ \hat\zeta_{9,3(2)}^{14} \zeta _{14} \varepsilon ^2
\notag\\&\hphantom{={}}
+ \left( \hat{\zeta }_{9,3{(3)}}^{3;12} \zeta _3 \zeta _{12}
+\frac{525}{4} \zeta _7 \zeta _8 +\frac{135}{2} \zeta _6 \zeta _9 +126 \zeta _5 \zeta _{10} \right) \varepsilon ^3
+ \hat{\zeta }_{9,3{(4)}}^{16}\zeta_{16} \varepsilon ^4
   +O\left(\varepsilon ^5\right),
\end{align}
\begin{align} \label{eq:zeta6411}
 \hat\zeta_{6,4,1,1} &{}= \zeta _{6,4,1,1}-\frac{3}{2} \zeta _4 \zeta _{5,3}-3 \zeta _2 \zeta _{7,3}+\frac{9}{4} \zeta _6 \zeta _3^2+\frac{1}{2} \zeta _4 \zeta _5 \zeta _3-10 \zeta _2 \zeta _7 \zeta _3-\frac{7}{2} \zeta _2 \zeta _5^2
 \notag\\
&\hphantom{={}} + \hat\zeta_{6,4,1,1(0)}^{12} \zeta _{12}
\notag\\
&\hphantom{={}} +\frac{1}{96} \left(48 \zeta _4 \zeta _3^3-4794 \zeta _{10} \zeta _3+16995 \zeta _6 \zeta _7+9744 \zeta _5 \zeta _8+14392  \zeta _4 \zeta _9\right) \varepsilon
\notag\\
&\hphantom{={}} +\left(\frac{21}{16} \zeta _8 \zeta _3^2+ \hat\zeta_{6,4,1,1(2)}^{14} \zeta _{14}\right) \varepsilon ^2
\notag\\
&\hphantom{={}}
+ \left( \hat{\zeta }_{6,4,1,1{(3)}}^{3;12} \zeta _3 \zeta _{12} -\frac{5}{6} \zeta _6 \zeta _3^3-\frac{39655}{64} \zeta _7 \zeta _8 -\frac{8995}{36} \zeta _6 \zeta _9 -609 \zeta _5 \zeta _{10} \right) \varepsilon ^3 
\notag\\
&\hphantom{={}}
+ \left( \hat{\zeta}_{6,4,1,1{(4)}}^{16} \zeta _{16} -\frac{33}{8} \zeta _3^2 \zeta _{10}\right) \varepsilon ^4 
+O\left(\varepsilon^5\right).
\end{align}

\textbf{Weight 13:}
\begin{align} \label{eq:zeta553}
\hat\zeta_{5,5,3} &{}= \left(\zeta _{5,5,3} +\frac{145}{12} \zeta _5 \zeta _8+25 \zeta _4 \zeta _9+\frac{275}{2} \zeta _2 \zeta _{11} \right)
+\left(\frac{75}{4} \zeta _4 \zeta _5^2+\hat\zeta_{5,5,3(1)}^{14} \zeta _{14}\right) \varepsilon 
\notag\\
&\hphantom{={}}
+ \left( \hat{\zeta }_{5,5,3{(2)}}^{3;12} \zeta _3 \zeta _{12} +\frac{14525}{376} \zeta _5 \zeta _{10} \right) \varepsilon ^2
+ \left( \hat{\zeta}_{5,5,3{(3)}}^{16} \zeta _{16} -\frac{125}{4} \zeta _5^2 \zeta _6 \right) \varepsilon^3
+O\left(\varepsilon ^4 \right),
\end{align}
\begin{align} \label{eq:zeta733}
 \hat\zeta_{7,3,3} &{}= \zeta _{7,3,3}-4 \zeta _6 \zeta _7+\frac{29}{2} \zeta _5 \zeta _8+28 \zeta _4 \zeta _9+\frac{407}{2} \zeta _2 \zeta _{11}
\notag\\ 
&\hphantom{={}}
+\left(27 \zeta _4 \zeta_5^2+ \hat\zeta_{7,3,3(1)}^{14} \zeta _{14}+\frac{3}{2} \zeta _4 \zeta _{7,3} \right) \varepsilon
\notag\\ 
&\hphantom{={}}
+ \left( \hat{\zeta }_{7,3,3{(2)}}^{3;12} \zeta _3 \zeta _{12} -\frac{147}{8} \zeta _7 \zeta _8+\frac{8715}{188} \zeta _5 \zeta _{10} \right) \varepsilon ^2 
\notag\\ 
&\hphantom{={}}
+\left( \hat{\zeta }_{7,3,3{(3)}}^{16} \zeta _{16} -\frac{5}{2} \zeta _6 \zeta _{7,3}-45 \zeta _6 \zeta _5^2\right) \varepsilon ^3 
+O\left(\varepsilon ^4\right).
\end{align}

\textbf{Weight 14:}
\begin{align} \label{eq:zeta95}
 \hat\zeta_{9,5} &{}= \zeta _{9,5} + \hat\zeta_{9,5(0)}^{14} \zeta _{14} 
+   \left( \hat{\zeta }_{9,5{(1)}}^{3;12}\zeta _3 \zeta _{12}
 -\frac{105}{2} \zeta _7 \zeta _8-\frac{345}{2} \zeta _6 \zeta _9 -\frac{495}{2} \zeta _4 \zeta_{11} \right) \varepsilon
\notag\\&\hphantom{={}}
 + \hat{\zeta }_{9,5{(2)}}^{16}\zeta _{16} \varepsilon ^2
 +O\left(\varepsilon^3\right),
\end{align}
\begin{align} \label{eq:zeta11,3}
 \hat\zeta_{11,3} &{}= \zeta _{11,3} + \hat\zeta_{11,3(0)}^{14} \zeta _{14}
+ \left( \hat{\zeta }_{11,3{(1)}}^{3;12} \zeta _3 \zeta _{12}
 -\frac{105}{2} \zeta _7 \zeta _8-70 \zeta _6 \zeta _9-27 \zeta _5 \zeta _{10}-66 \zeta _4 \zeta_{11}\right) \varepsilon  
\notag\\&\hphantom{={}}
+ \hat{\zeta }_{11,3{(2)}}^{16} \zeta _{16} \varepsilon ^2
  +O\left(\varepsilon^3\right),
\end{align}
\begin{align}\label{eq:zeta5333}
 \hat\zeta_{5,3,3,3} &{}= \zeta _{5,3,3,3}+\frac{1}{2} \zeta _6 \zeta _{5,3}+\frac{10}{3} \zeta _2 \zeta _{9,3}+20 \zeta _2 \zeta _5 \zeta _7 + \hat\zeta_{5,3,3,3(0)}^{14} \zeta_{14} 
\notag\\
&\hphantom{={}} 
+ \left( \hat{\zeta }_{5,3,3,3{(1)}}^{3;12} \zeta _3 \zeta _{12} +\frac{3}{2} \zeta _4 \zeta _{5,3,3}-\frac{693}{4} \zeta _7 \zeta _8-\frac{3105 }{8} \zeta _6 \zeta_9 -\frac{93}{8} \zeta _5 \zeta _{10} -\frac{45}{2}  \zeta _4 \zeta _{11} \right) \varepsilon
\notag\\
&\hphantom{={}}
+ \left( \hat{\zeta }_{5,3,3,3{(2)}}^{16} \zeta _{16} +\frac{21}{16} \zeta _8 \zeta _{5,3}\right) \varepsilon ^2 
+O(\varepsilon^3).
\end{align}

\textbf{Weight 15:}
\begin{align} \label{eq:zeta735}
 \hat\zeta_{7,3,5} &{}=  \zeta _{7,3,5}-\frac{287}{6} \zeta _7 \zeta _8-17 \zeta _6 \zeta _9-\frac{2520}{47} \zeta _5 \zeta _{10}+55 \zeta _4 \zeta _{11}+\frac{1001 }{2}\zeta _2 \zeta _{13}+ \hat{\zeta }_{7,3,5{(0)}}^{3;12} \zeta _3 \zeta _{12} 
 \notag\\ 
&\hphantom{={}}
 + \left( \hat{\zeta }_{7,3,5{(1)}}^{16} \zeta_{16} +\frac{5}{2} \zeta _6 \zeta _{7,3}-45 \zeta_6 \zeta _5^2-126 \zeta _4 \zeta _7 \zeta_5 \right) \varepsilon +O\left(\varepsilon ^2\right),
\end{align}
\begin{align} \label{eq:zeta933}
 \hat\zeta_{9,3,3} &{}=\zeta _{9,3,3}+\frac{157}{4} \zeta _7 \zeta _8+\frac{29}{2} \zeta _6 \zeta _9+\frac{2379}{47} \zeta _5 \zeta _{10}+126 \zeta_4 \zeta _{11}+\frac{1209}{2} \zeta _2 \zeta _{13}+\hat{\zeta}_{9,3,3{(0)}}^{3;12} \zeta _3 \zeta _{12}
 \notag\\ 
&\hphantom{={}}
 +\left(45 \zeta _5^2 \zeta _6+135 \zeta _4 \zeta _5 \zeta _7+\frac{3}{2} \zeta _4 \zeta _{9,3}+\hat{\zeta}_{9,3,3{(1)}}^{16} \zeta_{16}\right) \varepsilon 
 +O\left(\varepsilon ^2\right),
\end{align}
\begin{align}\label{eq:zeta64311}
 \hat\zeta_{6,4,3,1,1} &{}= \zeta _{6,4,3,1,1} + \hat{\zeta }_{6,4,3,1,1{(0)}}^{3;12} \zeta _3 \zeta _{12} -\frac{5}{2} \zeta _4 \zeta _3 \zeta _{5,3}-3 \zeta _2 \zeta _3 \zeta _{7,3}+\frac{21}{2} \zeta _2 \zeta _5 \zeta _{5,3}+10 \zeta _4 \zeta_{5,3,3}
\notag\\
&\hphantom{={}}
 -\frac{14}{5} \zeta _2 \zeta _{5,5,3}+2 \zeta _2 \zeta _{7,3,3}-\frac{3}{4} \zeta _6 \zeta _3^3-\frac{25}{4} \zeta _4 \zeta _5 \zeta _3^2-16 \zeta _2 \zeta _7 \zeta _3^2-\frac{9}{2} \zeta _2 \zeta_5^2 \zeta _3
 \notag\\
&\hphantom{={}}
-\frac{7811}{144} \zeta _7 \zeta _8+\frac{84185}{144} \zeta _6 \zeta _9-\frac{210685}{1128} \zeta _5 \zeta_{10}-\frac{27199}{24} \zeta _4 \zeta _{11}-\frac{56717}{120} \zeta _2 \zeta _{13} 
\notag\\
&\hphantom{={}}
 + \biggl( \hat{\zeta }_{6,4,3,1,1{(1)}}^{16} \zeta _{16} +70 \zeta _8 \zeta_{5,3}-\frac{35725}{448} \zeta _6 \zeta _{7,3}+65 \zeta _4 \zeta _{9,3}+15 \zeta _4 \zeta _{6,4,1,1}
\notag\\
&\hphantom{{}={} + \biggl( }
 -\frac{11}{8} \zeta _4 \zeta _3^4-\frac{3989}{32} \zeta _{10} \zeta _3^2-\frac{8225}{32} \zeta _6 \zeta _7 \zeta_3+\frac{245}{4} \zeta _5 \zeta _8 \zeta _3+\frac{541}{24} \zeta _4 \zeta _9 \zeta _3
\notag\\	
&\hphantom{{}={} + \biggl( }
 -\frac{169475}{448} \zeta_5^2 \zeta _6-\frac{4095}{16} \zeta _4 \zeta _5 \zeta _7 \biggr) \varepsilon
+O(\varepsilon^2).
\end{align}

\textbf{Weight 16:}
\begin{align} \label{eq:zeta11,5}
 \hat\zeta_{11,5} &{}= \zeta _{11,5} + \hat\zeta_{11,5(0)}^{16} \zeta _{16} +O\left(\varepsilon\right),
\end{align}
\begin{align} \label{eq:zeta13,3}
 \hat\zeta_{13,3} &{}= \zeta _{13,3} + \hat\zeta_{13,3(0)}^{16} \zeta _{16} +O\left(\varepsilon\right),
\end{align}
\begin{align}\label{eq:zeta5533}
 \hat\zeta_{5,5,3,3} &{}= \zeta _{5,5,3,3}+\hat{\zeta }_{5,5,3,3{(0)}}^{16} \zeta _{16} -\frac{10}{3} \zeta _4 \zeta _{9,3}+25
   \zeta _2 \zeta _{9,5}-\frac{175}{2} \zeta _2 \zeta _{11,3} 
\notag\\
&\hphantom{={}}
+\frac{25}{4} \zeta _6 \zeta _5^2-35 \zeta _4 \zeta _7 \zeta _5-750 \zeta _2 \zeta _9 \zeta _5-\frac{1665}{4} \zeta _2 \zeta _7^2+O(\varepsilon),
\end{align}
\begin{align}\label{eq:zeta7333}
 \hat\zeta_{7,3,3,3} &{}= \zeta _{7,3,3,3}+\hat{\zeta }_{7,3,3,3{(0)}}^{16} \zeta _{16} +\frac{1}{2} \zeta _6 \zeta_{7,3}+\frac{43}{3} \zeta _2 \zeta _{9,5}-\frac{89}{2} \zeta _2 \zeta _{11,3}+9 \zeta _6 \zeta _5^2-18 \zeta _4
   \zeta _7 \zeta _5
\notag\\
&\hphantom{={}}
   -537 \zeta _2 \zeta _9 \zeta _5-\frac{1031}{4} \zeta _2 \zeta _7^2+O(\varepsilon),
\end{align}
\begin{align}\label{eq:zeta8611}
 \hat\zeta_{8,6,1,1} &{}= \zeta _{8,6,1,1}+\hat{\zeta }_{8,6,1,1{(0)}}^{16} \zeta _{16} +\frac{9}{10} \zeta _8 \zeta_{5,3}+\frac{1}{4} \zeta _6 \zeta _{7,3}-\frac{35}{9} \zeta _4 \zeta _{9,3}-\frac{16}{3} \zeta _2 \zeta _{9,5}+8  \zeta _2 \zeta _{11,3}
\notag\\
&\hphantom{={}}
   +\frac{569}{188} \zeta _{10} \zeta _3^2+18 \zeta _6 \zeta _7 \zeta _3+10 \zeta _5 \zeta _8
   \zeta _3-\frac{5}{2} \zeta _4 \zeta _9 \zeta _3-\frac{329}{2} \zeta _2 \zeta _{11} \zeta _3+31 \zeta _2 \zeta_7^2-\zeta _5^2 \zeta _6
\notag\\
&\hphantom{={}}
-\frac{245}{6} \zeta _4 \zeta _5 \zeta _7-\frac{11}{2} \zeta _2 \zeta _5 \zeta _9+O(\varepsilon).
\end{align}

The MZVs with hats $\hat\zeta_{\boldsymbol{\sigma}}$ defined in eqs.~\eqref{eq:zeta53}--\eqref{eq:zeta8611} coincide with the results in \cite{Baikov:2018gap,Baikov:2019zmy} up to transcendental weight 12. As can be observed, $\hat\zeta_{\boldsymbol{\sigma}}$ contain $\pi$-dependent terms even at $\epsilon=0$, unlike the depth-1 MZVs $\hat\zeta_{2s-1}$. The likely reason is that more appropriate representations for the master integrals can be expressed in terms of so-called ``non-alternating'' MZVs \cite{Schnetz:2008mp,Broadhurst:1996az,Borinsky:2022lds}, whereas the $\pi$-dependent terms arise when transitioning from non-alternating to alternating MZVs. Indeed, for the Vasiliev integral \cite{VPH}, for instance, the result---when expressed in terms of non-alternating MZVs---turns out to be significantly simpler (see \cite{Broadhurst:1996yc} and the discussion therein).
 
In eqs.~\eqref{eq:zeta53}--\eqref{eq:zeta8611}, 46 parameters $\hat\zeta_{\boldsymbol{\sigma}(n)}^{12}$, $\hat\zeta_{\boldsymbol{\sigma}(n)}^{14}$, $\hat\zeta_{\boldsymbol{\sigma}(n)}^{3;12}$, and $\hat\zeta_{\boldsymbol{\sigma}(n)}^{16}$ remain unfixed. No choice of these parameters can simultaneously eliminate the $\zeta _{12}$ terms across all master integrals or their linear combinations, which is consistent with the observations reported in \cite{Baikov:2018gap,Baikov:2019zmy}. The same holds true for the $\zeta _{14}$ and $\zeta_3 \zeta_{12}$ terms. In contrast, the situation for the $\zeta _{16}$ terms is different: we can make all of them vanish by imposing 19 linear constraints on the free parameters, albeit at the cost of introducing peculiarly cumbersome rational numbers as coefficients in the constraints. The corresponding MZVs with hats are compiled in two auxiliary files---one for the most general expressions, and another for expressions where all parameters that do not contribute directly to the master integral remainders are set to zero.

Even when eliminating $\zeta _{16}$ from  the master integrals, we retain several free parameters---specifically, 2 $\hat\zeta_{\boldsymbol{\sigma}(n)}^{12}$, 10 $\hat\zeta_{\boldsymbol{\sigma}(n)}^{14}$, 13 $\hat\zeta_{\boldsymbol{\sigma}(n)}^{3;12}$, and  2 $\hat\zeta_{\boldsymbol{\sigma}(n)}^{16}$. In principle, these parameters can be used to further simplify the master integrals. For example, we can remove the $\zeta_3\zeta_{12}$ terms and set $c{}^{(15)}_{i,k}=0$ in the remainders of 14 out of the 19 master integrals~\eqref{eq:Mrem}.

It should be noted that a special role for $\zeta_{12}$ has already been observed in \cite{Schnetz:2022nsc,Borinsky:2022lds} as a result of direct analytical calculations of complicated convergent 7-loop p-integrals. Nevertheless, the distinct roles of $\zeta_{14}$, $\zeta_3\zeta_{12}$, and $\zeta_{16}$ require further investigation.

%==================================================
%==================================================
%==================================================

\section{\label{sec:PiConstraints}Constraints on the transcendental structure of RG functions}
\setcounter{equation}{0}

\subsection{Definitions of RG functions}

The renormalization procedure eliminates ultraviolet divergences and yields a finite expression for the Green's functions of a local quantum field theory. In theories belonging to the class of renormalizable ones, this procedure amounts to a multiplicative modification of the Lagrangian parameters, masses and coupling constants, and to multiplying the Green's function $\Gamma$ by a renormalization constant $Z_\Gamma$. The logic and notation of this subsection and the next one closely follow the brief yet comprehensive MIPT lectures by D.~I. Kazakov. Hence, we refer the reader to ref.~\cite{Kazakov:2008tr} for any technical nuances.

The relationship between the ``bare'' and renormalized Green's functions in dimensionally regularized ($d=4-2\varepsilon$) theory is given by:
\begin{align}\label{inv}
\Gamma_\text{bare} \left[ \mathscr{M}, \frac{1}{\varepsilon}, g_\text{bare} \right]
= 
Z_\Gamma^{-1} \left[ \frac{1}{\varepsilon} , g(\mu) \right]
\Gamma\left[ \mathscr{M}, \mu ^2, g(\mu) \right], 
\end{align}
where $\mathscr{M}$ is the set of all independent Mandelstam-like kinematic invariants that can be constructed from external momenta, $g$ is the coupling constant renormalized such that
\begin{align}
g_\text{bare} = g(\mu) Z_g \left[ \frac{1}{\varepsilon} , g(\mu) \right].
\end{align}
For brevity, we constrain ourselves to the case of a massless single-charge theory.

Moving the renormalization constant to the l.h.s.\ of eq.~\eqref{inv} and differentiating both sides with respect to the dimensional transmutation parameter $\mu^2$ yields a differential equation for the renormalized Green's function:
\begin{align}\label{rg}
\left[ \mu^2 \frac{\partial }{\partial \mu ^2} + \beta (g)\frac{\partial }{\partial  g} + \gamma_\Gamma (g) \right]
\Gamma \left[ \mathscr{M},\mu ^2,g(\mu) \right] =0.
\end{align}
Here, we have defined the beta function $\beta(g)$ as
\begin{align} \label{db}
\beta(g) = \left. \mu^2  \frac{d g}{d \mu ^2}\right\vert_{g_\text{bare}}
\end{align}
and the anomalous dimension $\gamma_\Gamma(g)$ of the Green's function as
\begin{align}\label{dg}
\gamma_\Gamma (g) = \left.-\mu^2 \frac{d \ln Z_\Gamma }{d \mu ^2} \right\vert_{g_\text{bare}}. 
\end{align}
Equation~(\ref{rg}) is the all-important Callan--Symanzik equation. Its practical significance lies in the fact that its solution is an improved expression for the Green's function, encapsulating an infinite series of Feynman diagrams.

\subsection{Pole equations for the renormalization constants in dimensional regularization}

Let us now consider the calculation of the $\beta$-function and anomalous dimensions within the framework of dimensional regularization and the minimal subtraction scheme. Moving to a fractional space-time dimension $4-2\ep$ shifts of the mass dimension of the ``bare'' coupling, $[g_\text{bare}]=2\ep$. Consequently, a factor of $(\mu^2)^{\ep}$ is inserted into the relation between the ``bare'' and renormalized coupling:
\begin{equation}\label{charged}
  g_\text{bare} = (\mu ^2)^{\ep} g(\mu) Z_g \left[ \frac{1}{\varepsilon} , g(\mu) \right].
\end{equation}
Differentiating the equation above with respect to $\ln (\mu^2)$, one can obtain that
\begin{align}\label{betad}
\beta_{4-2\ep}(g) =  -\ep g+\beta_4(g),
\end{align}
where $\displaystyle \beta_{4-2\ep}(g) = \frac{dg}{d\ln \mu^2}$ and $\displaystyle \beta_4(g) = \beta(g) = g\frac{d\ln Z_g}{d\ln \mu^2}$.

In the $\overline{\text{MS}}$ scheme, all renormalization constants are unity plus the principal part of a Laurent series in $\varepsilon$:
\begin{align}\label{zz}
Z_\Gamma &{}= 1+ \sum_{n \geqslant 1} \frac{c_n(g)}{\varepsilon ^n},
\\
\label{gg}
g Z_g &{}= g + \sum_{n \geqslant 1} \frac{a_n(g)}{\varepsilon^n},
\end{align}
where the pole coefficients are expanded in powers of the coupling constant,
\begin{align}\label{eq:cn&an.coupling.exp}
c_n(g)=\sum_{m \geqslant n} c_{nm}g^m,
\qquad
 a_n(g) = \sum_{m \geqslant n} a_{nm} g^{m+1}.
\end{align}

Differentiating the renormalization constant of the Green's function~(\ref{zz}) with respect to $\ln \mu^2 $ and recalling the
definitions~(\ref{db}) and (\ref{dg}), one finds the following relation:
\begin{equation}
-\left[1+ \sum_{n \geqslant 1} \frac{c_n(g)}{\varepsilon ^n}\right]
\gamma_\Gamma(g)=\left[-\varepsilon g+\beta(g)\right]\frac{d}{dg}
\sum_{n \geqslant 1} \frac{c_n(g)}{\varepsilon ^n}\,,
\label{Equ1}
\end{equation}
where we have used the fact that the l.h.s.\ of eq.~\eqref{zz} implies
\begin{equation}
\frac{dZ_\Gamma}{d\ln\mu^2} =-\gamma_\Gamma  Z_\Gamma=-\gamma_\Gamma \left[1+ \sum_{n \geqslant 1} \frac{c_n(g)}{\varepsilon ^n}\right],
\end{equation}
while the r.h.s.\ gives
\begin{equation}
\beta_{4-2\ep}(g)\frac{d}{dg} \sum_{n \geqslant 1} \frac{c_n(g)}{\varepsilon ^n}=\left[-\varepsilon g+\beta(g)\right]\frac{d}{dg} \sum_{n \geqslant 1} \frac{c_n(g)}{\varepsilon ^n}.
\end{equation}

Matching the coefficients of the same-order poles in eq.~(\ref{Equ1}) leads to the 't~Hooft pole equations for the anomalous dimension:
\begin{align}
g\frac{d}{dg}c_1(g)&{}= \gamma_\Gamma (g) , \label{polgam}
\\
g\frac{d}{dg}c_n(g)&{}=\left[\gamma_\Gamma(g) + \beta (g)\frac{d}{dg}\right]c_{n-1}(g),  \qquad n\geqslant 2 .
\label{polgam1}
\end{align}
It is clear from the recursive equations above that the higher-pole coefficients $c_n$ for $n\geqslant 2$ are determined by the $\beta$-function and the coefficient of the simple pole, $c_1$.

A similar recursion holds for the pole coefficients of the $\beta$-function. Matching the coefficients of equal powers of $\varepsilon$ in the result of differentiating eq.~(\ref{gg}) with respect to $\ln \mu ^2$, one obtains the pole equations
\begin{align}\label{polbet}
\left(g\frac{d}{dg}-1\right)a_1(g)&{}= \beta (g),
\\ \label{poln}
\left(g\frac{d}{dg}-1\right)a_n(g)&{}= \beta (g)\frac{d}{dg}a_{n-1}(g), \qquad n \geqslant 2\,.
\end{align}
Equations \eqref{polgam}--\eqref{poln} show that the simple-pole coefficients, $a_1$ and $c_1$, are sufficient to uniquely reconstruct all the coefficients of the higher-order poles.

Indeed, using $a_n(g)$ from eq.~(\ref{eq:cn&an.coupling.exp}) and eq.~(\ref{polbet}), we have
\begin{equation}
\beta(g)=\sum_{m \geqslant 1} \beta_m g^{m+1},
\qquad
\beta_m=ma_{1,m}\,.
\label{betaA}
\end{equation}
From eq.~(\ref{poln}), a straightforward algebraic evaluation gives the following recursive results for the coefficients $a_{n,m}$ of the higher-order poles:
\begin{equation}
m a_{n,m}=\sum_{l=n-1}^{m-1} \, (l+1)(m-l) a_{n-1,l}a_{1,m-l} = \sum_{l=n-1}^{m-1} \, (l+1)a_{n-1,l}\beta_{m-l}\,.
\label{poln.1}
\end{equation}

Analogously, with $c_n(g)$ in eq.~(\ref{eq:cn&an.coupling.exp}) and eq.~(\ref{polgam}), the anomalous dimension takes the form
\begin{equation}
\gamma_{\Gamma}(g)=\sum_{m \geqslant 1} \gamma_m g^{m},
\qquad
\gamma_m=mc_{1,m}\,.
\label{gammaA}
\end{equation}

Using eq.~(\ref{polgam1}), the higher-order pole coefficients $c_{n,m}$ can be rewritten as
\begin{equation}
m c_{n,m}=\sum_{l=n-1}^{m-1} \,(m-l)c_{n-1,l} \bigl[la_{1,m-l}+c_{1,m-l}\bigr] = \sum_{l=n-1}^{m-1} \, a_{n-1,l}\bigl[l\beta_{m-l}+\gamma_{m-l}\bigr]\,.
\label{polng.1}
\end{equation}

Equations~(\ref{poln.1}) and (\ref{polng.1}) allow one to determine all coefficients $a_{n,m}$ and $c_{n,m}$, thereby constructing the renormalization constants
$Z_g$ and $Z_\Gamma$. Following refs.~\cite{Baikov:2018wgs,Baikov:2018gap,Baikov:2019zmy}, we can consider $Z_g$ and $Z_\Gamma$ in two different schemes: one utilizing
standard zeta values and another utilizing hatted zeta values.\footnote{In the present study, we restrict ourselves to theories with a single coupling constant. Applications to multi-coupling theories can be found in \cite{Jack:2023pxz}.}
The schemes are related by a finite transformation and all coefficients in front of singularities should be zero. This requirement produces a multitude of relations between RG functions of lower orders
and the $\pi$-dependent terms of higher-order RG functions. These relations lead to predictions of the $\pi$-dependent terms of the RG function of higher orders.
Examples demonstrating the extraction of the simplest relations are provided in appendix~\ref{app:simprels}.
The explicit lists of constraints for $\beta$-functions and ADs up to 8 loops are presented in the two subsequent subsections.
The relations at the 9-loop level for $\beta$-functions and ADs are collected in appendix~\ref{app:9Lrels}, while complete lists of relations extending up to the 13-loop level are provided in the supplementary files included with the arXiv version of this paper.

\subsection{Notation for the rational prefactors of MZVs}

The RG functions for $n\geqslant 3$ are expanded over the same set of $\mathbb{Q}$-linearly independent products of MZVs used for the master integrals and hatted MZV expansions in section~\ref{sec:hatMZVs}~\cite{BBV,Kuipers:2011vqa}:
\begin{align}\label{eq:Fexpansatz}
	F_{n} = F^{\text{R}}_{n} + \sum_{\lvert\boldsymbol{\Lambda}\rvert \leqslant 2n-3} F_{n}^{\zeta_{\boldsymbol{\Lambda}}} \zeta_{\boldsymbol{\Lambda}} + \text{non-MZV transcendental terms},
	\qquad
	F = \beta, \; \gamma.
\end{align} 
Here, $F^{\text{R}}_{n} \in \mathbb{Q}$ is the rational part of the $n$-loop coefficient of the RG function $F$, and $F_{n}^{\zeta_{\boldsymbol{\Lambda}}} \in \mathbb{Q}$ is the rational prefactor of the MZV product $\zeta_{\boldsymbol{\Lambda}}$. The notation for multi-indices $\boldsymbol{\Lambda}$ and weights $\lvert\boldsymbol{\Lambda}\rvert$ was introduced in section~\ref{sec:MZVnotation}. The weight-28 minimal-depth basis of refs.~\cite{BBV,Kuipers:2011vqa} is sufficient to express the RG function expansions~\eqref{eq:Fexpansatz} up to the 15-loop approximation, $n=15$. 

The first two coefficients are purely rational, the third can contain $\zeta_3$, and the four-loop coefficient is the first to include an even zeta value, namely $\zeta_4$:
\begin{align}
	F_{1} &= F^{\text{R}}_{1},
	\\
	F_{2} &= F^{\text{R}}_{2},
	\\
	F_{3} &= F^{\text{R}}_{3} + F^{\zeta_3}_{3} \zeta_3,
	\\
	F_{4} &= F^{\text{R}}_{4} + F^{\zeta_3}_{4} \zeta_3 + F^{\zeta_4}_{4} \zeta_4 + F^{\zeta_5}_{4} \zeta_5.
\end{align} 
Consequently, the first constraint arises at four loops, for the rational prefactor $F^{\zeta_4}_{4}$.

Note also that the relations for the rational prefactors in the $\beta$-function can always be obtained by substituting $\gamma \to \beta$ in all terms of the corresponding relations for the AD:
\begin{align}
\beta_{n}^{\zeta_{2m;\boldsymbol{\Lambda}}} = \gamma_{n}^{\zeta_{2m;\boldsymbol{\Lambda}}} \bigr\rvert_{\gamma^{\zeta_{\boldsymbol{\Lambda}}}_n \to \beta^{\zeta_{\boldsymbol{\Lambda}}}_n,\,\gamma^{\text{R}}_n \to \beta^{\text{R}}_n}\,.
\end{align} 
As stated in the previous section, we were unable to find a choice of parameters for which the terms proportional to $\zeta_{12}$ and $\zeta_{14}$ can be fully absorbed into the circumflexed MZVs, leaving no remainder in the master integrals. This implies that there is, in principle, a wide range of possible $\varepsilon$ expansions for the MZVs. For this reason, from now on, we leave the corresponding parameters unfixed.

\subsection{Relations for the $\beta$-function up to eight loops}

\noindent\textbf{Four-loop constraint:}
\begin{align}
  \beta^{\zeta_4}_4 &= \beta^{\zeta_3}_3\beta_1\,.
  \label{b4}
\end{align}

\noindent\textbf{Five-loop constraints:}
\begin{align}
\beta^{\zeta_4}_5 &= \frac{9}{8} \,\beta^{\zeta_3}_4 \beta_1+\frac{1}{2} \,\beta^{\zeta_3}_3 \beta_2
\qquad
\beta^{\zeta_6}_5 = \frac{15}{8} \,\beta^{\zeta_5}_4 \beta_1,
\qquad
\beta^{\zeta_3\zeta_4}_5 =0.
\end{align}

\noindent\textbf{Six-loop constraints:}
\begin{align}
\beta^{\zeta_4}_6 &= \frac{6}{5} \,\beta^{\zeta_3}_5 \beta_1+\frac{3}{4} \,\beta^{\zeta_3}_4 \beta_2,~~\beta^{\zeta_6}_6 = 2 \,\beta^{\zeta_5}_5 \beta_1
+\frac{5}{4} \,\beta^{\zeta_5}_4 \beta_2-\beta^{\zeta_3}_3 \beta^3_1,~~\beta^{\zeta_3\zeta_4}_6 =\frac{12}{5} \,\beta^{\zeta^2_3}_5 \beta_1,~~\nonumber \\
\beta^{\zeta_8}_6 &+ \frac{29}{12} \, \beta^{\zeta_{5,3}}_6 = \frac{14}{5} \,\beta^{\zeta_7}_5 \beta_1,~~\beta^{\zeta_4\zeta_5}_6 =0,~~\beta^{\zeta_3\zeta_6}_6 =0.
\label{eq:6LbetaConst}
\end{align}

\noindent\textbf{Seven-loop constraints:}
\begin{align}
\beta_7^{\zeta_2\zeta_9} &= 45\beta_7^{\zeta_{5,3,3}}\,, \nonumber \\
\beta_7^{\zeta_4} &= \frac{5}{4}\beta_{1}\beta_6^{\zeta_3} + \frac{9}{10}\beta_{2}\beta_5^{\zeta_3} + \frac{3}{8}\beta^{\text{R}}_{3}\beta_4^{\zeta_3} - \frac{1}{2}\beta^{\text{R}}_{4}\beta_3^{\zeta_3}\,, \nonumber \\
\beta_7^{\zeta_3\zeta_4} &= \frac{5}{2}\beta_{1}\beta_6^{\zeta_3^2} + \frac{9}{5}\beta_{2}\beta_5^{\zeta_3^2} - \frac{1}{8}\beta_3^{\zeta_3}\beta_4^{\zeta_3}\,, \nonumber \\
\beta_7^{\zeta_3^2\zeta_4} &= \frac{15}{4}\beta_{1}\beta_6^{\zeta_3^3}\,, \qquad\qquad
\beta_7^{\zeta_4\zeta_7} = 3\beta_7^{\zeta_{5,3,3}}\,, \nonumber \\
\beta_7^{\zeta_4\zeta_5} &= -\frac{25}{4}\beta_{1}\beta_6^{\zeta_{5,3}} + \frac{5}{4}\beta_{1}\beta_6^{\zeta_3\zeta_5} - \frac{1}{2}\beta_3^{\zeta_3}\beta_4^{\zeta_5}\,, \nonumber \\
\beta_7^{\zeta_6} &= -2\beta_{1}^2\beta_{2}\beta_3^{\zeta_3} - \frac{5}{4}\beta_{1}^3\beta_4^{\zeta_3} + \frac{25}{12}\beta_{1}\beta_6^{\zeta_5} + \frac{3}{2}\beta_{2}\beta_5^{\zeta_5} + \frac{5}{8}\beta^{\text{R}}_{3}\beta_4^{\zeta_5}\,, \nonumber 
\\
\beta_7^{\zeta_3\zeta_6} &= \frac{25}{12}\beta_{1}\beta_6^{\zeta_3\zeta_5} + \frac{5}{8}\beta_3^{\zeta_3}\beta_4^{\zeta_5}\,, \nonumber \\
\beta_7^{\zeta_5\zeta_6} &= -\frac{5}{2}\beta_7^{\zeta_{5,3,3}}\,, \nonumber \\
\beta_7^{\zeta_8} &= -\frac{29}{12}\beta_7^{\zeta_{5,3}} - \frac{35}{8}\beta_{1}^3\beta_4^{\zeta_5} + \frac{7}{4}\beta_{1}^2\beta_5^{\zeta_3^2} - \frac{7}{24}\beta_{1}\left(\beta_3^{\zeta_3}\right)^2  + \frac{35}{12}\beta_{1}\beta_6^{\zeta_7}  + \frac{21}{10}\beta_{2}\beta_5^{\zeta_7}\,, \nonumber \\
\beta_7^{\zeta_3\zeta_8} &= -\frac{29}{12}\beta_7^{\zeta_3\zeta_{5,3}}\,, \nonumber \\
\beta_7^{\zeta_{10}} &= \frac{15}{4}\beta_{1}\beta_6^{\zeta_9} - \frac{793}{94}\beta_7^{\zeta_{7,3}}\,. \label{eq:7LbetaRels}
\end{align}

\noindent\textbf{Eight-loop constraints:}
\begin{align}
\beta_8^{\zeta_4} &= \frac{9}{7}\beta_{1}\beta_7^{\zeta_3} + \beta_{2}\beta_6^{\zeta_3} + \frac{3}{5}\beta^{\text{R}}_{3}\beta_5^{\zeta_3} - \beta^{\text{R}}_{5}\beta_3^{\zeta_3}\,, \nonumber \\
\beta_8^{\zeta_3\zeta_4} &= \frac{18}{7}\beta_{1}\beta_7^{\zeta_3^2} + 2\beta_{2}\beta_6^{\zeta_3^2} + \frac{6}{5}\beta^{\text{R}}_{3}\beta_5^{\zeta_3^2} - \frac{2}{5}\beta_3^{\zeta_3}\beta_5^{\zeta_3}\,, \nonumber \\
\beta_8^{\zeta_3^2\zeta_4} &= \frac{27}{7}\beta_{1}\beta_7^{\zeta_3^3} + 3\beta_{2}\beta_6^{\zeta_3^3} + \frac{1}{5}\beta_3^{\zeta_3}\beta_5^{\zeta_3^2}\,, \qquad\qquad
\beta_8^{\zeta_3^3\zeta_4} = 0\,, \nonumber \\
\beta_8^{\zeta_4\zeta_5} &= \frac{9}{7}\beta_{1}\beta_7^{\zeta_3\zeta_5} - \frac{45}{7}\beta_{1}\beta_7^{\zeta_{5,3}} + \beta_{2}\beta_6^{\zeta_3\zeta_5} - 5\beta_{2}\beta_6^{\zeta_{5,3}} - \beta_3^{\zeta_3}\beta_5^{\zeta_5}\,, \nonumber \\
\beta_8^{\zeta_3\zeta_4\zeta_5} &= \frac{18}{7}\beta_{1}\beta_7^{\zeta_3^2\zeta_5} - \frac{45}{7}\beta_{1}\beta_7^{\zeta_3\zeta_{5,3}} + \frac{1}{2}\beta_8^{\zeta_{6,4,1,1}}\,, \qquad\qquad
\beta_8^{\zeta_2\zeta_5^2} = -\frac{7}{2}\beta_8^{\zeta_{6,4,1,1}}\,, \nonumber \\
\beta_8^{\zeta_6} &= -\frac{10}{7}\beta_{1}^3\beta_5^{\zeta_3} - \frac{20}{7}\beta_{1}^2\beta_{2}\beta_4^{\zeta_3} - \frac{6}{7}\beta_{1}^2\beta^{\text{R}}_{3}\beta_3^{\zeta_3} - \frac{10}{7}\beta_{1}\beta_{2}^2\beta_3^{\zeta_3} \nonumber \\
&\hphantom{={}} + \frac{15}{7}\beta_{1}\beta_7^{\zeta_5} + \frac{5}{3}\beta_{2}\beta_6^{\zeta_5} + \beta^{\text{R}}_{3}\beta_5^{\zeta_5}\,, \nonumber 
\\
\beta_8^{\zeta_3\zeta_6} &= -\frac{20}{7}\beta_{1}^3\beta_5^{\zeta_3^2} - \frac{6}{7}\beta_{1}^2\left(\beta_3^{\zeta_3}\right)^2 + \frac{15}{7}\beta_{1}\beta_7^{\zeta_3\zeta_5} + \frac{5}{3}\beta_{2}\beta_6^{\zeta_3\zeta_5} + \beta_3^{\zeta_3}\beta_5^{\zeta_5}\,, \nonumber \\
\beta_8^{\zeta_3^2\zeta_6} &= \frac{15}{7}\beta_{1}\beta_7^{\zeta_3^2\zeta_5} + \frac{9}{4}\beta_8^{\zeta_{6,4,1,1}}\,, \qquad\qquad
\beta_8^{\zeta_2\zeta_{7,3}} = -3\beta_8^{\zeta_{6,4,1,1}}\,, \nonumber 
\\
\beta_8^{\zeta_5\zeta_6} &= \frac{30}{7}\beta_{1}\beta_7^{\zeta_5^2} - \frac{90}{7}\beta_{1}\beta_7^{\zeta_{7,3}} - \frac{5}{2}\beta_8^{\zeta_{5,3,3}}\,, \qquad\qquad
\beta_8^{\zeta_2\zeta_3\zeta_7} = -10\beta_8^{\zeta_{6,4,1,1}}\,, \nonumber 
\end{align}
\begin{align} 
\beta_8^{\zeta_4\zeta_7} &= \frac{9}{7}\beta_{1}\beta_7^{\zeta_3\zeta_7} - 18\beta_{1}\beta_7^{\zeta_{7,3}} - \beta_3^{\zeta_3}\beta_5^{\zeta_7} + 3\beta_8^{\zeta_{5,3,3}}\,, \nonumber 
\\
\beta_8^{\zeta_4\zeta_7} &= \frac{9}{7}\beta_{1}\beta_7^{\zeta_3\zeta_7} - 18\beta_{1}\beta_7^{\zeta_{7,3}} - \beta_3^{\zeta_3}\beta_5^{\zeta_7} + 3\beta_8^{\zeta_{5,3,3}}\,, \nonumber \\
\beta_8^{\zeta_4\zeta_{5,3}} &= \frac{9}{7}\beta_{1}\beta_7^{\zeta_3\zeta_{5,3}} + \frac{9}{7}\beta_{1}\beta_7^{\zeta_{5,3,3}} - \frac{3}{2}\beta_8^{\zeta_{6,4,1,1}}\,, \qquad\qquad
\beta_8^{\zeta_6\zeta_7} = -4\beta_8^{\zeta_{7,3,3}}\,, \nonumber \\
\beta_8^{\zeta_8} &= 3\beta_{1}^5\beta_3^{\zeta_3} - 5\beta_{1}^3\beta_5^{\zeta_5} - 10\beta_{1}^2\beta_{2}\beta_4^{\zeta_5} + \frac{15}{8}\beta_{1}^2\beta_6^{\zeta_3^2} - \frac{3}{4}\beta_{1}\beta_3^{\zeta_3}\beta_4^{\zeta_3} \nonumber \\
&\hphantom{={}} + \frac{11}{4}\beta_{1}\beta_{2}\beta_5^{\zeta_3^2} + 3\beta_{1}\beta_7^{\zeta_7} - \frac{7}{24}\beta_{2}\left(\beta_3^{\zeta_3}\right)^2 + \frac{7}{3}\beta_{2}\beta_6^{\zeta_7} + \frac{7}{5}\beta^{\text{R}}_{3}\beta_5^{\zeta_7} - \frac{29}{12}\beta_8^{\zeta_{5,3}}\,, \nonumber
\\
\beta_8^{\zeta_3\zeta_8} &= \frac{45}{8}\beta_{1}^2\beta_6^{\zeta_3^3} + 3\beta_{1}\beta_7^{\zeta_3\zeta_7} + \frac{7}{5}\beta_3^{\zeta_3}\beta_5^{\zeta_7} - \frac{29}{12}\beta_8^{\zeta_3\zeta_{5,3}}\,, \nonumber \\
\beta_8^{\zeta_5\zeta_8} &= -\frac{29}{12}\beta_8^{\zeta_5\zeta_{5,3}} + \frac{145}{12}\beta_8^{\zeta_{5,5,3}} + \frac{29}{2}\beta_8^{\zeta_{7,3,3}}\,, \qquad\qquad
\beta_8^{\zeta_2\zeta_9} = 45\beta_8^{\zeta_{5,3,3}}\,, \nonumber
\end{align}
\begin{align} 
\beta_8^{\zeta_4\zeta_9} &= 25\beta_8^{\zeta_{5,5,3}} + 28\beta_8^{\zeta_{7,3,3}}\,, \qquad\qquad
\beta_8^{\zeta_3\zeta_{10}} = -\frac{793}{94}\beta_8^{\zeta_3\zeta_{7,3}}\,, \nonumber \\
\beta_8^{\zeta_{10}} &= -12\beta_{1}^3\beta_5^{\zeta_7} + \frac{165}{56}\beta_{1}^2\beta_6^{\zeta_3\zeta_5} - \frac{2075}{376}\beta_{1}^2\beta_6^{\zeta_{5,3}} - \frac{33}{28}\beta_{1}\beta_3^{\zeta_3}\beta_4^{\zeta_5} \nonumber \\
&\hphantom{={}} + \frac{27}{7}\beta_{1}\beta_7^{\zeta_9} + 3\beta_{2}\beta_6^{\zeta_9} - \frac{793}{94}\beta_8^{\zeta_{7,3}}\,, \nonumber \\
\beta_8^{\zeta_2\zeta_{11}} &= \frac{275}{2}\beta_8^{\zeta_{5,5,3}} + \frac{407}{2}\beta_8^{\zeta_{7,3,3}}\,, \nonumber \\
\beta_8^{\zeta_{12}} &= \hat{\zeta}_{9,3(0)}^{{12}}\beta_8^{\zeta_{9,3}} + \hat{\zeta}_{6,4,1,1(0)}^{{12}} \beta_8^{\zeta_{6,4,1,1}} + \frac{33}{7}\beta_{1}\beta_7^{\zeta_{11}} - \frac{37323}{11056}\beta_{1}\beta_7^{\zeta_3\zeta_{5,3}} \nonumber \\
&\hphantom{={}} + \frac{375375}{22112}\hat{\zeta}_{5,3,3(1)}^{12}\beta_1\beta_7^{\zeta_{5,3,3}}\,, \label{b8_relations}
\end{align}
where
\begin{align}
\hat{\zeta}_{6,4,1,1(0)}^{12} &= \frac{347709}{103040}\hat{\zeta}_{7,3(2)}^{12} + \frac{171929}{220800}\hat{\zeta}_{5,3(4)}^{12} + \frac{2439715279523}{10252892160}\,, \label{eq:hz_6411_12} \\
\hat{\zeta}_{9,3(0)}^{12} &= \frac{9}{140}\hat{\zeta}_{7,3(2)}^{12} + \frac{3}{100}\hat{\zeta}_{5,3(4)}^{12} - \frac{24682093}{1547840}\,, \label{eq:hz_93_12} \\
\hat{\zeta}_{5,3,3(1)}^{12} &= \frac{32502567}{141015875}\hat{\zeta}_{7,3(2)}^{12} + \frac{52039901}{906530625}\hat{\zeta}_{5,3(4)}^{12} + \frac{34010353331}{1107615600}\,. \label{eq:hz_533_12}
\end{align}

The results (\ref{b4})--(\ref{b8_relations}) agree with the corresponding constraints obtained in ref.~\cite{Baikov:2019zmy}. There are only the following differences:
  $\beta^{\zeta_{8}}_n +(29/12)  \,\beta^{\zeta_{5,3}}_n$ with $n=6,7,8$, $\beta^{\zeta_{10}}_n +(793/94)  \,\beta^{\zeta_{7,3}}_n$ with $n=7,8$, and 
  $\beta^{\zeta_{3}\zeta_{8}}_n +(29/12)  \,\beta^{\zeta_{3}\zeta_{5,3}}_n$ with $n=7,8$ correspond to $\beta^{\zeta_{8}}_n$, $\beta^{\zeta_{10}}_n$, and $\beta^{\zeta_{3}\zeta_{8}}_n$ in
  ref.~\cite{Baikov:2019zmy}, respectively. Moreover, the results with $\beta^{\zeta_{a_1}...\zeta_{a_m}}_8$ with $\sum_{i=1}^{m} a_i\geqslant 12$ are new.

\subsection{Relations for the AD up to eight loops}

\noindent\textbf{Four-loop constraint:}
\begin{align}
  \gamma^{\zeta_4}_4 &= \frac{3}{2}\,\gamma^{\zeta_3}_3\beta_1-\frac{1}{2}\,\beta^{\zeta_3}_3\gamma_1.\label{c4a}
  \end{align}
  
\noindent\textbf{Five-loop constraints:}
\begin{align}
  \gamma^{\zeta_4}_5 &= 
\frac{3}{2} \,\sum_{i=3}^4 \gamma^{\zeta_3}_i \beta_{5-i}
  -\frac{3}{8} \,\beta^{\zeta_3}_4 \gamma_1- \,\beta^{\zeta_3}_3 \gamma_2,~~
 \gamma^{\zeta_6}_5 = \frac{5}{2} \,\gamma^{\zeta_5}_4 \beta_1-\frac{5}{8} \,\beta^{\zeta_5}_4 \gamma_1,~~ 
\gamma^{\zeta_3\zeta_4}_5 =0.
\end{align}

\noindent\textbf{Six-loop constraints:}
\begin{align}
\gamma^{\zeta_4}_6 &= \frac{3}{2} \,\sum_{i=3}^5 \gamma^{\zeta_3}_i \beta^{\text{R}}_{6-i}
-\frac{3}{10} \,\beta^{\zeta_3}_5 \gamma_1-\frac{3}{4} \,\beta^{\zeta_3}_4 \gamma_2-\frac{3}{2} \,\beta^{\zeta_3}_3 \gamma^{\text{R}}_{3},\nonumber \\
\gamma^{\zeta_6}_6 &= \frac{5}{2} \, \sum_{i=4}^5 \gamma^{\zeta_5}_i \beta_{6-i}-\frac{5}{2} \,\gamma^{\zeta_3}_3 \beta^3_1
-\frac{1}{2} \,\beta^{\zeta_5}_5 \gamma_1-\frac{5}{4} \,\beta^{\zeta_5}_4 \gamma_2+\frac{3}{2} \,\beta^{\zeta_3}_3 \beta^2_1\gamma_{1},\nonumber \\
\gamma^{\zeta_3\zeta_4}_6 &= 3 \,\gamma^{\zeta^2_3}_5 \beta_1-\frac{3}{5}\beta^{\zeta^2_3}_5 \gamma_1,~~
\gamma^{\zeta_8}_6+\frac{29}{12}\gamma^{\zeta_{5,3}}_6 = \frac{7}{2} \,\gamma^{\zeta_7}_5 \beta_1-\frac{7}{10} \,\beta^{\zeta_7}_5 \gamma_1,~~\nonumber \\
\gamma^{\zeta_4\zeta_5}_6 &=0,~~\gamma^{\zeta_3\zeta_6}_6 =0,
\label{eq:6LgamRels}
\end{align}
where $\beta_{\text{\text{R}}i}=\beta_{i}$ and $\gamma_{\text{\text{R}}i}=\gamma_{i}$ for $i=1,2$.

\noindent\textbf{Seven-loop constraints:}
\begin{align}
\gamma_7^{\zeta_4} &= \frac{3}{2} \,\sum_{i=3}^6 \gamma^{\zeta_3}_i \beta_{\text{R}(7-i)} - \frac{1}{4}\gamma_1\beta_6^{\zeta_3} - \frac{3}{5}\gamma_2\beta_5^{\zeta_3} - \frac{9}{8}\gamma^{\text{R}}_{3}\beta_4^{\zeta_3} - 2\gamma^{\text{R}}_{4}\beta_3^{\zeta_3}\,, \nonumber \\
\gamma_7^{\zeta_3\zeta_4} &= 3\beta_1\gamma_6^{\zeta_3^2} + 3\beta_2\gamma_5^{\zeta_3^2} + \frac{3}{8}\beta_4^{\zeta_3}\gamma_3^{\zeta_3} - \frac{1}{2}\beta_3^{\zeta_3}\gamma_4^{\zeta_3} - \frac{1}{2}\gamma_1\beta_6^{\zeta_3^2} - \frac{6}{5}\gamma_2\beta_5^{\zeta_3^2}\,, \nonumber \\
\gamma_7^{\zeta_3^2\zeta_4} &= \frac{9}{2}\beta_1\gamma_6^{\zeta_3^3} - \frac{3}{4}\gamma_1\beta_6^{\zeta_3^3}\,, \nonumber \\
\gamma_7^{\zeta_4\zeta_5} &= \frac{3}{2}\beta_1\gamma_6^{\zeta_3\zeta_5} - \frac{15}{2}\beta_1\gamma_6^{\zeta_{5,3}} + \frac{3}{2}\beta_4^{\zeta_5}\gamma_3^{\zeta_3} - 2\beta_3^{\zeta_3}\gamma_4^{\zeta_5} - \frac{1}{4}\gamma_1\beta_6^{\zeta_3\zeta_5} + \frac{5}{4}\gamma_1\beta_6^{\zeta_{5,3}}\,, \nonumber \\
\gamma_7^{\zeta_6} &= \frac{5}{2}  \,\sum_{i=4}^6 \gamma^{\zeta_5}_i \beta_{\text{R}(7-i)} -\frac{5}{2}\beta_1^3\gamma_4^{\zeta_3} + \frac{5}{4}\beta_1^2\gamma_1\beta_4^{\zeta_3} - \frac{15}{2}\beta_1^2\beta_2\gamma_3^{\zeta_3} + 3\beta_1^2\gamma_2\beta_3^{\zeta_3} + \frac{5}{2}\beta_1\beta_2\gamma_1\beta_3^{\zeta_3} \nonumber \\
&\hphantom{={}}  - \frac{5}{12}\gamma_1\beta_6^{\zeta_5} - \gamma_2\beta_5^{\zeta_5} - \frac{15}{8}\gamma^{\text{R}}_{3}\beta_4^{\zeta_5}\,, \nonumber \\
\gamma_7^{\zeta_3\zeta_6} &= \frac{5}{2}\beta_1\gamma_6^{\zeta_3\zeta_5} + \frac{5}{2}\beta_3^{\zeta_3}\gamma_4^{\zeta_5} - \frac{15}{8}\beta_4^{\zeta_5}\gamma_3^{\zeta_3} - \frac{5}{12}\gamma_1\beta_6^{\zeta_3\zeta_5}\,, \nonumber \\
\gamma_7^{\zeta_5\zeta_6} &= -\frac{5}{2}\gamma_7^{\zeta_{5,3,3}}\,, \qquad\qquad
\gamma_7^{\zeta_4\zeta_7} = 3\gamma_7^{\zeta_{5,3,3}}\,, \nonumber \\
\gamma_7^{\zeta_8} &= -\frac{35}{4}\beta_1^3\gamma_4^{\zeta_5} + \frac{35}{8}\beta_1^2\gamma_1\beta_4^{\zeta_5} + \frac{21}{8}\beta_1^2\gamma_5^{\zeta_3^2} - \frac{7}{8}\beta_1\gamma_1\beta_5^{\zeta_3^2} - \frac{7}{8}\beta_1\beta_3^{\zeta_3}\gamma_3^{\zeta_3} + \frac{7}{2}\beta_1\gamma_6^{\zeta_7} \nonumber \\
&\hphantom{={}} + \frac{7}{12}\gamma_1\left(\beta_3^{\zeta_3}\right)^2 - \frac{7}{12}\gamma_1\beta_6^{\zeta_7} + \frac{7}{2}\beta_2\gamma_5^{\zeta_7} - \frac{7}{5}\gamma_2\beta_5^{\zeta_7} - \frac{29}{12}\gamma_7^{\zeta_{5,3}}\,, \nonumber \\
\gamma_7^{\zeta_3\zeta_8} &= -\frac{29}{12}\gamma_7^{\zeta_3\zeta_{5,3}}\,, \qquad\qquad
\gamma_7^{\zeta_2\zeta_9} = 45\gamma_7^{\zeta_{5,3,3}}\,, \nonumber \\
\gamma_7^{\zeta_{10}} &= \frac{9}{2}\beta_1\gamma_6^{\zeta_9} - \frac{3}{4}\gamma_1\beta_6^{\zeta_9} - \frac{793}{94}\gamma_7^{\zeta_{7,3}}\,. 
\label{eq:7LgamRels}
\end{align}

\noindent\textbf{Eight-loop constraints:}
\begin{align}
\gamma_8^{\zeta_4} &=  \frac{3}{2} \,\sum_{i=3}^7 \gamma^{\zeta_3}_i \beta_{\text{R}(8-i)}  - \frac{3}{14}\gamma_1\beta_7^{\zeta_3} - \frac{1}{2}\gamma_2\beta_6^{\zeta_3} - \frac{9}{10}\gamma^{\text{R}}_{3}\beta_5^{\zeta_3} - \frac{3}{2}\gamma^{\text{R}}_{4}\beta_4^{\zeta_3} - \frac{5}{2}\gamma^{\text{R}}_{5}\beta_3^{\zeta_3}\,, \nonumber \\
\gamma_8^{\zeta_3\zeta_4} &= 3  \sum_{i=5}^7 \gamma^{\zeta^2_3}_i \beta_{\text{R}(8-i)}  + \frac{3}{5}\beta_5^{\zeta_3}\gamma_3^{\zeta_3} - \beta_3^{\zeta_3}\gamma_5^{\zeta_3}  - \frac{3}{7}\gamma_1\beta_7^{\zeta_3^2} - \gamma_2\beta_6^{\zeta_3^2} - \frac{9}{5}\gamma^{\text{R}}_{3}\beta_5^{\zeta_3^2}\,, \nonumber 
\\
\gamma_8^{\zeta_3^2\zeta_4} &= \frac{9}{2}\beta_1\gamma_7^{\zeta_3^3} + \frac{9}{2}\beta_2\gamma_6^{\zeta_3^3} + \frac{1}{2}\beta_3^{\zeta_3}\gamma_5^{\zeta_3^2} - \frac{3}{10}\beta_5^{\zeta_3^2}\gamma_3^{\zeta_3} - \frac{9}{14}\gamma_1\beta_7^{\zeta_3^3} - \frac{3}{2}\gamma_2\beta_6^{\zeta_3^3}\,, \nonumber \\
\gamma_8^{\zeta_4\zeta_5} &= \frac{3}{2}\beta_1\gamma_7^{\zeta_3\zeta_5} - \frac{15}{2}\beta_1\gamma_7^{\zeta_{5,3}} + \frac{3}{2}\beta_2\gamma_6^{\zeta_3\zeta_5} - \frac{15}{2}\beta_2\gamma_6^{\zeta_{5,3}} + \frac{3}{2}\beta_5^{\zeta_5}\gamma_3^{\zeta_3} + \frac{3}{2}\beta_4^{\zeta_5}\gamma_4^{\zeta_3} \nonumber \\
&\hphantom{={}} - \frac{3}{2}\beta_4^{\zeta_3}\gamma_4^{\zeta_5} - \frac{5}{2}\beta_3^{\zeta_3}\gamma_5^{\zeta_5} - \frac{3}{14}\gamma_1\beta_7^{\zeta_3\zeta_5} + \frac{15}{14}\gamma_1\beta_7^{\zeta_{5,3}} - \frac{1}{2}\gamma_2\beta_6^{\zeta_3\zeta_5} + \frac{5}{2}\gamma_2\beta_6^{\zeta_{5,3}}\,, \nonumber \\
\gamma_8^{\zeta_3\zeta_4\zeta_5} &= 3\beta_1\gamma_7^{\zeta_3^2\zeta_5} - \frac{15}{2}\beta_1\gamma_7^{\zeta_3\zeta_{5,3}} - \frac{3}{7}\gamma_1\beta_7^{\zeta_3^2\zeta_5} + \frac{15}{14}\gamma_1\beta_7^{\zeta_3\zeta_{5,3}} + \frac{1}{2}\gamma_8^{\zeta_{6,4,1,1}}\,, \nonumber \\
\gamma_8^{\zeta_2\zeta_5^2} &= -\frac{7}{2}\gamma_8^{\zeta_{6,4,1,1}}\,, \qquad\qquad
\gamma_8^{\zeta_2\zeta_3\zeta_7} = -10\gamma_8^{\zeta_{6,4,1,1}}\,, \nonumber 
\\
\gamma_8^{\zeta_6} &= -\frac{5}{2}\beta_1^3\gamma_5^{\zeta_3} + \frac{15}{14}\beta_1^2\gamma_1\beta_5^{\zeta_3} - \frac{15}{2}\beta_1^2\beta_2\gamma_4^{\zeta_3} + \frac{5}{2}\beta_1^2\gamma_2\beta_4^{\zeta_3} - \frac{15}{2}\beta_1^2\beta^{\text{R}}_{3}\gamma_3^{\zeta_3} + \frac{9}{2}\beta_1^2\gamma^{\text{R}}_{3}\beta_3^{\zeta_3} \nonumber \\
&\hphantom{={}} - \frac{15}{2}\beta_1\beta_2^2\gamma_3^{\zeta_3} + \frac{15}{7}\beta_1\beta_2\gamma_1\beta_4^{\zeta_3} + 5\beta_1\beta_2\gamma_2\beta_3^{\zeta_3} + \frac{15}{7}\beta_1\beta^{\text{R}}_{3}\gamma_1\beta_3^{\zeta_3} + \frac{15}{14}\beta_2^2\gamma_1\beta_3^{\zeta_3} \nonumber \\
&\hphantom{={}}  - \frac{5}{14}\gamma_1\beta_7^{\zeta_5} - \frac{5}{6}\gamma_2\beta_6^{\zeta_5} - \frac{3}{2}\gamma^{\text{R}}_{3}\beta_5^{\zeta_5} - \frac{5}{2}\gamma^{\text{R}}_{4}\beta_4^{\zeta_5} + \frac{5}{2} \sum_{i=4}^7 \gamma^{\zeta_5}_i \beta^{\text{R}}_{8-i} \,, \nonumber \\
\gamma_8^{\zeta_3\zeta_6} &= -5\beta_1^3\gamma_5^{\zeta_3^2} + \frac{15}{7}\beta_1^2\gamma_1\beta_5^{\zeta_3^2} - 3\beta_1^2\beta_3^{\zeta_3}\gamma_3^{\zeta_3} + \frac{15}{7}\beta_1\gamma_1\left(\beta_3^{\zeta_3}\right)^2 + \frac{5}{2}\beta_1\gamma_7^{\zeta_3\zeta_5} + \frac{5}{2}\beta_2\gamma_6^{\zeta_3\zeta_5} \nonumber \\
&\hphantom{={}} + \frac{5}{2}\beta_3^{\zeta_3}\gamma_5^{\zeta_5} + \frac{5}{2}\beta_4^{\zeta_3}\gamma_4^{\zeta_5} - \frac{5}{2}\beta_4^{\zeta_5}\gamma_4^{\zeta_3} - \frac{3}{2}\beta_5^{\zeta_5}\gamma_3^{\zeta_3} - \frac{5}{14}\gamma_1\beta_7^{\zeta_3\zeta_5} - \frac{5}{6}\gamma_2\beta_6^{\zeta_3\zeta_5}\,, \nonumber \\
\gamma_8^{\zeta_3^2\zeta_6} &= \frac{5}{2}\beta_1\gamma_7^{\zeta_3^2\zeta_5} - \frac{5}{14}\gamma_1\beta_7^{\zeta_3^2\zeta_5} + \frac{9}{4}\gamma_8^{\zeta_{6,4,1,1}}\,, \qquad\qquad
\gamma_8^{\zeta_2\zeta_9} = 45\gamma_8^{\zeta_{5,3,3}}\,, \nonumber \\
\gamma_8^{\zeta_5\zeta_6} &= 5\beta_1\gamma_7^{\zeta_5^2} - 15\beta_1\gamma_7^{\zeta_{7,3}} - \frac{5}{7}\gamma_1\beta_7^{\zeta_5^2} + \frac{15}{7}\gamma_1\beta_7^{\zeta_{7,3}} - \frac{5}{2}\gamma_8^{\zeta_{5,3,3}}\,, \nonumber \\
\gamma_8^{\zeta_4\zeta_7} &= \frac{3}{2}\beta_1\gamma_7^{\zeta_3\zeta_7} - 21\beta_1\gamma_7^{\zeta_{7,3}} + \frac{3}{2}\beta_5^{\zeta_7}\gamma_3^{\zeta_3} - \frac{5}{2}\beta_3^{\zeta_3}\gamma_5^{\zeta_7} - \frac{3}{14}\gamma_1\beta_7^{\zeta_3\zeta_7} + 3\gamma_1\beta_7^{\zeta_{7,3}} + 3\gamma_8^{\zeta_{5,3,3}}\,,
\nonumber \\
\gamma_8^{\zeta_4\zeta_{5,3}} &= \frac{3}{2}\beta_1\gamma_7^{\zeta_3\zeta_{5,3}} + \frac{3}{2}\beta_1\gamma_7^{\zeta_{5,3,3}} - \frac{3}{14}\gamma_1\beta_7^{\zeta_3\zeta_{5,3}} - \frac{3}{14}\gamma_1\beta_7^{\zeta_{5,3,3}} - \frac{3}{2}\gamma_8^{\zeta_{6,4,1,1}}\,, \nonumber \\
\gamma_8^{\zeta_2\zeta_{7,3}} &= -3\gamma_8^{\zeta_{6,4,1,1}}\,, \qquad\qquad
\gamma_8^{\zeta_2\zeta_3\zeta_{5,3}} = 0\,, \qquad\qquad
\gamma_8^{\zeta_6\zeta_7} = -4\gamma_8^{\zeta_{7,3,3}}\,, \nonumber \\
\gamma_8^{\zeta_8} &= \frac{21}{2}\beta_1^5\gamma_3^{\zeta_3} - \frac{15}{2}\beta_1^4\gamma_1\beta_3^{\zeta_3} - \frac{35}{4}\beta_1^3\gamma_5^{\zeta_5} - \frac{105}{4}\beta_1^2\beta_2\gamma_4^{\zeta_5} + \frac{15}{4}\beta_1^2\gamma_1\beta_5^{\zeta_5} + \frac{35}{4}\beta_1^2\gamma_2\beta_4^{\zeta_5} \nonumber \\
&\hphantom{={}} + \frac{21}{8}\beta_1^2\gamma_6^{\zeta_3^2} + \frac{15}{2}\beta_1\beta_2\gamma_1\beta_4^{\zeta_5} + \frac{21}{4}\beta_1\beta_2\gamma_5^{\zeta_3^2} - \frac{7}{4}\beta_1\beta_3^{\zeta_3}\gamma_4^{\zeta_3} + \gamma_1\beta_3^{\zeta_3}\beta_4^{\zeta_3} - \frac{3}{4}\beta_1\gamma_1\beta_6^{\zeta_3^2} \nonumber \\
&\hphantom{={}} - \frac{7}{4}\beta_1\gamma_2\beta_5^{\zeta_3^2}- \frac{3}{4}\beta_2\gamma_1\beta_5^{\zeta_3^2} - \frac{7}{4}\beta_2\beta_3^{\zeta_3}\gamma_3^{\zeta_3} + \frac{35}{24}\gamma_2\left(\beta_3^{\zeta_3}\right)^2 - \frac{1}{2}\gamma_1\beta_7^{\zeta_7} - \frac{7}{6}\gamma_2\beta_6^{\zeta_7} \nonumber \\
&\hphantom{={}} - \frac{21}{10}\gamma^{\text{R}}_{3}\beta_5^{\zeta_7} - \frac{29}{12}\gamma_8^{\zeta_{5,3}} + \frac{7}{2} \,\sum_{i=5}^7\gamma^{\zeta_7}_i \beta^{\text{R}}_{8-i}\,, \nonumber 
\end{align}
\begin{align} 
\gamma_8^{\zeta_3\zeta_8} &= \frac{63}{8}\beta_1^2\gamma_6^{\zeta_3^3} - \frac{9}{4}\beta_1\gamma_1\beta_6^{\zeta_3^3} + \frac{7}{2}\beta_1\gamma_7^{\zeta_3\zeta_7} + \frac{7}{2}\beta_3^{\zeta_3}\gamma_5^{\zeta_7} - \frac{21}{10}\beta_5^{\zeta_7}\gamma_3^{\zeta_3} - \frac{1}{2}\gamma_1\beta_7^{\zeta_3\zeta_7} \nonumber \\
&\hphantom{={}}  - \frac{29}{12}\gamma_8^{\zeta_3\zeta_{5,3}}\,, \nonumber \\
\gamma_8^{\zeta_5\zeta_8} &= -\frac{29}{12}\gamma_8^{\zeta_5\zeta_{5,3}} + \frac{145}{12}\gamma_8^{\zeta_{5,5,3}} + \frac{29}{2}\gamma_8^{\zeta_{7,3,3}}\,, \qquad\qquad
\gamma_8^{\zeta_3\zeta_{10}} = -\frac{793}{94}\gamma_8^{\zeta_3\zeta_{7,3}}\,, \nonumber 
\\
\gamma_8^{\zeta_4\zeta_9} &= 25\gamma_8^{\zeta_{5,5,3}} + 28\gamma_8^{\zeta_{7,3,3}}\,, \qquad\qquad \gamma_8^{\zeta_2\zeta_{11}} = \frac{275}{2}\gamma_8^{\zeta_{5,5,3}} + \frac{407}{2}\gamma_8^{\zeta_{7,3,3}}\,, \nonumber\\
\gamma_8^{\zeta_{10}} &= -21\beta_1^3\gamma_5^{\zeta_7} + 9\beta_1^2\gamma_1\beta_5^{\zeta_7} + \frac{33}{8}\beta_1^2\gamma_6^{\zeta_3\zeta_5} - \frac{2905}{376}\beta_1^2\gamma_6^{\zeta_{5,3}} - \frac{33}{28}\beta_1\beta_3^{\zeta_3}\gamma_4^{\zeta_5} \nonumber \\
&\hphantom{={}}  + \frac{11}{7}\beta_1\gamma_1\beta_3^{\zeta_3}\beta_4^{\zeta_5} - \frac{33}{28}\beta_1\gamma_1\beta_6^{\zeta_3\zeta_5} + \frac{415}{188}\beta_1\gamma_1\beta_6^{\zeta_{5,3}} + \frac{9}{2}\beta_1\gamma_7^{\zeta_9} + \frac{9}{2}\beta_2\gamma_6^{\zeta_9} \nonumber \\
&\hphantom{={}} - \frac{9}{14}\gamma_1\beta_7^{\zeta_9} - \frac{3}{2}\gamma_2\beta_6^{\zeta_9} - \frac{793}{94}\gamma_8^{\zeta_{7,3}}\,, \nonumber
\\
\gamma_8^{\zeta_{12}} &= \frac{11}{2}\beta_1\gamma_7^{\zeta_{11}} - \frac{87087}{22112}\beta_1\gamma_7^{\zeta_3\zeta_{5,3}} - \frac{11}{14}\gamma_1\beta_7^{\zeta_{11}} + \frac{12441}{22112}\gamma_1\beta_7^{\zeta_3\zeta_{5,3}} + \hat{\zeta}_{9,3(0)}^{12}\gamma_8^{\zeta_{9,3}} \nonumber \\
&\hphantom{={}} + \hat{\zeta}_{6,4,1,1(0)}^{12}\gamma_8^{\zeta_{6,4,1,1}} + \frac{125125}{44224}\hat{\zeta}_{5,3,3(1)}^{12}\left(7\beta_1\gamma_7^{\zeta_{5,3,3}} - \gamma_1\beta_7^{\zeta_{5,3,3}}\right)\,, \label{c8}
\end{align}
where the coefficients $\hat{\zeta}_{6,4,1,1(0)}^{12}$, $\hat{\zeta}_{5,3,3(1)}^{12}$, and $\hat{\zeta}_{9,3(0)}^{12}$ are defined in eqs.~\eqref{eq:hz_6411_12}--\eqref{eq:hz_533_12}.

The constraints (\ref{c4a})--(\ref{c8}) coincide with the corresponding results obtained in ref.~\cite{Baikov:2019zmy}. Similar to the previous subsection,
  there are only a few differences:
  $\gamma^{\zeta_{8}}_n +(29/12)  \,\gamma^{\zeta_{5,3}}_n$ for $n=6,7,8$, $\gamma^{\zeta_{10}}_n +(793/94)  \,\gamma^{\zeta_{7,3}}_n$ for $n=7,8$, and 
  $\gamma^{\zeta_{3}\zeta_{8}}_n +(29/12)  \,\gamma^{\zeta_{3}\zeta_{5,3}}_n$ for $n=7,8$ correspond to $\gamma^{\zeta_{8}}_n$, $\gamma^{\zeta_{10}}_n$, and $\gamma^{\zeta_{3}\zeta_{8}}_n$ in
  ref.~\cite{Baikov:2019zmy}, respectively. Furthermore, all relations for $\gamma^{\zeta_{a_1}...\zeta_{a_m}}_8$ with $\sum_{i=1}^{m}a_i\geqslant 12$ are entirely new.

\subsection{\label{sec:ident}Higher-loop relations and their closed forms}

At the next (nine-loop) order, there are 55 nontrivial relations. Because their explicit forms are rather lengthy, we list them in Appendix~\ref{app:9Lrels}.

Formally, one can continue computing these relations at higher perturbative orders up to at least the fifteenth order (where the complete expansion in terms of a minimal-depth MZV basis is still known) and, for certain sequences of relations, even beyond. This computation helps identify combinatorial patterns and suggests an algorithm for predicting relations from the $\varepsilon$-expansions of hatted MZVs.

We constructed all relations up to $n=15$ and empirically found general all-order expressions for the subset that does not depend on unfixed parameters in the $\varepsilon$-expansions of hatted MZVs. This subset contains 52 sequences of relations (indexed by $n$), excluding eight trivial relations for vanishing prefactors, namely $F^{\zeta_2 \zeta_{\boldsymbol{\Lambda}}}_n = 0$ for $\boldsymbol{\Lambda} = [3]$, $[5]$, $[7]$, $[5,3]$, $[3;5]$, $[5,3,3]$, $[3;5,3]$, and $[6,4,1,1]$. Every sequence we considered can be represented as the following sum:
\begin{align}\label{eq:FviaR}
F_{n}^{\zeta_{2m} \zeta_{\boldsymbol{\Lambda}}} = \sum Q \left( \prod_{i=1}^s  \hat{\zeta}_{{\boldsymbol{\Sigma}_i}(g_i)}^{{2a_i; \boldsymbol{\Sigma}'_i}} \right) R^{F,n}_{(r;t \vert s)}(\zeta_{\boldsymbol{\Lambda}_1}, \dots, \zeta_{\boldsymbol{\Lambda}_t} \vert \zeta_{\boldsymbol{\Sigma}_1}, \dots, \zeta_{\boldsymbol{\Sigma}_s}),
\end{align}
where $F = \beta$ or $\gamma$, and $Q = \left( \prod \zeta_{2a_i} \right)/\zeta_{2m}$ (with $m = \sum_{i=1}^s a_i$) is a rational number. The degree-$p$ form $R^{F,n}_{(r;t \vert s)}$ ($r+t+s=p$) is a sum of all possible products of $r$ rational parts of AD or $\beta$-function coefficients and $t+s$ transcendental parts. The $s$ MZVs to the right of the vertical bar are related to the hatted MZVs in eqs.~\eqref{eq:KTformula} and \eqref{eq:zeta53}--\eqref{eq:zeta8611}; consequently, the corresponding forms are accompanied by factors of the expansion coefficients $\hat{\zeta}_{{\boldsymbol{\Sigma}_i}(n)}^{2a_i;{\boldsymbol{\Sigma}'_i}}$ ($i=1,\dots, s$) defined in eq.~\eqref{eq:zetaexpansatz}. When the MZV $\varepsilon$-expansions are known, a given relation~\eqref{eq:FviaR} can be obtained via a straightforward combinatorial enumeration as a sum over all possible forms, subject to the conditions that every term in eq.~\eqref{eq:FviaR} satisfies $p = r + t + s = 1 + \sum_{i=1}^s g_i$ and $\zeta_{\boldsymbol{\Lambda}} = \zeta_{\boldsymbol{\Lambda}_1;\dots;\boldsymbol{\Lambda}_t;\boldsymbol{\Sigma}'_1;\dots;\boldsymbol{\Sigma}'_s}$. Given sufficiently many terms of a sequence, one can then attempt to reconstruct the required forms explicitly. In the cases we considered, the forms $R^{F,n}_{(r;t \vert s)}$ proved to be universal, repeating themselves across all 55 sequences of relations.

The 55 relations and corresponding forms are relegated to appendix~\ref{app:rels}. Here, we permit ourselves to write out only the simplest relation as an example:
\begin{align}
F_{\text{T},n}^{\zeta_4} =
 \frac32 \sum_{m=1}^{n-3} \beta^{\text{R}}_{m} F_{n-m}^{\zeta_3}
- \frac32 \sum_{m=1}^{n-3} \frac{m}{n-m} F^{\text{R}}_{m} \beta_{n-m}^{\zeta_3}, \qquad n \geqslant 4, \qquad F = \beta, \; \gamma.
\end{align}

As a final remark, it is interesting to compare the number of relations with the total number of MZV products at $n$-loop order; the corresponding counts are listed in Table~\ref{tab:numofrels}. In principle, approximately $56\%$ of the rational prefactors could be constrained at every loop order from $n=5$ to $15$. However, at most $60$ of these constraints are independent of unfixed parameters in the hatted MZV expansions, accounting for only $30\%$ of the prefactors already at ten loops.

\begin{table}[t]
\centering
\setlength{\tabcolsep}{4pt} 
\begin{tabular}{lccccccccc}
\toprule
$n$     & 4 & 5 & 6 & 7 & 8  & 9  & 10  & 11  & 12   \\
\midrule
\makecell[l]{Number of relations without free parameters} & 2 & 5 & 10 & 19 & 34 & 60 & 60  & 60  & 60  \\
\makecell[l]{Number of relations with free parameters} & 0 & 0 & 0 & 0  & 1  & 3 & 52  & 138 & 289  \\
\makecell[l]{Total number of relations} & 2 & 5 & 10 & 19 & 35 & 63 & 112 & 198 & 349  \\
\makecell[l]{Total number of $F^{\zeta_{\boldsymbol{\Lambda}}}_n$} & 4 & 9 & 18 & 34 & 62 & 111 & 197  & 348 & 613   \\
\bottomrule
\end{tabular}
\caption{\label{tab:numofrels}Number of constrained rational prefactors $F_{n}^{\zeta_{\boldsymbol{\Lambda}}}$ (with $\zeta_{\boldsymbol{\Lambda}} = \zeta_{2 m}$ or $\zeta_{2 m} \zeta_{\boldsymbol{\Lambda}'}$) compared with the total number of prefactors. These counts also include the trivially vanishing prefactors for $m=1$ and $\boldsymbol{\Lambda}' = [3]$, $[5]$, $[7]$, $[5,3]$, $[3;5]$, $[5,3,3]$, $[3;5,3]$, and $[6,4,1,1]$.}
\end{table}

\section{\label{sec:PredictionsPhi4}Predictions for RG functions in $\text{O}(n)$-symmetric $\phi^4$ model}

Here, we apply the general formulae of the previous section to predict higher-loop $\pi^2$-dependent terms in the specific example of the self-interacting scalar field model $\phi^4$, where $\phi$ is an $n$-component field in the defining representation of the global symmetry group $\text{O}(n)$. To this end, we invoke the state-of-the-art seven- and eight-loop calculations of refs.~\cite{Schnetz:2016fhy,Schnetz:2022nsc}.

All $\pi^2$-dependent terms in the RG functions from refs.~\cite{Schnetz:2016fhy,Schnetz:2022nsc} satisfy the relations considered in this work, which confirms (and, in the case of the AD, extends by one perturbative order) the same conclusion of refs.~\cite{Baikov:2018wgs,Baikov:2018gap,Baikov:2019zmy}.

Below, the predictions for the $\pi^2$-dependent terms at the next yet-to-be-calculated orders are given explicitly. It should be noted that the seven-loop $\beta$-function contains a non-vanishing coefficient $\beta_7^{\zeta_2 \zeta_9}$, whereas all other coefficients of the form $\beta_n^{\zeta_2 \zeta_{\boldsymbol{\Lambda}}}$ are identically zero in all the RG functions. Therefore, in this section, we use the general-form constraints on the $\pi^2$-coefficients, i.e.\ without assuming $\beta_n^{\zeta_2 \zeta_{\boldsymbol{\Lambda}}} = 0$. Furthermore, it is interesting to note that the seven-loop contribution to the $\beta$-function is the first to include a period, $P_{7,11}$ (of transcendental weight 11), that conjecturally does not reduce to MZVs. By contrast, the seven-loop AD and eight-loop mass AD still contain nothing other than MZVs (of weight at most $10$).

\subsection{$\pi^2$-dependent terms in the $\beta$-function at eight loops}

\begin{gather}
\beta _8^{\zeta _2 \zeta _5^2}= -\frac{7}{2} \beta _8^{\zeta _{6,4,1,1}},
\qquad
\beta _8^{\zeta _2 \zeta _3 \zeta _7}= -10 \beta _8^{\zeta _{6,4,1,1}},
\qquad
\beta_8^{\zeta _2 \zeta _{11}}= \frac{275}{2} \beta _8^{\zeta _{5,5,3}}+\frac{407}{2} \beta _8^{\zeta _{7,3,3}},
\\
\beta _8^{\zeta _2 \zeta _{7,3}}= -3 \beta _8^{\zeta _{6,4,1,1}},
\qquad
\beta _8^{\zeta _2 \zeta _9}= 45 \beta _8^{\zeta _{5,3,3}},
\qquad
\beta _8^{\zeta _3^3 \zeta _4}= 0,
\\
\beta _8^{\zeta _6 \zeta _7}= -4 \beta_8^{\zeta _{7,3,3}},
\qquad
\beta _8^{\zeta _5 \zeta _8}= -\frac{29}{12} \beta _8^{\zeta _5 \zeta _{5,3}}+\frac{145}{12} \beta _8^{\zeta _{5,5,3}}+\frac{29}{2} \beta_8^{\zeta _{7,3,3}},
\\
\beta _8^{\zeta _4 \zeta _9}= 25 \beta _8^{\zeta _{5,5,3}}+28 \beta _8^{\zeta _{7,3,3}},
\qquad
\beta _8^{\zeta _3 \zeta _{10}}= -\frac{793}{94} \beta _8^{\zeta _3 \zeta _{7,3}},
\end{gather}
\begin{align}
\beta _8^{\zeta _4}&{}= \frac{1}{10032906240} \bigl( 25 n^7-162864 n^6+3588532 n^5+647836272 n^4
		\notag\\ &\hphantom{{}= \frac{1}{10032906240}  \bigl(  }
		{}+13171922000 n^3+105886259936 n^2+381817121664 n
		\notag\\ &\hphantom{{}= \frac{1}{10032906240}  \bigl(  }
		{}+515658063360 \bigr),
\\
\beta _8^{\zeta _3 \zeta _4}&{}= \frac{1}{313528320} \bigl( -14820 n^6-121883 n^5+25094418 n^4
		\notag\\ &\hphantom{{}= \frac{1}{313528320}  \bigl(  }
		{}+630956516 n^3+5837380344 n^2+23783217568 n
		\notag\\ &\hphantom{{}= \frac{1}{313528320}  \bigl(  }
		{}+36365733632 \bigr),
\\
\beta _8^{\zeta _3^2 \zeta _4}&{}= \frac{1}{6531840} \bigl( 3260 n^5+200385 n^4+4003176 n^3+35022944 n^2+140342184 n
		\notag\\ &\hphantom{{}= \frac{1}{6531840}  \bigl(  }
		{}+212633696 \bigr),
\\
\beta _8^{\zeta _4 \zeta _5}&{}= \frac{1}{15676416}  \bigl( -130 n^6-48311 n^5-2922150 n^4-57419798 n^3-488246872 n^2
		\notag\\ &\hphantom{{}= \frac{1}{15676416}  \bigl(  }
		{}-1855758408 n-2550049440 \bigr),
\end{align}
\begin{align}
\beta _8^{\zeta _3 \zeta _4 \zeta _5}&{}= \frac{1}{2} \beta _8^{\zeta _{6,4,1,1}}-\frac{(n+8) \left(1151 n^3+125646 n^2+1393004 n+3630584\right)}{290304},
\\
\beta _8^{\zeta _6}&{}= \frac{1}{752467968} \bigl( -45 n^7+80210 n^6+6199060 n^5+165370686 n^4
		\notag\\ &\hphantom{{}= \frac{1}{752467968}  \bigl(  }
		{}+2635159500 n^3+22125482888 n^2+85734817296 n
		\notag\\ &\hphantom{{}= \frac{1}{752467968}  \bigl(  }
		{}+120831531520 \bigr),
\\
\beta _8^{\zeta _3 \zeta _6}&{}= \frac{1}{23514624} \bigl( -280 n^6-8690 n^5+4283440 n^4+132215395 n^3+1429947692 n^2
		\notag\\ &\hphantom{{}= \frac{1}{23514624}  \bigl(  }
		{}+6496275812 n+10470998448 \bigr),
\\
\beta _8^{\zeta _3^2 \zeta _6}&{}= \frac{9}{4} \beta _8^{\zeta _{6,4,1,1}}+\frac{25 (n+8) \left(2693 n^3+84138 n^2+672548 n+1517288\right)}{1741824},
\\
\beta _8^{\zeta _5 \zeta _6}&{}= \frac{25 (n+8)}{11757312}  \bigl( 442 n^4-82916 n^3-1583539 n^2-9244298 n-17059232 \bigr)
		\notag\\ &\hphantom{{}={} }
		{} -\frac{5}{2} \beta _8^{\zeta _{5,3,3}},
\\
\beta _8^{\zeta _4 \zeta _7}&{}= 3 \beta _8^{\zeta _{5,3,3}}+\frac{1}{10450944}  \bigl( -14344 n^5-350676 n^4-1660877 n^3+22919374 n^2
		\notag\\ &\hphantom{{}= 3 \beta _8^{\zeta _{5,3,3}}+\frac{1}{10450944}  \bigl(  }
		{}+241458024 n+567464096 \bigr),
\\
\beta _8^{\zeta _8}&{}= \frac{1}{537477120}  \bigl( 61260 n^6+8039755 n^5+321407133 n^4+5257026590 n^3
		\notag\\ &\hphantom{{}= \frac{1}{537477120} \bigl( }
		{}+41944230764 n^2+159988064888 n+228403741760 \bigr)
		\notag\\ &\hphantom{{}={} }
		{} -\frac{29}{12} \beta _8^{\zeta _{5,3}},
\\
\beta _8^{\zeta _3 \zeta _8}&{}= \frac{1}{22394880}  \bigl( -55320 n^5-424980 n^4+24562735 n^3+452396838 n^2
		\notag\\ &\hphantom{{}= \frac{1}{22394880} \bigl( }
		{}+2698147848 n+5175905312 \bigr)
		 -\frac{29}{12} \beta _8^{\zeta _3 \zeta _{5,3}},
\\
\beta _8^{\zeta _{10}}&{}= \frac{1}{2210374656} \bigl( 57530734 n^5+2837539311 n^4+51607635845 n^3
		\notag\\ &\hphantom{{}= \frac{1}{2210374656} \bigl( }
		{}+438295844964 n^2+1752369839516 n+2625134636128 \bigr) 
		\notag\\ &\hphantom{{}={} }
		{}-\frac{793}{94} \beta _8^{\zeta _{7,3}},
\\
\beta _8^{\zeta _4 \zeta _{5,3}}&{}= -\frac{3}{2} \beta _8^{\zeta  _{6,4,1,1}}-\frac{(n+8) \left(16253 n^3+755698 n^2+6897508 n+16652168\right)}{1451520},
\\
\beta_8^{\zeta_{12}} &= -\frac{n+8}{11725959168}\left(30869 n^3 + 1302034 n^2 + 11653252 n + 27869192\right) \nonumber \\
&\hphantom{={}} \qquad\qquad\times \left(29683 \hat{\zeta}_{9,3(0)}^{12} + 6720 \hat{\zeta}_{6,4,1,1(0)}^{12}\right) + \hat{\zeta}_{9,3(0)}^{12} \beta_8^{\zeta_{9,3}} + \hat{\zeta}_{6,4,1,1(0)}^{12} \beta_8^{\zeta_{6,4,1,1}} \nonumber \\
&\hphantom{={}} + \frac{n+8}{108035170467840}\bigl(78447945055349 n^3 + 2367604318251534 n^2 \nonumber \\
&\hphantom{={}} \qquad\qquad\qquad\qquad+ 18531986866437464 n + 41348499168792944\bigr)\,.
\label{eq:phi4_b8z12}
\end{align}
The parameters $\hat{\zeta}_{6,4,1,1(0)}^{{12}}$ and $\hat{\zeta}_{9,3(0)}^{{12}}$ are defined in eqs.~\eqref{eq:hz_6411_12} and \eqref{eq:hz_93_12}.

All coefficients $\beta_8^{\zeta_2 \zeta_{\boldsymbol{\Lambda}}}$ not listed explicitly above vanish identically.

\subsection{$\pi^2$-dependent terms in the AD at nine loops}

This is the first example of an RG function to which one can apply the nine-loop relations listed explicitly in their general form in appendix~\ref{app:9Lrels}.

\begin{gather}
\gamma_9^{\zeta_2 \zeta_9} = 45 \gamma_9^{\zeta_{5,3,3}} \,, \quad
\gamma_9^{\zeta_2 \zeta_{13}} = \frac{1001}{2} \gamma_9^{\zeta_{7,3,5}} + \frac{1209}{2} \gamma_9^{\zeta_{9,3,3}} - \frac{56717}{120} \gamma_9^{\zeta_{6,4,3,1,1}} \,, \\
\gamma_9^{\zeta_2 \zeta_3 \zeta_7} = -10 \gamma_9^{\zeta_{6,4,1,1}} \,, \qquad
\gamma_9^{\zeta_2 \zeta_{7,3}} = -3 \gamma_9^{\zeta_{6,4,1,1}} \,, \qquad
\gamma_9^{\zeta_2 \zeta_5^2} = -\frac{7}{2} \gamma_9^{\zeta_{6,4,1,1}} \,, \\
\gamma_9^{\zeta_2 \zeta_{11}} = \frac{275}{2} \gamma_9^{\zeta_{5,5,3}} + \frac{407}{2} \gamma_9^{\zeta_{7,3,3}} \,, \qquad
\gamma_9^{\zeta_2 \zeta_3 \zeta_9} = 45 \gamma_9^{\zeta_3 \zeta_{5,3,3}} \,, \qquad
\gamma_9^{\zeta_2 \zeta_{9,3}} = \frac{10}{3} \gamma_9^{\zeta_{5,3,3,3}} \,, \\
\gamma_9^{\zeta_2 \zeta_5 \zeta_7} = 20 \gamma_9^{\zeta_{5,3,3,3}} \,, \qquad
\gamma_9^{\zeta_2 \zeta_3 \zeta_{7,3}} = -3 \gamma_9^{\zeta_3 \zeta_{6,4,1,1}} - 3 \gamma_9^{\zeta_{6,4,3,1,1}} \,, \\
\gamma_9^{\zeta_2 \zeta_3 \zeta_5^2} = -\frac{7}{2} \gamma_9^{\zeta_3 \zeta_{6,4,1,1}} - \frac{9}{2} \gamma_9^{\zeta_{6,4,3,1,1}} \,, \qquad
\gamma_9^{\zeta_2 \zeta_3^2 \zeta_7} = -10 \gamma_9^{\zeta_3 \zeta_{6,4,1,1}} - 16 \gamma_9^{\zeta_{6,4,3,1,1}} \,, 
\end{gather}
\begin{gather}
\gamma_9^{\zeta_2 \zeta_5 \zeta_{5,3}} = \frac{21}{2} \gamma_9^{\zeta_{6,4,3,1,1}} \,, \qquad
\gamma_9^{\zeta_2 \zeta_{7,3,3}} = 2 \gamma_9^{\zeta_{6,4,3,1,1}} \,, \qquad
\gamma_9^{\zeta_2 \zeta_{5,5,3}} = -\frac{14}{5} \gamma_9^{\zeta_{6,4,3,1,1}} \,, \\
\gamma_9^{\zeta_3 \zeta_4 \zeta_{5,3}} = -\frac{3}{2} \gamma_9^{\zeta_3 \zeta_{6,4,1,1}} - \frac{5}{2} \gamma_9^{\zeta_{6,4,3,1,1}} \,, \qquad
\gamma_9^{\zeta_3^2 \zeta_4 \zeta_5} = \frac{1}{2} \gamma_9^{\zeta_3 \zeta_{6,4,1,1}} - \frac{25}{4} \gamma_9^{\zeta_{6,4,3,1,1}} \,, \\
\gamma_9^{\zeta_3^3 \zeta_6} = \frac{9}{4} \gamma_9^{\zeta_3 \zeta_{6,4,1,1}} - \frac{3}{4} \gamma_9^{\zeta_{6,4,3,1,1}} \,, \qquad
\gamma_9^{\zeta_4 \zeta_{5,3,3}} = 10 \gamma_9^{\zeta_{6,4,3,1,1}} \,, \\
\gamma_9^{\zeta_4 \zeta_{11}} = 55 \gamma_9^{\zeta_{7,3,5}} + 126 \gamma_9^{\zeta_{9,3,3}} - \frac{27199}{24} \gamma_9^{\zeta_{6,4,3,1,1}} \,, \\
\gamma_9^{\zeta_5 \zeta_{10}} = -\frac{793}{94} \gamma_9^{\zeta_5 \zeta_{7,3}} - \frac{2520}{47} \gamma_9^{\zeta_{7,3,5}} + \frac{2379}{47} \gamma_9^{\zeta_{9,3,3}} - \frac{210685}{1128} \gamma_9^{\zeta_{6,4,3,1,1}} \,, \\
\gamma_9^{\zeta_6 \zeta_9} = -17 \gamma_9^{\zeta_{7,3,5}} + \frac{29}{2} \gamma_9^{\zeta_{9,3,3}} + \frac{84185}{144} \gamma_9^{\zeta_{6,4,3,1,1}} \,, \\
\gamma_9^{\zeta_7 \zeta_8} = -\frac{29}{12} \gamma_9^{\zeta_7 \zeta_{5,3}} - \frac{287}{6} \gamma_9^{\zeta_{7,3,5}} + \frac{157}{4} \gamma_9^{\zeta_{9,3,3}} - \frac{7811}{144} \gamma_9^{\zeta_{6,4,3,1,1}} \,,
\end{gather}
\begin{gather}
\gamma_9^{\zeta_3^2 \zeta_8} = -\frac{29}{12} \gamma_9^{\zeta_3^2 \zeta_{5,3}} \, , \qquad
\gamma_9^{\zeta_5 \zeta_8} = -\frac{29}{12} \gamma_9^{\zeta_5 \zeta_{5,3}} + \frac{145}{12} \gamma_9^{\zeta_{5,5,3}} + \frac{29}{2} \gamma_9^{\zeta_{7,3,3}} \, , \\
\gamma_9^{\zeta_4 \zeta_9} = 25 \gamma_9^{\zeta_{5,5,3}} + 28 \gamma_9^{\zeta_{7,3,3}} \, , \qquad
\gamma_9^{\zeta_3 \zeta_{10}} = -\frac{793}{94} \gamma_9^{\zeta_3 \zeta_{7,3}} \, , \qquad
\gamma_9^{\zeta_3 \zeta_4 \zeta_7} = 3 \gamma_9^{\zeta_3 \zeta_{5,3,3}} \, , \\
\gamma_9^{\zeta_3 \zeta_5 \zeta_6} = -\frac{5}{2} \gamma_9^{\zeta_3 \zeta_{5,3,3}} \, , \qquad
\gamma_9^{\zeta_6 \zeta_7} = -4 \gamma_9^{\zeta_{7,3,3}} \, , \qquad
\gamma_9^{\zeta_6 \zeta_{5,3}} = \frac{1}{2} \gamma_9^{\zeta_{5,3,3,3}} \, ,
\end{gather}

\begin{align}
\gamma_9^{\zeta_4} &= -\frac{2+n}{96315899904} \big( 3125477120 + 1786210944 n - 235594848 n^2\nonumber \\
&\hphantom{{}=11}\qquad - 159632408 n^3- 10266864 n^4 + 413326 n^5 + 898 n^6 + 91 n^7 \big) , \\
\gamma_9^{\zeta_3 \zeta_4} &= \frac{2+n}{6019743744} \big( 1102443968 + 854327872 n + 246074048 n^2 + 31444696 n^3 \nonumber \\
&\hphantom{{}= \frac{2+n}{6019743744} \big(} + 1617656 n^4 + 17480 n^5 - 350 n^6 + 7 n^7 \big) \, , \\
\gamma_9^{\zeta_3^2 \zeta_4} &= -\frac{2+n}{62705664} \left( 20930960 + 11432432 n + 2110250 n^2 + 146313 n^3 + 2442 n^4 \right) , \\
\gamma_9^{\zeta_4 \zeta_5} &= \frac{2+n}{376233984} \big( 278214912 + 171260576 n + 36429350 n^2 + 3180989 n^3  \nonumber \\
&\hphantom{{}=\frac{2+n}{376233984} \big(} + 95508 n^4 + 264 n^5 \big) , \\
\gamma_9^{\zeta_3 \zeta_4 \zeta_5} &= \frac{2+n}{10450944} \left( 3630584 + 1393004 n + 125646 n^2 + 1151 n^3 \right) + \frac{1}{2} \gamma_9^{\zeta_{6,4,1,1}} \, , \\
\gamma_9^{\zeta_6} &= -\frac{5(2+n)}{36118462464} \big( 6482548800 + 4137531728 n + 896988944 n^2 + 76585242 n^3 \nonumber \\
&\hphantom{{}=-\frac{5(2+n)}{36118462464} \big(} + 2302671 n^4 - 38591 n^5 + 618 n^6 + 7 n^7 \big) \, , \\
\gamma_9^{\zeta_3 \zeta_6} &= \frac{5(2+n)}{2257403904} \big( -439327344 - 212966296 n - 31613016 n^2 - 1443830 n^3 \nonumber \\
&\hphantom{{}= \frac{5(2+n)}{2257403904} \big(} + 7031 n^4 + 1078 n^5 \big) \, ,
\\
\gamma_9^{\zeta_3^2 \zeta_6} &= -\frac{25(2+n)}{62705664} \left( 1517288 + 672548 n + 84138 n^2 + 2693 n^3 \right) + \frac{9}{4} \gamma_9^{\zeta_{6,4,1,1}} \, , \\
\gamma_9^{\zeta_5 \zeta_6} &= -\frac{5(2+n)}{141087744} \left( 10366352 + 4984112 n + 837610 n^2 + 67595 n^3 + 2489 n^4 \right)  \nonumber \\
&\hphantom{{}=}- \frac{5}{2} \gamma_9^{\zeta_{5,3,3}} \, , \\
\gamma_9^{\zeta_4 \zeta_7} &= \frac{2+n}{376233984} \left( 304840 + 15441454 n + 6000167 n^2 + 580408 n^3 + 13336 n^4 \right) \nonumber \\
&\hphantom{{}=}  + 3 \gamma_9^{\zeta_{5,3,3}} \, , \\
\gamma_9^{\zeta_8} &= -\frac{2+n}{2579890176} \big( 9285699888 + 5582386984 n + 1179565588 n^2 + 106543916 n^3 \nonumber \\
&\hphantom{{}=} + 4053889 n^4 + 68998 n^5 \big) - \frac{29}{12} \gamma_9^{\zeta_{5,3}} \, , \\
\gamma_9^{\zeta_3 \zeta_8} &= \frac{2+n}{161243136} \left( -94285112 - 35364434 n - 2966041 n^2 + 5752 n^3 + 5464 n^4 \right) \nonumber \\
&\hphantom{{}=}  - \frac{29}{12} \gamma_9^{\zeta_3 \zeta_{5,3}} \, ,
\\
\gamma_9^{\zeta_{10}} &= -\frac{2+n}{212195966976} \big( 1492968631552 + 823363096672 n + 156237741394 n^2 \nonumber \\
&\hphantom{{}=} + 12280298787 n^3 + 321422286 n^4 \big) - \frac{793}{94} \gamma_9^{\zeta_{7,3}} \, , \\
\gamma_9^{\zeta_4 \zeta_{5,3}} &= \frac{2+n}{52254720} \left( 16652168 + 6897508 n + 755698 n^2 + 16253 n^3 \right) - \frac{3}{2} \gamma_9^{\zeta_{6,4,1,1}} \, ,
\end{align}
\begin{align}
\gamma_9^{\zeta_{12}} &= \frac{n+2}{422134530048}\left(30869 n^3 + 1302034 n^2 + 11653252 n + 27869192\right) \nonumber \\
&\hphantom{={}} \qquad\qquad \times \left(29683 \hat{\zeta}_{9,3(0)}^{12} + 6720 \hat{\zeta}_{6,4,1,1(0)}^{12}\right) + \hat{\zeta}_{9,3(0)}^{12} \gamma_9^{\zeta_{9,3}} + \hat{\zeta}_{6,4,1,1(0)}^{12} \gamma_9^{\zeta_{6,4,1,1}} \nonumber \\
&\hphantom{={}} - \frac{n+2}{3889266136842240}\Bigl(78447945055349 n^3 + 2367604318251534 n^2 \nonumber \\
&\hphantom{={}} \qquad\qquad\qquad\qquad + 18531986866437464 n + 41348499168792944\Bigr)\,, \label{g9z12} \\[6pt]
\gamma_9^{\zeta_3\zeta_{12}} &= \hat{\zeta}_{9,3(0)}^{12} \gamma_9^{\zeta_3\zeta_{9,3}} + \hat{\zeta}_{7,3,5(0)}^{3;12}  \gamma_9^{\zeta_{7,3,5}}+ \hat{\zeta}_{9,3,3(0)}^{3;12} \gamma_9^{\zeta_{9,3,3}} + \hat{\zeta}_{6,4,1,1(0)}^{12} \gamma_9^{\zeta_3\zeta_{6,4,1,1}} \nonumber \\
&\hphantom{={}} + \hat{\zeta}_{6,4,3,1,1(0)}^{3;12} \gamma_9^{\zeta_{6,4,3,1,1}}\,, \label{g9z3z12} \\[6pt]
\gamma_9^{\zeta_{14}} &= \hat{\zeta}_{9,5(0)}^{{14}} \gamma_9^{\zeta_{9,5}} + \hat{\zeta}_{11,3(0)}^{{14}}  \gamma_9^{\zeta_{11,3}}+ \hat{\zeta}_{5,3,3,3(0)}^{{14}} \gamma_9^{\zeta_{5,3,3,3}}\,. \label{g9z14}
\end{align}
Here, the parameters $\hat{\zeta}_{6,4,1,1(0)}^{{12}}$ and $\hat{\zeta}_{9,3(0)}^{{12}}$ are defined in eqs.~\eqref{eq:hz_6411_12} and \eqref{eq:hz_93_12}. All other parameters $\hat{\zeta}_{\boldsymbol{\lambda}(0)}^{\boldsymbol{\Lambda}}$ are unfixed. All coefficients $\gamma_9^{\zeta_2 \zeta_{\boldsymbol{\Lambda}}}$ not listed explicitly above vanish identically.

\subsection{$\pi^2$-dependent terms in mass AD at eight loops}

\begin{align}
\gamma_{m8}^{\zeta_2 \zeta_3^3} &= 0 \, , &
\gamma_{m8}^{\zeta_2 \zeta_5^2} &= -\frac{7}{2} \gamma_{m8}^{\zeta_{6,4,1,1}} \, , \\
\gamma_{m8}^{\zeta_4 \zeta_3^3} &= 0 \, , &
\gamma_{m8}^{\zeta_2 \zeta_3 \zeta_7} &= -10 \gamma_{m8}^{\zeta_{6,4,1,1}} \, , \\
\gamma_{m8}^{\zeta_2 \zeta_3^2 \zeta_5} &= 0 \, , &
\gamma_{m8}^{\zeta_6 \zeta_7} &= -4 \gamma_{m8}^{\zeta_{7,3,3}} \, , \\
\gamma_{m8}^{\zeta_2 \zeta_3 \zeta_{5,3}} &= 0 \, , &
\gamma_{m8}^{\zeta_3 \zeta_{10}} &= -\frac{793}{94} \gamma_{m8}^{\zeta_3 \zeta_{7,3}} \, , \\
\gamma_{m8}^{\zeta_2 \zeta_{5,3,3}} &= 0 \, , &
\gamma_{m8}^{\zeta_4 \zeta_9} &= 25 \gamma_{m8}^{\zeta_{5,5,3}} + 28 \gamma_{m8}^{\zeta_{7,3,3}} \, , \\
\gamma_{m8}^{\zeta_2 \zeta_9} &= 45 \gamma_{m8}^{\zeta_{5,3,3}} \, , &
\gamma_{m8}^{\zeta_2 \zeta_{11}} &= \frac{275}{2} \gamma_{m8}^{\zeta_{5,5,3}} + \frac{407}{2} \gamma_{m8}^{\zeta_{7,3,3}} \, , \\
\gamma_{m8}^{\zeta_2 \zeta_{7,3}} &= -3 \gamma_{m8}^{\zeta_{6,4,1,1}} \, , &
\gamma_{m8}^{\zeta_5 \zeta_8} &= -\frac{29}{12} \gamma_{m8}^{\zeta_5 \zeta_{5,3}} + \frac{145}{12} \gamma_{m8}^{\zeta_{5,5,3}} + \frac{29}{2} \gamma_{m8}^{\zeta_{7,3,3}} \, ,
\end{align}

\begin{align}
\gamma_{m8}^{\zeta_4} &= \frac{2+n}{30098718720} \big( {-52380905984} - 32845225280 n - 9631374800 n^2 \nonumber \\
&\hphantom{={}}  - 1363823072 n^3 - 63578464 n^4 + 265350 n^5 + 165 n^6 \big) \, , \\
\gamma_{m8}^{\zeta_3 \zeta_4} &= \frac{2+n}{1881169920} \big( -9686326656 - 6056089184 n - 1398597024 n^2 \nonumber \\
&\hphantom{={}}  - 144239556 n^3- 5885470 n^4 - 46230 n^5 + 315 n^6 \big) \, , \\
\gamma_{m8}^{\zeta_3^2 \zeta_4} &= \frac{2+n}{3919104} \left( 14910440 + 7702712 n + 1342374 n^2 + 85100 n^3 + 1065 n^4 \right) , \\
\gamma_{m8}^{\zeta_4 \zeta_5} &= -\frac{2+n}{15676416} \big( 56691752 + 36404188 n + 8291118 n^2 \nonumber \\
&\hphantom{={}-\frac{2+n}{15676416} \big(}   + 838705 n^3 + 34468 n^4 + 244 n^5 \big) , 
\end{align}
\begin{align}
\gamma_{m8}^{\zeta_3 \zeta_4 \zeta_5} &= -\frac{2+n}{870912} \left( 3630584 + 1393004 n + 125646 n^2 + 1151 n^3 \right) + \frac{1}{2} \gamma_{m8}^{\zeta_{6,4,1,1}} \, , \\
\gamma_{m8}^{\zeta_6} &= -\frac{2+n}{2257403904} \big( {-13460809264} - 7908356696 n - 1459221256 n^2 \nonumber \\
&\hphantom{={}}  - 77197020 n^3+ 1638490 n^4 + 318150 n^5 + 1095 n^6 \big) \, , \\
\gamma_{m8}^{\zeta_3 \zeta_6} &= -\frac{5(2+n)}{23514624} \big( {-29981372} - 13550254 n - 1961800 n^2 - 118023 n^3 \nonumber \\
&\hphantom{={}-\frac{5(2+n)}{23514624} \big( } - 3651 n^4 + 112 n^5 \big) \, , \\
\gamma_{m8}^{\zeta_3^2 \zeta_6} &= \frac{25(2+n)}{5225472} \left( 1517288 + 672548 n + 84138 n^2 + 2693 n^3 \right) + \frac{9}{4} \gamma_{m8}^{\zeta_{6,4,1,1}} \, , \\
\gamma_{m8}^{\zeta_5 \zeta_6} &= \frac{5(2+n)}{5878656} \left( 16547816 + 8111796 n + 1297480 n^2 + 81800 n^3 + 1717 n^4 \right)  \nonumber \\
&\hphantom{={}} - \frac{5}{2} \gamma_{m8}^{\zeta_{5,3,3}} \, , \\
\gamma_{m8}^{\zeta_4 \zeta_7} &= -\frac{2+n}{15676416} \left( {-20475796} - 2330300 n + 1452541 n^2 + 208031 n^3 + 5982 n^4 \right)  \nonumber \\
&\hphantom{={}} + 3 \gamma_{m8}^{\zeta_{5,3,3}} \, , \\
\gamma_{m8}^{\zeta_8} &= \frac{2+n}{1612431360} \big( 57382138392 + 31968329228 n + 6022465938 n^2 \nonumber \\
&\hphantom{={}\frac{2+n}{1612431360} \big( }  + 452059655 n^3 + 11986625 n^4 + 233880 n^5 \big) - \frac{29}{12} \gamma_{m8}^{\zeta_{5,3}} \, , \\
\gamma_{m8}^{\zeta_3 \zeta_8} &= -\frac{2+n}{6718464} \big( {-67770772} - 27733244 n - 3030563 n^2 - 79297 n^3 \nonumber \\
&\hphantom{={} -\frac{2+n}{6718464} \big( } + 2046 n^4 \big) - \frac{29}{12} \gamma_{m8}^{\zeta_3 \zeta_{5,3}} \, , \\
\gamma_{m8}^{\zeta_{10}} &= \frac{2+n}{26524495872} \big( 1851833038064 + 959875285480 n + 166749159458 n^2 \nonumber \\
&\hphantom{={}} + 12040016455 n^3 + 304762620 n^4 \big) - \frac{793}{94} \gamma_{m8}^{\zeta_{7,3}} \, , \\
\gamma_{m8}^{\zeta_4 \zeta_{5,3}} &= -\frac{2+n}{4354560} \left( 16652168 + 6897508 n + 755698 n^2 + 16253 n^3 \right) - \frac{3}{2} \gamma_{m8}^{\zeta_{6,4,1,1}} \, ,
\\
\gamma_{m8}^{\zeta_{12}} &= -\frac{n+2}{35177877504}\left(30869 n^3 + 1302034 n^2 + 11653252 n + 27869192\right) \nonumber \\
&\hphantom{={}} \qquad  \qquad \times \left(29683 \hat{\zeta}_{9,3(0)}^{12} - 6720 \hat{\zeta}_{6,4,1,1(0)}^{12}\right) + \hat{\zeta}_{9,3(0)}^{12} \gamma_{m8}^{\zeta_{9,3}} + \hat{\zeta}_{6,4,1,1(0)}^{12} \gamma_{m8}^{\zeta_{6,4,1,1}} \nonumber \\
&\hphantom{={}} + \frac{n+2}{324105511403520}\bigl(78447945055349 n^3 + 2367604318251534 n^2 \nonumber \\
&\hphantom{={}} \qquad  \qquad\qquad\qquad  + 18531986866437464 n + 41348499168792944\bigr)\,.
\label{eq:phi4_g8z12}
\end{align}
The parameters $\hat{\zeta}_{6,4,1,1(0)}^{{12}}$ and $\hat{\zeta}_{9,3(0)}^{{12}}$ are defined in eqs.~\eqref{eq:hz_6411_12} and \eqref{eq:hz_93_12}. All coefficients $\gamma_{m8}^{\zeta_2 \zeta_{\boldsymbol{\Lambda}}}$ not listed explicitly above vanish identically.

%============================================================
%============================================================
%============================================================

\section{\label{sec:PredictionsYM}Predictions for the gauge model with fermions}

All RG functions in Yang--Mills (YM) theory coupled to $N_f$ copies of a massive Dirac fermion in a general representation of a semisimple Lie group were computed at five loops in an arbitrary covariant gauge and the $\overline{\text{MS}}$ scheme in refs.~\cite{Herzog:2017ohr,Ruijl:2018poj,Chetyrkin:2017bjc,Chetyrkin:2017bjcAddendum,Luthe:2016xec,Baikov:2017ujl} (see also the references therein for earlier partial results and independent cross-checks). The six-loop renormalization of QED (with $N_f$ equal-mass fermions) was completed in a general covariant gauge in the $\overline{\text{MS}}$ scheme in ref.~\cite{Li:2026bck}.

Consistency checks on the $\pi$-dependent parts of these results in the Landau gauge were carried out in refs.~\cite{Baikov:2018wgs,Baikov:2018gap,Baikov:2019zmy,Li:2026bck}, confirming complete agreement between the constraints on $\pi$-terms and the explicit multi-loop calculations. Moreover, ref.~\cite{Baikov:2018wgs} derived predictions for the six-loop $\beta$-function and mass AD for a generic gauge group. All of these conclusions are confirmed by our calculations.

In this section, for completeness, we combine the five- and six-loop results mentioned above to present explicit predictions for the $\beta$-function and three independent ADs\footnote{Due to gauge invariance, only four ADs are independent, while all remaining ADs are linearly related to the $\beta$-function and these four independent ADs via the Slavnov--Taylor identities (see, e.g., ref.~\cite{Luthe:2017ttc}). In the Landau gauge, the ghost--gluon vertex AD vanishes identically to all perturbative orders, which reduces the number of independent ADs to three.} at six loops in general YM theory with fermions, as well as at seven loops in QED. These predictions apply in the Landau gauge, except for the $\beta$-function and the mass AD, which are gauge-invariant.

For group-theory objects, we adopt the notation of refs.~\cite{Luthe:2016xec,Luthe:2017ttc,Luthe:2017ttg}. The generators $T^a$ and antisymmetric structure constants $f^{abc}$ of a semisimple Lie algebra satisfy the commutation relation $[T^a,T^b]=if^{abc}T^c$, with the traces of the generators normalized such that $\operatorname{Tr}(T^aT^b)=T_F \delta^{ab}$. The fundamental (adjoint) representation has dimension $N_F$ ($N_A$), while the corresponding quadratic Casimir operator is given by $T^aT^a= C_F \hat{1}$ ($f^{acd}f^{bcd}=C_A \delta^{ab}$, where $\delta^{ab}$ is the Kronecker delta). Note also that the useful relation $N_A T_F = N_F C_F$ holds for a general Lie algebra \cite{vanRitbergen:1998pn,Georgi:1999wka}. At higher perturbative orders, traces of four generators also appear \cite{vanRitbergen:1998pn}:
\begin{align}
d_F^{abcd} &= \frac{1}{4!} \operatorname{Tr} \left( T^a T^b T^c T^d + \text{all permutations of the indices} \right),
\\
d_A^{abcd} &= \frac{1}{4!} \operatorname{Tr} \left( F^a F^b F^c F^d + \text{all permutations of the indices} \right).
\end{align}
Here, $[F^a]_{bc}=-if^{abc}$ are the generators of the adjoint representation. Finally, in what follows, all RG functions are expressed in terms of the group invariants
\begin{gather}\label{cf}
n_f=\frac{N_fT_F}{C_A}, \qquad c_f=\frac{C_F}{C_A}, 
\quad
d_1=\frac{d_F^{abcd}d_F^{abcd}}{N_AT_F^2C_A^2},
\quad
d_2=\frac{d_F^{abcd}d_A^{abcd}}{N_AT_FC_A^3},
\quad
d_3=\frac{d_A^{abcd}d_A^{abcd}}{N_AC_A^4}.
\end{gather}
In the special case of the group $\text{SU}(N)$ with $T_F = 1/2$ and $C_A = N$, the normalized invariants in eq.~\eqref{cf} become \cite{vanRitbergen:1998pn}
\begin{equation}
n_f=\frac{N_f}{2N}, \quad c_f=\frac{N^2-1}{2N^2}, \quad d_1=\frac{N^4-6N^2+18}{24N^4}, \quad d_2=\frac{N^2+6}{24N^2}, \quad d_3=\frac{N^2+36}{24N^2}.
\label{di}
\end{equation}
Hence, for QCD with gauge group $\text{SU}(3)$, we obtain
\begin{equation}
n_f=\frac{N_f}{6}, \qquad c_f=\frac{4}{9}, \qquad d_1=\frac{5}{216}, \qquad d_2=\frac{5}{72}, \qquad d_3=\frac{5}{24}.
\label{diQCD}
\end{equation}
The Abelian reduction to QED corresponds to the substitutions 
\begin{equation}
C_A n_f = N_f, \qquad C_A c_f = 1, \qquad C_A^2 d_1=1, \qquad d_2=0, \qquad d_3=0.
\label{diQED}
\end{equation}

\subsection{$\pi^2$-dependent terms in the $\beta$-function at six and seven loops}

The $\beta$-function of general YM theory coupled to fermions can be written as
\begin{equation}
\beta = a \sum_{i \geqslant 1} \beta_i a^i.
    \label{beta}
\end{equation}
The $\pi$-dependent terms in the six-loop coefficient can be predicted from the lower-loop results given in refs.~\cite{Herzog:2017ohr,Ruijl:2018poj,Luthe:2017ttg,Chetyrkin:2017bjc,Chetyrkin:2017bjcAddendum} together with the six-loop relations in eqs.~\eqref{eq:6LbetaConst}. The result reads
\begin{align}
\beta_6 \overset{\pi}{=} \beta_6^{\zeta_4} \zeta_4 + \beta_6^{\zeta_6} \zeta_6 - \frac{29}{12} \beta_6^{\zeta_{5,3}} \zeta_8,
\end{align}
Here, the coefficients can be decomposed as
\begin{equation}
\beta_6^{\zeta_4}= \frac{2C_A^6}{405}\,\sum_{i=0}^{5}\,\beta_{6,i}^{\zeta_4} n_f^i= \,\sum_{i=0}^{5}\,\overline{\beta}_{6,i}^{\zeta_4} N_f^i,
\qquad
\beta_6^{\zeta_6}= -\frac{10C_A^6}{27}\,\sum_{i=0}^{4}\,\beta_{6,i}^{\zeta_6} n_f^i=\,\sum_{i=0}^{5}\,\overline{\beta}_{6,i}^{\zeta_6} N_f^i , 
   \label{beta6N}
\end{equation}
where
\begin{align}
\beta_{6,0}^{\zeta_4}={}&{-26345} + 5962572 d_3, \nonumber \\
\beta_{6,1}^{\zeta_4}={}&205055- 13600584 d_2 - 4313376 d_3
+ (7596 + 380160 d_2- 142560 d_3 )\,c_f \nonumber \\
&- 1601853 c_f^2 + 1691712 c_f^3 - 114048 c_f^4, \nonumber \\
\beta_{6,2}^{\zeta_4}={}&563638 +4287024 d_1+ 9135936 d_2+ 747072 d_3
+(957552 + 1520640 d_1  \nonumber \\
&{} + 198720 d_2)\,c_f- 467712 c_f^2 - 1275696 c_f^3 + 41472 c_f^4, \nonumber \\ 
\beta_{6,3}^{\zeta_4}={}&{-353648}- 2404224 d_1- 1445760 d_2
- (570384 +  656640 d_1)\,c_f \nonumber \\
&+ 643824 c_f^2 + 279072 c_f^3, \nonumber \\ 
\beta_{6,4}^{\zeta_4}={}&31904  + 283392 d_1 + 63936 c_f - 91392 c_f^2, \nonumber \\
\beta_{6,5}^{\zeta_4}={}&2560 + 1536 c_f,
\label{beta64iN}
\end{align}
   and
\begin{align}
\beta_{6,0}^{\zeta_6}={}&2299+101640 d_3, \nonumber \\
\beta_{6,1}^{\zeta_6}={}&{-25190} - 148544 d_2 - 45760 d_3 + (57552 -42240 d_2) c_f- 84612 c_f^2 + 49368 c_f^3, \nonumber \\
\beta_{6,2}^{\zeta_6}={}&23376- 4928 d_1+ 56832 d_2+3200 d_3
   - (31488- 84480 d_1 - 15360 d_2)\,c_f \nonumber \\
&{}+ 59808 c_f^2 - 48576 c_f^3,  \nonumber \\  
\beta_{6,3}^{\zeta_6}={}&{-5984}-6656 d_1- 1024 d_2 +(3840 -30720 d_1)\,c_f
   - 10560 c_f^2 + 11136 c_f^3,  \nonumber \\
\beta_{6,4}^{\zeta_6}={}&256 + 3072 d_1.
\label{beta66iN}
\end{align}

The coefficient $\beta_{6,5}^{\zeta_4}$ is in full agreement with the large-$N_f$ expression in ref.~\cite{Gracey:2018ame}. Indeed, from eqs.~(\ref{beta6N}) and (\ref{beta64iN}), we obtain
   \begin{equation}
   \overline{\beta}_{6,5}^{\zeta_4}=\frac{C_A}{2^4}\,\frac{1}{405}\,\beta_{6,5}^{\zeta_4}=\frac{32}{405} \Bigl[5C_A+3C_F\Bigr]\,.
   \label{beta64.5}
   \end{equation}
   
The seven-loop $\pi$-dependent terms in the QED $\beta$-function predicted from eqs.~\eqref{eq:7LbetaRels} and the six-loop calculation in ref.~\cite{Li:2026bck} read as follows:
\begin{align}
\beta_7 \overset{\pi}{=}{}& \beta_7^{\zeta_4} \zeta_4 + \beta_7^{\zeta_3 \zeta_4} \zeta_3 \zeta_4 + \beta_7^{\zeta_6} \zeta_6 + \beta_7^{\zeta_8} \zeta_8 - \frac{793}{94} \beta_7^{\zeta_{7,3}} \zeta_{10} -\frac{29}{12} \beta_7^{\zeta_3 \zeta_{5,3}} \zeta_3 \zeta_8
\notag\\
 &{}+ \beta_7^{\zeta_{5,3,3}} \left( 3 \zeta_4 \zeta_7 - \frac{5}{2} \zeta_5 \zeta_6 + 45 \zeta_2 \zeta_9\right),
\end{align}
where
\begin{align}
\beta_7^{\zeta_4} &= -\frac{16}{3645} N_f^2 \left( 3520 N_f^4 + 624032 N_f^3 - 4595130 N_f^2 - 466371 N_f - 2340576 \right) , \\
\beta_7^{\zeta_3 \zeta_4} &= -\frac{1024}{27} N_f^3 \left( 24 N_f^2 + 961 N_f + 30 \right) , \\
\beta_7^{\zeta_6} &= \frac{640}{243} N_f^2 \left( 2056 N_f^3 - 16282 N_f^2 + 42159 N_f - 13500 \right) , \\
\beta_7^{\zeta_8} &= \frac{6272}{81} N_f^3 \left( 400 N_f - 3769 \right) - \frac{29}{12} \beta_7^{\zeta_{5,3}} \, .
\end{align}

\subsection{$\pi^2$-dependent terms in the ADs at six and seven loops}

The ADs in general YM theory are expanded in terms of the coupling as 
\begin{equation}
\gamma = \sum_{i \geqslant 0} \gamma_i a^i.
\end{equation}

The five-loop fermion-mass AD, a gauge independent physical quantity, was first calculated in ref.~\cite{Luthe:2016xec,Baikov:2017ujl}. The other five-loop ADs can be obtained in the Landau gauge from the results in ref.~\cite{Chetyrkin:2017bjc,Chetyrkin:2017bjcAddendum} that computed them in an arbitrary covariant gauge. Below, we combine these five-loop expressions for ADs and $\beta$-function \cite{Herzog:2017ohr,Ruijl:2018poj,Luthe:2017ttg,Chetyrkin:2017bjc,Chetyrkin:2017bjcAddendum} with the six-loop relations in eqs.~\eqref{eq:6LgamRels} to predict the $\pi$-dependent terms in the six-loop ADs for the general YM theory coupled to fermions. Analogously, we use seven-loop relations in eqs.~\eqref{eq:7LgamRels} together with six-loop ADs from ref.~\cite{Li:2026bck} to infer the $\pi$-dependent terms in the seven-loop ADs of QED.

The five-loop fermion-mass AD, a gauge-independent physical quantity, was first calculated for a general gauge group in ref.~\cite{Luthe:2016xec} (and independently confirmed in ref.~\cite{Baikov:2017ujl}). The remaining five-loop ADs in the Landau gauge can be extracted from the results of ref.~\cite{Chetyrkin:2017bjc,Chetyrkin:2017bjcAddendum}, which computed them in an arbitrary covariant gauge. Below, we combine these five-loop expressions for the ADs and the $\beta$-function~\cite{Herzog:2017ohr,Ruijl:2018poj,Luthe:2017ttg,Chetyrkin:2017bjc,Chetyrkin:2017bjcAddendum} with the six-loop relations in eqs.~\eqref{eq:6LgamRels} to predict the $\pi$-dependent terms in the six-loop ADs of general YM theory coupled to fermions. Analogously, we use the seven-loop relations in eqs.~\eqref{eq:7LgamRels} together with the six-loop ADs from ref.~\cite{Li:2026bck} to infer the $\pi$-dependent terms in the seven-loop ADs of QED with $N_f$ fermions.

\subsubsection{Fermion-mass AD}

The nonvanishing $\pi$-dependent terms read
\begin{align}
\gamma_{m,6} \overset{\pi}{=} \gamma_{m,6}^{\zeta_4} \zeta_4 + \gamma_{m,6}^{\zeta_3 \zeta_4} \zeta_3 \zeta_4 + \gamma_{m,6}^{\zeta_6} \zeta_6 +  \gamma_{m,6}^{\zeta_8} \zeta_8,
\end{align}
where
   %   Quark Mass AD
\begin{eqnarray}
&&\gamma_{m,6}^{\zeta_4}= \frac{c_fC_A^6}{810}\,\sum_{i=0}^{5}\,\gamma_{6,i}^{\zeta_4} n_f^i\,=\sum_{i=0}^{5}\,\overline{\gamma}_{6,i}^{\zeta_4} N_f^i\,,~~
\gamma_{m,6}^{\zeta_6}= -\frac{10c_fC_A^6}{27}\,\sum_{i=0}^{4}\,\gamma_{6,i}^{\zeta_6} n_f^i=\sum_{i=0}^{4}\,\overline{\gamma}_{6,i}^{\zeta_6} N_f^i\,,\nonumber \\
&&\gamma_{m,6}^{\zeta_3 \zeta_4}= -\frac{16c_fC_A^6}{3}\,\sum_{i=0}^{3}\,\gamma_{6,i}^{\zeta_3 \zeta_4} n_f^i\,,~~ 
\gamma_{m,6}^{\zeta_8}+\frac{29}{12}\gamma_{m,6}^{\zeta_3 \zeta_5}= -\frac{196c_fC_A^6}{3}\,\sum_{i=0}^{2}\,\gamma_{6,i}^{\zeta_3 \zeta_5} n_f^i.
\label{gammaQM}
\end{eqnarray}
Here, the coefficients in the expansion in powers of $n_f$ are given by
\begin{align}
\gamma_{6,0}^{\zeta_4}={}&25273705  + 1510920 d_2+ 554472 d_3
- (77291955 - 17107200 d_2  + 213840 d_3)\,c_f\nonumber \\
&{}+ 112046490 c_f^2 - 62809020 c_f^3 + 3777840 c_f^4, \nonumber \\
\gamma_{6,1}^{\zeta_4}={}&{-78396460}- 31438800 d_1- 7653744 d_2- 537408 d_3
+ (72269226 - 21954240 d_1 \nonumber \\
&{} - 7192800 d_2)\,c_f - 41250168 c_f^2 + 51609312 c_f^3 +  352512 c_f^4, \nonumber \\   
\gamma_{6,2}^{\zeta_4}={}& 45525028+ 18831744 d_1+3680640 d_2
- (13964184-8916480 d_1)\,c_f\nonumber \\  
&{}- 21903264 c_f^2 - 13132368 c_f^3, \nonumber \\ 
\gamma_{6,3}^{\zeta_4}={}&{-5385776} - 2965248 d_1+ 272160 c_f + 7613568 c_f^2, \nonumber \\ 
\gamma_{6,4}^{\zeta_4}={}&{-609280} - 255744 c_f,\nonumber \\
\gamma_{6,5}^{\zeta_4}={}&25600;
\label{gammaQM4}
   \end{align}

\begin{align}
\gamma_{6,0}^{\zeta_6}={}&564531-  574200 d_2 + 108240 d_3
- (1062655- 629640 d_2) \,c_f\nonumber \\
&{}+ 817905 c_f^2 - 244200 c_f^3 - 51480 c_f^4 , \nonumber \\
\gamma_{6,1}^{\zeta_6}={}&{-637499}+294360 d_1 + 210144 d_2- 45120 d_3\nonumber \\
&{}+ (604206 - 293040 d_1 - 220320 d_2) \,c_f
- 370608 c_f^2 + 383952 c_f^3 + 18720 c_f^4, \nonumber \\   
\gamma_{6,2}^{\zeta_6}={}&195976- 87552 d_1 + 9984 d_2
- (48096 c_f  - 89280 d_1)\,c_f- 40800 c_f^2 - 117696 c_f^3, \nonumber \\ 
\gamma_{6,3}^{\zeta_6}={}&{-3760}  - 4992 d_1- 10336 c_f + 24000 c_f^2, \nonumber \\ 
\gamma_{6,4}^{\zeta_6}={}&{-3328} + 768 c_f;
   \label{gammaQM6}
   \end{align}
   
\begin{align}
\gamma_{6,0}^{\zeta_3\zeta_4}={}&{-1331}  + 7986 d_2 - 7986 d_3 + 1331 c_f, \nonumber \\
\gamma_{6,1}^{\zeta_3\zeta_4}={}&363  - 2904 d_1 + 2904 d_3 + 3267 c_f - 3630 c_f^2, \nonumber \\   
\gamma_{6,2}^{\zeta_3\zeta_4}={}&{-2904}  + 1056 d_1 - 1056 d_2 + 3432 c_f - 528 c_f^2, \nonumber \\ 
\gamma_{6,3}^{\zeta_3\zeta_4}={}&1072 - 1744 c_f + 672 c_f^2;
   \label{gammaQM34}
   \end{align}

\begin{align}
\gamma_{6,0}^{\zeta_3\zeta_5}={}&{-605}  - 242 d_2 + 242 d_3- 1210 c_f + 1815 c_f^2, \nonumber \\
\gamma_{6,1}^{\zeta_3\zeta_5}={}&990  + 88 d_2 - 88 d_3+ 330 c_f - 1320 c_f^3, \nonumber \\   
\gamma_{6,2}^{\zeta_3\zeta_5}={}&{-280} + 40 c_f - 240 c_f^2 + 480 c_f^3\,.
   \label{gammaQM35}
   \end{align}

The results for $\gamma_{6,5}^{\zeta_4}$  and   $\gamma_{6,4}^{\zeta_6}$ agree with the large-$N_f$ expressions in ref.~\cite{Gracey:2018ame}. Specifically, from eqs. (\ref{gammaQM}), (\ref{gammaQM4}), and (\ref{gammaQM6}),  we obtain
   \begin{eqnarray}
   &&\overline{\gamma}_{6,5}^{\zeta_4}=\frac{C_A}{2^5}\,\frac{1}{810}\,\gamma_{6,5}^{\zeta_4}=\frac{80C_F}{81};\,~~\label{gamma64.5}\\
   &&\overline{\gamma}_{6,4}^{\zeta_6}=-\frac{C^2_A}{2^4}\,\frac{10}{27}\,\gamma_{6,4}^{\zeta_6}=\frac{160C_F}{27} \Bigl[13C_A-3C_F\Bigr]\,.
   \label{gamma66.4}
   \end{eqnarray}

The nonvanishing $\pi$-dependent terms in the seven-loop fermion-mass AD of QED are given by
\begin{align}
\gamma_{m,7} \overset{\pi}{=}{}& \gamma_{m,7}^{\zeta_4} \zeta_4 + \gamma_{m,7}^{\zeta_3 \zeta_4} \zeta_3 \zeta_4 + \gamma_{m,7}^{\zeta_4 \zeta_5} \zeta_4 \zeta_5 + \gamma_{m,7}^{\zeta_6} \zeta_6  + \gamma_{m,7}^{\zeta_3 \zeta_6} \zeta_3 \zeta_6 +  \gamma_{m,7}^{\zeta_8} \zeta_8 +  \gamma_{m,7}^{\zeta_{10}} \zeta_{10}
\notag\\
&{}+\gamma_{m,7}^{\zeta_{5,3,3}} \left( 3 \zeta_4 \zeta_7 - \frac{5}{2} \zeta_5 \zeta_6 + 45 \zeta_2 \zeta_9  \right) - \frac{29}{12} \gamma_{m,7}^{\zeta_3 \zeta_{5,3}} \zeta_3 \zeta_8,
\end{align}
where
\begin{align}
\gamma_{m,7}^{\zeta_4} &= \frac{16 N_f}{3645} \big( 11200 N_f^5 + 308400 N_f^4 - 1763298 N_f^3 \nonumber \\
&\hphantom{{}= \frac{16 N_f}{3645} \big(} + 49920030 N_f^2 + 66411819 N_f - 120065571 \big) , \\
\gamma_{m,7}^{\zeta_3 \zeta_4} &= -\frac{128}{9} N_f \left( 128 N_f^4 - 1512 N_f^3 - 3311 N_f^2 - 38388 N_f - 14904 \right) , \\
\gamma_{m,7}^{\zeta_4 \zeta_5} &= 51200 N_f^3 \, , \\
\gamma_{m,7}^{\zeta_6} &= -\frac{80}{243} N_f \big( 512 N_f^5 - 2880 N_f^4 + 34704 N_f^3 \nonumber \\
&\hphantom{{}= -\frac{80}{243} N_f  \big(} - 713994 N_f^2 - 1490499 N_f - 4771305 \big) , \\
\gamma_{m,7}^{\zeta_3 \zeta_6} &= \frac{256000}{3} N_f^3 \, , \\
\gamma_{m,7}^{\zeta_8} &= \frac{112}{9} N_f \left( 464 N_f^3 - 76111 N_f^2 - 182035 N_f - 113421 \right) - \frac{29}{12} \gamma_{m,7}^{\zeta_{5,3}} \, , \\
\gamma_{m,7}^{\zeta_{10}} &= 725760 N_f^2 - \frac{793}{94} \gamma_{m,7}^{\zeta_{7,3}} \, .
\end{align}

%   GF AD
\subsubsection{Gauge-field AD in the Landau gauge}

The nonvanishing $\pi$-dependent terms are
\begin{align}
\gamma_{3,6} \overset{\pi}{=} \gamma_{3,6}^{\zeta_4} \zeta_4 + \gamma_{3,6}^{\zeta_3 \zeta_4} \zeta_3 \zeta_4 + \gamma_{3,6}^{\zeta_6} \zeta_6 +  \gamma_{3,6}^{\zeta_8} \zeta_8,
\end{align}
where
\begin{align}
\gamma_{3,6}^{\zeta_4} &{}= -\frac{C_A^6}{2^{13} 3^4 5}\,\sum_{i=0}^{5}\,\gamma_{6,i}^{\zeta_4} n_f^i, \\
\gamma_{3,6}^{\zeta_6} &{}= \frac{5 C_A^6}{2^{13} 3^3}\,\sum_{i=0}^{4}\,\gamma_{6,i}^{\zeta_6} n_f^i, \\
\gamma_{3,6}^{\zeta_3 \zeta_4} &{}= \frac{C_A^6}{2^{12} 3} (-11+4n_f)\sum_{i=0}^{2}\,\gamma_{6,i}^{\zeta_3 \zeta_4} n_f^i, \\
\gamma_{3,6}^{\zeta_8} &{}= \frac{7^2 C_A^6}{2^{16} 3^2} (-11+4n_f)\sum_{i=0}^{1}\,\gamma_{6,i}^{\zeta_8} n_f^i - \frac{29}{12}\gamma_{3,6}^{\zeta_3 \zeta_5};
\end{align}
\begin{align}
\gamma_{6,0}^{\zeta_4} &= -16448954095 + 52387084152 d_3 \, , \\
\gamma_{6,1}^{\zeta_4} &= 70867522660 - 174759727104 d_2 - 42436887840 d_3 \nonumber \\
&\hphantom{{}=} - c_f \left( 37011939264 - 8471347200 d_2 - 695070720 d_3 \right) \nonumber \\
&\hphantom{{}=} - 64805826048 c_f^2 + 33398341632 c_f^3 - 2059665408 c_f^4 \, , \\
\gamma_{6,2}^{\zeta_4} &= -36931669312 + 73306865664 d_1 + 125195894784 d_2 + 8596781568 d_3 \nonumber \\
&\hphantom{{}=} + c_f \left( 31414824960 + 27462205440 d_1 + 760872960 d_2 \right) \nonumber \\
&\hphantom{{}=} + 42012278784 c_f^2 - 20295843840 c_f^3 + 679477248 c_f^4 \, , 
\end{align}
\begin{align}
\gamma_{6,3}^{\zeta_4} &= 61024256 - 39475740672 d_1 - 21847080960 d_2 \nonumber \\
&\hphantom{{}=} - c_f \left( 11088101376 + 10758389760 d_1 \right) - 2071461888 c_f^2 + 4572315648 c_f^3 \, , \\
\gamma_{6,4}^{\zeta_4} &= 1516896256 + 4643094528 d_1 + 1524105216 c_f - 1497366528 c_f^2 \, , \\
\gamma_{6,5}^{\zeta_4} &= -10485760 + 25165824 c_f \, ;
\end{align}
\begin{align}
\gamma_{6,0}^{\zeta_6} &= -1495359537 - 222639600 d_3 \, , \\
\gamma_{6,1}^{\zeta_6} &= 1896574556 - 971730944 d_2 - 118419520 d_3 \nonumber \\
&\hphantom{{}=} + c_f \left( 1956537696 - 381665280 d_2 + 70398720 d_3 \right) \nonumber \\
&\hphantom{{}=} - 2387116032 c_f^2 - 300613632 c_f^3 \, , \\
\gamma_{6,2}^{\zeta_6} &= -342549568 - 283639808 d_1 + 9584640 d_2 + 105950720 d_3 \nonumber \\
&\hphantom{{}=} - c_f \left( 1300070400 - 1525678080 d_1 - 59965440 d_2 \right) \nonumber \\
&\hphantom{{}=} + 1896087552 c_f^2 - 307494912 c_f^3 \, , \\
\gamma_{6,3}^{\zeta_6} &= -141840384 - 52428800 d_1 + 81526784 d_2 \nonumber \\
&\hphantom{{}=} + c_f \left( 213123072 - 503316480 d_1 \right) - 369623040 c_f^2 + 182452224 c_f^3 \, , \\
\gamma_{6,4}^{\zeta_6} &= 29884416 + 50331648 d_1 - 6291456 c_f \, ;
\end{align}
\begin{align}
\gamma_{6,0}^{\zeta_3 \zeta_4} &= -1702449 + 405288 d_3 \, , \\
\gamma_{6,1}^{\zeta_3 \zeta_4} &= 28384 - 10973184 d_2 + 1549056 d_3 + c_f \left( 10186752 + 3538944 d_2 \right) \nonumber \\
&\hphantom{{}=}  - 10813440 c_f^2 \, , \\
\gamma_{6,2}^{\zeta_3 \zeta_4} &= -5439488 - 3145728 d_1 - 1179648 d_2 + 11403264 c_f - 5505024 c_f^2 \, ;
\end{align}
\begin{align}
\gamma_{6,0}^{\zeta_8} &= -27827669 - 21379992 d_3 \, , \\
\gamma_{6,1}^{\zeta_8} &= 32852992 + 11010048 d_2 - 18554880 d_3 - c_f \left( 19521536 - 6684672 d_2 \right) \nonumber \\
&\hphantom{{}=} + 47185920 c_f^2 - 47185920 c_f^3 \, .
\end{align}

\subsubsection{Fermion-field AD in the Landau gauge}

The nonvanishing $\pi$-dependent terms are as follows:
\begin{align}
\gamma_{2,6} \overset{\pi}{=} \gamma_{2,6}^{\zeta_4} \zeta_4 + \gamma_{2,6}^{\zeta_3 \zeta_4} \zeta_3 \zeta_4 + \gamma_{2,6}^{\zeta_6} \zeta_6 +  \gamma_{2,6}^{\zeta_8} \zeta_8,
\end{align}
where
\begin{align}
\gamma_{2,6}^{\zeta_4} &{}= \frac{c_fC_A^6}{2^9 3^3}\,\sum_{i=0}^{5}\,\gamma_{6,i}^{\zeta_4} n_f^i, \\
\gamma_{2,6}^{\zeta_6} &{}= \frac{5c_fC_A^6}{2^{11} 3^3}\,\sum_{i=0}^{3}\,\gamma_{6,i}^{\zeta_6} n_f^i, \\
\gamma_{2,6}^{\zeta_3 \zeta_4} &{}= \frac{c_f C_A^6}{2^7 3} (11-4n_f)\sum_{i=0}^{2}\,\gamma_{6,i}^{\zeta_3 \zeta_4} n_f^i, \\
\gamma_{2,6}^{\zeta_8} &{}= \frac{7^2 c_f C_A^6}{2^{13} 3^3} (-11+4n_f)\sum_{i=0}^{2}\,\gamma_{6,i}^{\zeta_8} n_f^i - \frac{29}{12}\gamma_{m,6}^{\zeta_3 \zeta_5};
\end{align}
\begin{align}
\gamma_{6,0}^{\zeta_4} &= 476219581 - 1332062280 d_2 + 38543076 d_3\nonumber \\
&\hphantom{{}=}  - c_f \left( 2942231856 - 1334665728 d_2 - 3649536 d_3 \right) \nonumber \\
&\hphantom{{}=} + 5966005632 c_f^2 - 4578911232 c_f^3 + 1216512000 c_f^4 \, , \\
\gamma_{6,1}^{\zeta_4} &= -178099396 + 683679744 d_1 + 450436320 d_2 - 14679216 d_3 \nonumber \\
&\hphantom{{}=} + c_f \left( 1045019856 - 919683072 d_1 - 505498752 d_2 \right) \nonumber \\
&\hphantom{{}=} - 2439482688 c_f^2 + 1806987264 c_f^3 - 409190400 c_f^4 \, , \\
\gamma_{6,2}^{\zeta_4} &= 90208480 - 241827840 d_1 + 11243520 d_2 - c_f \left( 46745088 - 337084416 d_1 \right) \nonumber \\
&\hphantom{{}=} + 99366912 c_f^2 - 65015808 c_f^3 \, , \\
\gamma_{6,3}^{\zeta_4} &= -38432768 - 2949120 d_1 + 28753920 c_f - 196608 c_f^2 \, , \\
\gamma_{6,4}^{\zeta_4} &= 1310720 - 2555904 c_f \, , \\
\gamma_{6,5}^{\zeta_4} &= 262144 \, ;
\end{align}
\begin{align}
\gamma_{6,0}^{\zeta_3 \zeta_4} &= 70665 + 1625760 d_2 + 21462 d_3 - c_f \left( 1248656 + 1875456 d_2 \right) \nonumber \\
&\hphantom{{}=} + 3474624 c_f^2 - 3152896 c_f^3 + 958464 c_f^4 \, , \\
\gamma_{6,1}^{\zeta_3 \zeta_4} &= 260288 - 890880 d_1 + 17472 d_2 - c_f \left( 466176 - 1400832 d_1 \right) \nonumber \\
&\hphantom{{}=} - 24576 c_f^2 + 131072 c_f^3 \, , \\
\gamma_{6,2}^{\zeta_3 \zeta_4} &= -16384 + 32768 c_f \, ;
\end{align}
\begin{align}
\gamma_{6,0}^{\zeta_6} &= -564670377 + 610064160 d_2 - 385135080 d_3 + c_f \left( 2326685408 + 102896640 d_2 \right) \nonumber \\
&\hphantom{{}=} - 2046155520 c_f^2 - 2603745280 c_f^3 + 2291097600 c_f^4 \, , \\
\gamma_{6,1}^{\zeta_6} &= 342590444 - 1643642880 d_1 - 716891520 d_2 + 140049120 d_3 \nonumber \\
&\hphantom{{}=} - c_f \left( 1903959296 - 1832878080 d_1 + 25597440 d_2 \right) \nonumber \\
&\hphantom{{}=} + 2742850560 c_f^2 + 532971520 c_f^3 - 903905280 c_f^4 \, , \\
\gamma_{6,2}^{\zeta_6} &= 958464 + 662568960 d_1 + 182753280 d_2 + c_f \left( 239796224 - 666501120 d_1 \right) \nonumber \\
&\hphantom{{}=} - 706805760 c_f^2 + 192020480 c_f^3 \, , \\
\gamma_{6,3}^{\zeta_6} &= -15245312 - 23592960 d_1 + 46399488 c_f - 11796480 c_f^2 \, ;
\end{align}
\begin{align}
\gamma_{6,0}^{\zeta_8} &= 44688713 - 62049744 d_2 - 21442368 d_3 - c_f \left( 248900928 - 82579968 d_2 \right) \nonumber \\
&\hphantom{{}=} + 588858624 c_f^2 - 652898304 c_f^3 + 250822656 c_f^4 \, , \\
\gamma_{6,1}^{\zeta_8} &= 16106752 - 41484288 d_1 - 24029184 d_2 - c_f \left( 74084352 - 45711360 d_1 \right) \nonumber \\
&\hphantom{{}=} + 73433088 c_f^2 \, , \\
\gamma_{6,2}^{\zeta_8} &= 1161216 \, .
\end{align}

The nonvanishing $\pi$-dependent terms for the seven-loop fermion-field AD in QED are given by
\begin{align}
\gamma_{2,7} \overset{\pi}{=}{}& \gamma_{2,7}^{\zeta_4} \zeta_4 + \gamma_{2,7}^{\zeta_3 \zeta_4} \zeta_3 \zeta_4 + \gamma_{2,7}^{\zeta_3^2 \zeta_4} \zeta_3^2 \zeta_4 + \gamma_{2,7}^{\zeta_4 \zeta_5} \zeta_4 \zeta_5 + \gamma_{2,7}^{\zeta_6} \zeta_6  + \gamma_{2,7}^{\zeta_3 \zeta_6} \zeta_3 \zeta_6 +  \gamma_{2,7}^{\zeta_8} \zeta_8 +  \gamma_{2,7}^{\zeta_{10}} \zeta_{10}
\notag\\
&{}+\gamma_{2,7}^{\zeta_{5,3,3}} \left( 3 \zeta_4 \zeta_7 - \frac{5}{2} \zeta_5 \zeta_6 + 45 \zeta_2 \zeta_9  \right) - \frac{29}{12} \gamma_{2,7}^{\zeta_3 \zeta_{5,3}} \zeta_3 \zeta_8,
\end{align}
where
\begin{align}
\gamma_7^{\zeta_4} &= -\frac{4}{1215} N_f \Big( {-155195838} + 144830781 N_f + 19294650 N_f^2 \nonumber \\
&\hphantom{= -\frac{4}{1215} N_f \Big(} + 702936 N_f^3 - 27072 N_f^4 + 6400 N_f^5 \Big) \, , \\
\gamma_7^{\zeta_3 \zeta_4} &= \frac{64}{27} N_f \left( 402219 - 449136 N_f + 55196 N_f^2 + 432 N_f^3 \right) , \\
\gamma_7^{\zeta_3^2 \zeta_4} &= -12288 N_f \left( 71 + 5 N_f + N_f^2 \right) , \\
\gamma_7^{\zeta_4 \zeta_5} &= \frac{128}{3} N_f \left( 13956 + 31545 N_f + 1448 N_f^2 \right) , \\
\gamma_7^{\zeta_6} &= \frac{32}{81} N_f \left( 4296960 - 2749404 N_f + 1229203 N_f^2 + 47720 N_f^3 + 480 N_f^4 \right) , \\
\gamma_7^{\zeta_3 \zeta_6} &= \frac{128}{9} N_f \left( 36840 + 23967 N_f + 436 N_f^2 \right) , \\
\gamma_7^{\zeta_8} &= -\frac{28}{27} N_f \left( -10661427 - 3231630 N_f + 171872 N_f^2 \right) - \frac{29}{12} \gamma_7^{\zeta_{5,3}} \, , \\
\gamma_7^{\zeta_{10}} &= -\frac{32}{3} N_f \left( 2025587 + 98396 N_f + 76804 N_f^2 \right) - \frac{793}{94} \gamma_7^{\zeta_{7,3}} \, .
\end{align}

\section{Summary}
\label{sec:summary}

In this work, we investigated the transcendental structure of multi-loop massless two-point correlators. We reviewed the derivation of the so-called ``hatted'' Riemann zeta-function values presented in ref.~\cite{Kotikov:2019bqo}; this derivation is based on the Landau--Khalatnikov--Fradkin transformation and yields the complete all-order expansion of hatted depth-1 multiple zeta values (MZVs). Furthermore, we found and proved a new compact integral representation for these results.

By performing a deep $\varepsilon$-expansion of the four-loop two-point master integrals, we extended the results of refs.~\cite{Baikov:2018wgs,Baikov:2018gap,Baikov:2019zmy} to include MZVs up to weight 16. We derived $\varepsilon$-expansions for the hatted MZVs and found that the coefficients of $\zeta_{12}$, $\zeta_{14}$, and $\zeta_3\zeta_{12}$ cannot be uniquely determined from the $\varepsilon$-expansion of the four-loop two-point master integrals. For $\zeta_{12}$, this observation is entirely consistent with earlier findings in refs.~\cite{Baikov:2018gap,Baikov:2019zmy}. In contrast, all $\zeta_{16}$ terms in the master integrals can be completely absorbed into the hatted-MZV expansions, although their rational prefactors in those expansions take an extremely cumbersome form. Consequently, $\zeta_{12}$, $\zeta_{14}$, and $\zeta_3\zeta_{12}$ are contained within the hatted MZVs $\hat{\zeta}_{\boldsymbol{\sigma}}$ while simultaneously appearing as independent terms in the expansions of the master integrals. For $\zeta_{12}$, this behavior is in full agreement with the work of O.~Schnetz and collaborators (see refs.~\cite{Schnetz:2016fhy,Borinsky:2022lds} and the discussions therein). The status of $\zeta_{16}$, however, remains unclear and requires further investigation.

Drawing on the framework of refs.~\cite{Baikov:2018wgs,Baikov:2018gap,Baikov:2019zmy}, we obtained expressions for the $\pi$-dependent terms of the general anomalous dimensions (ADs) and $\beta$-functions in single-charge models up to the 15th order of perturbation theory. Up to and including the 8th order, our results are in complete agreement with those reported in refs.~\cite{Baikov:2018gap,Baikov:2019zmy} and are provided in the main body of the paper. The $\pi$-dependent terms at the 9th order are considerably more lengthy and are therefore presented in appendix~\ref{app:9Lrels}. Given the complexity of the expressions beyond the 9th order, we provide them in separate supplementary files accompanying the arXiv submission of this paper.

Using these general expressions, we predicted the $\pi$-dependent terms in the eight- and nine-loop renormalization-group (RG) functions of $\text{O}(n)$ $\phi^4$ theory. The seven-loop RG functions (and the eight-loop field AD) calculated in refs.~\cite{Schnetz:2016fhy,Schnetz:2022nsc} are in complete agreement with our constraints on the $\pi$-dependent structure.

For gauge theories coupled to fermions, we also inferred the $\pi$-dependent terms for all RG functions at six loops for a general non-Abelian semisimple gauge group and at seven loops in QED. To this end, we extracted the rational prefactors of MZV products from the existing five- and six-loop calculations in refs.~\cite{Herzog:2017ohr,Ruijl:2018poj,Chetyrkin:2017bjc,Chetyrkin:2017bjcAddendum,Luthe:2016xec,Baikov:2017ujl,Li:2026bck}. We further verified the mutual consistency between these lower-loop results and the constraints on the transcendental structure derived in this work, finding complete agreement that corroborates the conclusions of refs.~\cite{Baikov:2018wgs,Baikov:2018gap,Baikov:2019zmy,Li:2026bck}. Finally, the leading-$N_f$ terms in our six- and seven-loop predictions agree with all-order large-$N_f$ calculations (see, e.g., ref.~\cite{Gracey:2018ame}).

\acknowledgments

N.I.V.\ gratefully acknowledges Prof.\ Xurong Chen and colleagues for their hospitality during his stay at the Huizhou campus of IMP CAS, where part of this work was conducted.
A.V.K.\ is grateful to Prof.\ Pengming Zhang for inviting him to the Sun Yat-sen University in Zhuhai, where he completed his part of the draft of this paper.
A.V.K.\ and I.A.K.\ were supported in part by the Russian Science Foundation grant No. 25-22-00576.

\appendix
\section{\label{app:LeeIntegral}Integral representation for $\hat{\zeta}_{2n+1}$}
\def\theequation{A\arabic{equation}}
\setcounter{equation}{0}

Taking the integral representation for the $\zeta$-values
\begin{equation}
   \zeta_{n+1}=\frac{1}{n!} \, \int_0^1 \, \frac{dt\,}{1-t} \, \ln^{n}\left(\frac{1}{t}\right)\,,
     \label{Liz}
\end{equation}
we have
\begin{equation}
\zeta_{2m} =-\frac{1}{(2m-1)!} \, \int_0^1 \, \frac{dt }{1-t} \, \ln^{2m-1}t\,,~~
\zeta_{2m+1} =\frac{1}{(2m)!} \, \int_0^1 \, \frac{dt }{1-t} \, \ln^{2m}t\,.  \label{zeta2m1}
\end{equation}

It is now convenient to rewrite eq.~(\ref{hZeta}) by substituting $s \to m+1$:
\begin{equation}
\hat{\zeta}_{2m+1} = \zeta_{2m+1} +  \sum_{k=m+1}^{\infty}\,\zeta_{2k} \, \hat{C}_{2k,2m+1} \, \ep^{2(k-m)-1}\,. 
\label{hZeta1}
\end{equation}

The integral expression for the first term on the r.h.s.\ is shown in eq.~(\ref{zeta2m1}). Following eq.~(\ref{zeta2m1}), the second term on the r.h.s.\ can be represented as
\begin{equation}
-\sum_{k=m+1}^{\infty}\, \hat{C}_{2k,2m+1} \, \frac{\ep^{2(k-m)-1}}{(2k-1)!} \, \int_0^1 \, \frac{dt }{1-t} \,  \ln^{2k-1}t\,. 
\label{hZeta2}
\end{equation}

Using eq.~(\ref{hC}), the integrand can be rewritten as follows (with $k=m+l$):
\begin{eqnarray}
&& \hspace{-1cm} -\sum_{k=m+1}^{\infty}\, \hat{C}_{2k,2m+1} \, \frac{\ep^{2(k-m)-1}}{(2k-1)!} \,  \ln^{2k-1}t
= -\sum_{k=m+1}^{\infty}\,  \frac{b_{2(k-m)-1}\,\ep^{2(k-m)-1}}{(2m)!(2(k-m)-1)!} \,  \ln^{2k-1}t \nonumber \\
&& =-2 \,\frac{\ln^{m}t}{(2m)!} \, \sum_{l=1}^{\infty}\, B_{2l} \, \frac{2^{2l}-1}{(2l)!} \, (\ep\,\ln t)^{2l-1}
\,. \label{zeta2m3}
\end{eqnarray}

We then use the property of the Bernoulli numbers $B_{n}$,
\begin{equation}
\frac{y}{e^y-1}=1-\frac{y}{2}+\sum_{n=2}^{\infty} \frac{B_n}{n!} \,y^n=1-\frac{y}{2}+\sum_{l=1}^{\infty} \frac{B_{2l}}{(2l)!} \,y^{2l}
\,, \label{BerNum}
\end{equation}
since $B_{2m+1}=0$ for $m\geq 1$.

Thus, we obtain
\begin{equation}
\sum_{l=1}^{\infty} \frac{2^{2l}-1}{(2l)!}\,B_{2l} \,y^{2l}=\frac{y}{2}-\frac{y}{e^y+1}
\label{BerNum1}
\end{equation}
and
\begin{equation}
2\sum_{l=1}^{\infty} \frac{2^{2l}-1}{(2l)!}\,B_{2l} \,(\ep \ln t)^{2l-1}=1-\frac{2}{e^{\ep \ln t}+1}=1-\frac{2}{t^{\ep}+1}
\,.  \label{BerNum2}
\end{equation}

Combining eq.~(\ref{BerNum2}) with the expression for $\zeta_{2m+1}$ in eq.~(\ref{zeta2m1}) yields the final result given in eq.~(\ref{hzeta2m1}) of the main text.

\section{Representations of homogeneous transcendentality for master integrals}
\def\theequation{B\arabic{equation}}
\setcounter{equation}{0}
\label{app:homrepMI}

In this appendix, we list the linear combinations $M'_{i,k}$ of the master integrals in the form of eq.~\eqref{eq:Relatiion} whose expansions in the $G$-scheme\footnote{In contrast to section~\ref{sec:hat-zetas}, here we employ the $G$-scheme because it is more commonly used to construct a homogeneously transcendental representation of the master integrals.} are homogeneously transcendental,
\begin{align}\label{eq:homrep1a}
M'_{i,k} = M_{i,k}^{\prime \text{R}} + \sum_{n \geqslant 3}  \varepsilon^n \sum_{\forall \boldsymbol{\Lambda}: \lvert \boldsymbol{\Lambda} \rvert = n } M_{i,k}^{\prime \boldsymbol{\Lambda}} \zeta_{\boldsymbol{\Lambda}},
\end{align}
where $M_{i,k}^{\prime \text{R}}$ and $M_{i,k}^{\prime \boldsymbol{\Lambda}}$ are rational numbers. The multi--multi-index notation for MZV products $\zeta_{\boldsymbol{\Lambda}}$ and their transcendental weights $\lvert \boldsymbol{\Lambda} \rvert$ was introduced in eqs.~\eqref{eq:zetaexpansatz} and \eqref{eq:zetaexpansatz2}. In the series~\eqref{eq:homrep1a}, each order of the $\varepsilon$-expansion is a sum of MZV products of a single weight, with the weight increasing by 1 at each subsequent order.

The expressions below (except for complexity level 0) were derived in ref.~\cite{Lee:2011jt}\footnote{\label{ftnt:homreps}To be precise, the homogeneous representations in ref.~\cite{Lee:2011jt} differ from those used in this work by a shift in the powers of the regularization parameter; i.e., they have the form $M'_{i,k} / \varepsilon^{\alpha_{i,k}}$ for an integer $\alpha_{i,k}$.}. For the last three master integrals $M_{6,k}$ ($k=1, \dots, 3$), such representations had not been constructed previously. However, ref.~\cite{Lee:2011jt} suggested a normalization in which MZVs of a given weight appear only in a limited number of orders in $\varepsilon$.

All combinations $M'_{i,k}$ in ref.~\cite{Lee:2011jt} have the structure
\begin{align}\label{eq:homrep2}
M'_{i,k} = \varepsilon^{\alpha_{i,k}} \sum_{p,r} R_{i,k}^{p,r} (\varepsilon) M_{p,r},
\end{align}
where $\alpha_{i,k}$ is an integer and the sum runs over integrals obtained from $M_{i,k}$ by contracting internal lines. In eq.~\eqref{eq:homrep2}, the rational functions $R_{i,k}^{p,r} (\varepsilon)$ are finite and nonvanishing at $\varepsilon = 0$, satisfying $0 < \lvert R_{i,k}^{p,r} (0) \rvert < \infty$. For the three master integrals $M_{6,k}$, homogeneous representations of the form~\eqref{eq:homrep2} do not exist at weight 16. However, homogeneous representations can be obtained if the rational functions in eq.~\eqref{eq:homrep2} are allowed to be singular or vanishing at $\varepsilon=0$.

Below, we present the transcendentally homogeneous integrals ordered by their complexity level. The expressions $M'_{i,k}$ for $i=1,\dots,5$ are taken from ref.~\cite{Lee:2011jt} (up to a shift in the powers of $\varepsilon$; see remark~\ref{ftnt:homreps}). The completely homogeneous forms \eqref{eq:homrep1a} for the integrals $M'_{6,k}$ are given in eqs.~\eqref{eq:M61}--\eqref{eq:M63} for the first time.

\subsection{Master integrals of the zeroth complexity level}
\begin{equation}
M'_{0,1} =
%M01
 -\frac{8 (1-5 \varepsilon ) (2-5 \varepsilon ) (3-5 \varepsilon ) (4-5 \varepsilon ) (1-4 \varepsilon ) (3-4 \varepsilon )}{(1-2 \varepsilon )^3 \varepsilon^3}M_{0,1}
%M01
 ,
\label{eq:M01}
\end{equation}
\begin{equation}
M'_{1,1} = 
%M11
\frac{6 (1-3 \varepsilon ) (2-3 \varepsilon ) (1-4 \varepsilon ) (3-4 \varepsilon )}{ (1-2 \varepsilon )^2 \varepsilon^2} M_{1,1}
%M11/
,
\end{equation}
\begin{equation}
M'_{1,2} = 
%M12
\frac{4 (1-3 \varepsilon )^2 (2-3 \varepsilon )^2}{(1-2 \varepsilon )^2 \varepsilon^2}  M_{1,2}
%M12/
,
\end{equation}
\begin{equation}
M'_{1,3} = 
%M13
\frac{8 (2-3 \varepsilon ) (1-5 \varepsilon ) (2-5 \varepsilon ) (3-5 \varepsilon ) }{3 (1-2 \varepsilon )^2 \varepsilon^2} M_{1,3}
%M13/
,
\end{equation}
\begin{equation}
M'_{1,4} = 
%M14
\frac{4 (1-4 \varepsilon ) (1-5 \varepsilon ) (2-5 \varepsilon ) (3-5 \varepsilon ) }{(1-2 \varepsilon ) (1-3 \varepsilon ) \varepsilon^2} M_{1,4}
%M14/
,
\end{equation}
\begin{equation}
M'_{2,3} = 
%M23
-\frac{2 (1-3 \varepsilon ) (2-3 \varepsilon ) }{(1-2 \varepsilon ) \varepsilon} M_{2,3}
%M23/
,
\end{equation}
\begin{equation}
M'_{2,4} = 
%M24
\frac{3 (4 \varepsilon-1)}{\varepsilon} M_{2,4}
%M24
,
\end{equation}
\begin{equation}
M'_{2,5} = 
%M25
-\frac{4 ( 1-5 \varepsilon ) ( 2-5 \varepsilon )}{3 (1-2 \varepsilon ) \varepsilon} M_{2,5}
%M25/
,
\end{equation}
\begin{equation}
M'_{3,1} = 
%M31
M_{3,1}
%M31/
.
\end{equation}

\subsection{Master integrals of the first complexity level}
\begin{multline}
M'_{2,1} = 
%M21
\frac{1-4 \varepsilon }{\varepsilon} \biggl[ M_{2,1}+\frac{(3-7 \varepsilon ) (3-5 \varepsilon ) (4-5 \varepsilon ) (3-4 \varepsilon )}{(1-3 \varepsilon ) (1-2 \varepsilon )^3}M_{0,1}\\
-\frac{(3-5 \varepsilon ) (2-3 \varepsilon ) (1-\varepsilon )}{2 (1-4 \varepsilon ) (1-2 \varepsilon )^2}M_{1,3} \biggr]
%M21/
,
\end{multline}
\begin{multline}
M'_{2,2} = 
%M22
\frac{(1-5 \varepsilon ) (2-5 \varepsilon )}{2 (1-2 \varepsilon ) \varepsilon} \Biggl[ M_{2,2}-\frac{4 (3-5 \varepsilon ) (4-5 \varepsilon ) (3-4 \varepsilon )}{(1-3 \varepsilon ) (1-2 \varepsilon )^2}M_{0,1}
\\+\frac{(2-3 \varepsilon )^2}{(2-5 \varepsilon ) (1-2 \varepsilon )}M_{1,2} \Biggr]
%M22/
,
\end{multline}
\begin{equation}
M'_{2,6} = 
%M26
\frac{(1-5 \varepsilon ) (2-5 \varepsilon ) (1-3 \varepsilon )}{2 (1-2 \varepsilon )^2 \varepsilon} \left[ M_{2,6}+\frac{4 (3-5 \varepsilon ) (4-5 \varepsilon ) (3-4 \varepsilon )}{(1-3 \varepsilon )^2 (1-2 \varepsilon )}M_{0,1} \right]
%M26/
,
\end{equation}
\begin{equation}
M'_{2,7} = 
%M27
\frac{(1-4 \varepsilon ) (2-3 \varepsilon ) }{2 (1-2 \varepsilon ) \varepsilon} \left[ M_{2,7}-\frac{(3-5 \varepsilon ) (4-5 \varepsilon ) (3-4 \varepsilon ) \left(7-41 \varepsilon +58 \varepsilon ^2\right)}{(1-4 \varepsilon ) (1-3 \varepsilon) (2-3 \varepsilon ) (1-2 \varepsilon )^2}M_{0,1} \right]
%M27/
,
\end{equation}
\begin{equation}
M'_{3,2} = 
%M32
\frac{ 1-3 \varepsilon }{1-2 \varepsilon } \left[ M_{3,2}+\frac{4 (3-4 \varepsilon ) (2-3 \varepsilon )}{(1-3 \varepsilon ) (1-2 \varepsilon )}M_{1,1} \right]
%M32/
,
\end{equation}
\begin{equation}
M'_{3,3} = 
%M33
\frac{1-4 \varepsilon}{1-2 \varepsilon } \left[ M_{3,3}+\frac{4 (2-5 \varepsilon ) (3-5 \varepsilon )}{(1-4 \varepsilon ) (1-3 \varepsilon )}M_{1,4} \right]
%M33/
,
\end{equation}
\begin{equation}
M'_{3,4} = 
%M34
\frac{(1-4 \varepsilon ) (1-3 \varepsilon )}{(1-2 \varepsilon )^2} \left[ M_{3,4}+\frac{2 (2-5 \varepsilon ) (3-5 \varepsilon ) (1-2 \varepsilon )}{(1-4 \varepsilon ) (1-3 \varepsilon )^2}M_{1,4} \right]
%M34/
,
\end{equation}
\begin{equation}
M'_{3,6} = 
%M36
\frac{ (1-5 \varepsilon ) \varepsilon^2}{(1-2 \varepsilon )^3} M_{3,6}
%M36/
.
\end{equation}

\subsection{Master integrals of the second complexity level}
\begin{equation}
M'_{3,5} = 
%M35
\frac{(1-5 \varepsilon ) (1-4 \varepsilon ) \varepsilon}{(1-2 \varepsilon )^3} M_{3,5}
%M35/
,
\end{equation}
\begin{equation}
M'_{4,1} = 
%M41
\frac{\varepsilon^2}{(1-2 \varepsilon )^2} M_{4,1}
%M41/
,
\end{equation}
\begin{equation}
M'_{4,2} = 
%M42
\frac{\varepsilon^2}{(1-2 \varepsilon )^2} M_{4,2}
%M42/
,
\end{equation}
\begin{equation}
M'_{4,3} = 
%M43
\frac{ (1+3 \varepsilon) \varepsilon^2 }{(1-2 \varepsilon )^3} M_{4,3}
%M43/
,
\end{equation}
\begin{equation}
M'_{4,4} = 
%M44
\frac{\varepsilon^3}{ (1-2 \varepsilon )^3} M_{4,4}
%M44/
,
\end{equation}
\begin{equation}
M'_{5,2} = 
%M52
\frac{\varepsilon^2}{(1-2 \varepsilon )^2} M_{5,2}
%M52/
.
\end{equation}

\subsection{Master integrals of the third complexity level}
\begin{equation}
M'_{4,5} = 
%M45
\frac{ (1-6 \varepsilon) \varepsilon^2 } {(1-2 \varepsilon )^3} M_{4,5}
%M45/
,
\end{equation}
\begin{equation}
M'_{5,1} = 
%M51
\frac{ (1+3 \varepsilon) \varepsilon^2 }{(1-2 \varepsilon )^3} M_{5,1}
%M51/
,
\end{equation}
\begin{multline}\label{eq:M61}
M'_{6,1} =
%M61
-\frac{\varepsilon^2\, (3 \varepsilon +1)^2}{(2 \varepsilon -1)^3 (5 \varepsilon +1)} \Biggl[ M_{6,1} 
- \frac{4 (4 \varepsilon -3) (5 \varepsilon -4) (5 \varepsilon -3)}{9 \varepsilon ^4\, (3 \varepsilon -1) (3 \varepsilon +1)^3} \bigl(91572 \varepsilon ^5
\\
-73433 \varepsilon ^4+8685 \varepsilon^3+6715 \varepsilon ^2-2193 \varepsilon +190\bigr) M_{0,1}
\\
   - \frac{8 (3 \varepsilon -2) (3 \varepsilon -1) (4 \varepsilon -3) (4 \varepsilon -1) (21 \varepsilon +1) (2 \varepsilon -1)}{3 \varepsilon ^3 (3 \varepsilon +1)^3} M_{1,1} 
\\
   -\frac{2 (3 \varepsilon -2)^2 \left(384 \varepsilon^3-163 \varepsilon ^2+18 \varepsilon +1\right) (2 \varepsilon -1)}{3 \varepsilon ^3 (3 \varepsilon +1)^3} M_{1,2} 
\\
   +\frac{4 (3 \varepsilon -2) (5 \varepsilon -3) \left(2489 \varepsilon ^3-958 \varepsilon ^2-109 \varepsilon +42\right) (2 \varepsilon -1)}{3 \varepsilon ^3 (3 \varepsilon +1)^3} M_{1,3} 
\\
+\frac{4 (5 \varepsilon -3) (5 \varepsilon -2) \left(840 \varepsilon ^3+385 \varepsilon ^2-276 \varepsilon +35\right) (2 \varepsilon -1)^2}{9 \varepsilon ^3\, (3 \varepsilon -1) (3 \varepsilon +1)^3} M_{1,4} 
\\
+\frac{32 (4 \varepsilon -1) (9 \varepsilon +1) (2 \varepsilon -1)^3}{3 \varepsilon ^2 (3 \varepsilon +1)^3} M_{2,1} 
-\frac{14 (5 \varepsilon -2) (5 \varepsilon -1) (2 \varepsilon -1)^2}{3 \varepsilon ^2 (3 \varepsilon +1)^2} M_{2,2} 
\\
+\frac{8 (3 \varepsilon -2) (3 \varepsilon -1) (2 \varepsilon -1)^2}{\varepsilon ^2 (3 \varepsilon +1)^2} M_{2,3} 
-\frac{80 (4 \varepsilon -1) (2 \varepsilon -1)^3}{3 \varepsilon ^2 (3 \varepsilon +1)^2} M_{2,4} 
\\
-\frac{10 (3 \varepsilon -1) (5 \varepsilon -2) (5 \varepsilon -1) (5 \varepsilon +1) (2 \varepsilon -1)}{\varepsilon ^2 (3 \varepsilon +1)^3} M_{2,6} 
\\
+\frac{40 (3 \varepsilon -2) (4 \varepsilon -1) (2 \varepsilon -1)^2}{3 \varepsilon ^2 (3 \varepsilon +1)^2} M_{2,7} 
-\frac{16 (4 \varepsilon -1) (2 \varepsilon -1)^2}{3 (3 \varepsilon +1)^3} M_{3,3} 
\\
+\frac{8 (3 \varepsilon -1) (4 \varepsilon -1) (2 \varepsilon -1)}{\varepsilon  (3 \varepsilon +1)^2} M_{3,4} 
-\frac{20 \varepsilon \, (5 \varepsilon -1) (5 \varepsilon +1)}{3 (3 \varepsilon +1)^3} M_{3,6} 
\\
-\frac{8 \varepsilon ^2}{(3 \varepsilon +1)^2} M_{4,4} \Biggr]
%M61/
.
\end{multline}

\subsection{Master integrals of the fourth complexity level}\label{app:MI_cl4}

\begin{multline}
M'_{6,2} =
%M62
	-\frac{(3 \varepsilon +1)^2 \varepsilon ^3}{(2 \varepsilon -1)^3 (4 \varepsilon +1) (5 \varepsilon +1)} M_{6,2} 
	\\
	+ \frac{4 (4 \varepsilon -3) (5 \varepsilon -4) (5 \varepsilon -3) }{9 (2 \varepsilon -1)^3 (3 \varepsilon -1) (3 \varepsilon +1) (4 \varepsilon +1)^2 (5 \varepsilon +1) \varepsilon ^2} \bigl(553164 \varepsilon ^7-531063 \varepsilon ^6
	\\
	\qquad\qquad\qquad {} +59585 \varepsilon ^5+70286 \varepsilon ^4-19638 \varepsilon ^3-729 \varepsilon ^2
	+697 \varepsilon -62 \bigr) M_{0,1}
	\\
	- \frac{4 (3 \varepsilon -2) (3 \varepsilon -1) (4 \varepsilon -3) (4 \varepsilon -1) \left(696 \varepsilon ^2+253 \varepsilon +19\right)}{9 (2 \varepsilon -1)^2 (3 \varepsilon +1) (4 \varepsilon +1) (5 \varepsilon +1) \varepsilon } M_{1,1}
	\\
	-\frac{2 (3 \varepsilon -2)^2 (3 \varepsilon -1)^2 \left(829 \varepsilon ^2+435 \varepsilon +56\right)}{3 (2 \varepsilon -1)^2 (3 \varepsilon +1) (4 \varepsilon +1)^2 (5 \varepsilon +1)} M_{1,2} 
	\\
	+\frac{8 (3 \varepsilon -2) (5 \varepsilon -3) \left(31215 \varepsilon ^5-4284 \varepsilon ^4-4379 \varepsilon ^3+619 \varepsilon ^2+179 \varepsilon -10\right)}{9 (2 \varepsilon -1)^2 (3 \varepsilon +1) (4 \varepsilon +1)^2 (5 \varepsilon +1) \varepsilon} M_{1,3} 
	\\
	-\frac{4 (5 \varepsilon -3) (5 \varepsilon -2) \left(57840 \varepsilon ^5-4180 \varepsilon ^4-7245 \varepsilon ^3+1034 \varepsilon ^2+225 \varepsilon -34\right)}{9 (2 \varepsilon -1) (3 \varepsilon -1) (3 \varepsilon +1) (4 \varepsilon +1)^2 (5 \varepsilon +1) \varepsilon } M_{1,4} 
	\\
	+\frac{32 (4 \varepsilon -1) \left(195 \varepsilon ^2+83 \varepsilon +8\right)}{9 (3 \varepsilon +1) (4 \varepsilon +1) (5 \varepsilon +1)} M_{2,1} 
	\\
	-\frac{10 (3 \varepsilon -1) (5 \varepsilon -2) (5 \varepsilon -1) (6 \varepsilon +1)}{3 (2 \varepsilon -1)^2 (3 \varepsilon +1) (4 \varepsilon +1)} M_{2,6} 
	+\frac{8 (3 \varepsilon -2) (4 \varepsilon -1) (31 \varepsilon +7)}{3 (2 \varepsilon -1) (4 \varepsilon +1) (5 \varepsilon +1)} M_{2,7} 
	\\
	-\frac{16 (4 \varepsilon -1) \varepsilon ^2}{3 (2 \varepsilon -1) (3 \varepsilon +1) (5 \varepsilon +1)} M_{3,3} 
	-\frac{4 (3 \varepsilon -1) (4 \varepsilon -1) (7 \varepsilon +1) \varepsilon }{(2 \varepsilon -1)^2 (4 \varepsilon +1) (5 \varepsilon +1)} M_{3,4} 
	\\
	+\frac{8 (4 \varepsilon -1) (5 \varepsilon -1) (56 \varepsilon +11) \varepsilon ^2}{9 (2 \varepsilon -1)^3 (4 \varepsilon +1) (5 \varepsilon +1)} M_{3,5} 
	-\frac{20 (5 \varepsilon -1) (13 \varepsilon +4) \varepsilon ^4}{3 (2 \varepsilon-1)^3 (3 \varepsilon +1) (4 \varepsilon +1)^2} M_{3,6} 
	\\
	+ \frac{20 \varepsilon ^3}{(2 \varepsilon -1)^2 (4 \varepsilon +1)} M_{4,1}
	+ \frac{10 (3 \varepsilon +1) \varepsilon^3}{(2 \varepsilon -1)^3 (4 \varepsilon +1)} M_{4,3} 
	\\
	-\frac{8 (6 \varepsilon -1) (6 \varepsilon +1) \varepsilon ^3}{(2 \varepsilon -1)^3 (4 \varepsilon +1) (5 \varepsilon+1)} M_{4,5}
%M62/
	,
\end{multline}

\begin{multline}\label{eq:M63}
M'_{6,3} = 
%M63
- \frac{(3 \varepsilon +1) \varepsilon ^3}{(2 \varepsilon -1)^3 (5 \varepsilon +1)} M_{6,3}
\\ 
+ \frac{4 (4 \varepsilon -3) (5 \varepsilon -4) (5 \varepsilon -3)}{9 (2 \varepsilon -1)^3 (3 \varepsilon -1) (4 \varepsilon +1) (5 \varepsilon +1) \varepsilon ^2} \bigl(46056 \varepsilon ^5-44850 \varepsilon ^4+11387 \varepsilon ^3
\\
+1396 \varepsilon ^2-899 \varepsilon +86\bigr) M_{0,1}
\\
+ \frac{4 (3 \varepsilon -2) (3 \varepsilon -1) (4 \varepsilon -3) (4 \varepsilon -1) (140 \varepsilon +23)}{3 (2 \varepsilon -1)^2 (4 \varepsilon +1) (5 \varepsilon +1) \varepsilon } M_{1,1}
\\
- \frac{2 (3 \varepsilon -2)^2 \left(90 \varepsilon ^3+83 \varepsilon^2-26 \varepsilon +1\right)}{3 (2 \varepsilon -1)^2 (4 \varepsilon +1) (5 \varepsilon +1) \varepsilon } M_{1,2}
\\
- \frac{16 (3 \varepsilon -2) (5 \varepsilon -3) \left(465 \varepsilon ^3-192 \varepsilon ^2-16 \varepsilon +7\right)}{3 (2 \varepsilon -1)^2 (4 \varepsilon +1) (5 \varepsilon +1) \varepsilon } M_{1,3}
\\
+ \frac{56 (4 \varepsilon -1) (5 \varepsilon -3) (5 \varepsilon -2) (5 \varepsilon -1) (6 \varepsilon +1)}{9 (2 \varepsilon -1) (3 \varepsilon -1) (4 \varepsilon +1) (5 \varepsilon +1) \varepsilon } M_{1,4}
+ \frac{16 (4 \varepsilon -1)}{3 (4 \varepsilon +1)} M_{2,1}
\\
- \frac{4 (5 \varepsilon -2) (5 \varepsilon -1) (18 \varepsilon +5)}{3 (2 \varepsilon -1) (4 \varepsilon +1) (5 \varepsilon +1)} M_{2,2}
+ \frac{8 (3 \varepsilon -2) (3 \varepsilon -1) \varepsilon}{(2 \varepsilon -1) (4 \varepsilon +1) (5 \varepsilon +1)} M_{2,3}
\\
- \frac{16 (3 \varepsilon -1) (4 \varepsilon -1)}{3 (4 \varepsilon +1) (5 \varepsilon +1)} M_{2,4}
+ \frac{16 (3 \varepsilon -2) (4 \varepsilon -1) (19 \varepsilon +4)}{3 (2 \varepsilon -1) (4 \varepsilon +1) (5 \varepsilon +1)} M_{2,7}
\\
+ \frac{8 (4 \varepsilon -1) (5 \varepsilon -1) (32 \varepsilon +7) \varepsilon ^2}{3 (2 \varepsilon -1)^3 (4 \varepsilon +1) (5 \varepsilon +1)} M_{3,5}
- \frac{20 (5 \varepsilon -1) \varepsilon ^3}{3 (2 \varepsilon -1)^3 (5 \varepsilon +1)} M_{3,6}
\\
- \frac{20 \varepsilon ^3}{(2 \varepsilon -1)^2 (4 \varepsilon +1)} M_{4,2}
+ \frac{16 \varepsilon ^5}{(2 \varepsilon -1)^3 (4 \varepsilon +1) (5 \varepsilon +1)} M_{4,4}
\\
- \frac{8 (6 \varepsilon -1) (6 \varepsilon +1) \varepsilon ^3}{(2 \varepsilon -1)^3 (4 \varepsilon +1) (5 \varepsilon +1)} M_{4,5}
+ \frac{(3 \varepsilon +1) \varepsilon ^3}{(2 \varepsilon -1)^3 (5 \varepsilon +1)} M_{5,1}
%M63/
.
\end{multline}

In the MV-scheme, all coefficients multiplying $M_{p,r}$ in eqs.~(\ref{eq:M01})--(\ref{eq:M63}) must be supplemented by a factor of $(1-2\ep)^4$. Furthermore, in the MV-scheme, the basis integrals $M'_{p,r}$ acquire an additional factor of $\left[\Gamma^3(1-\ep)\Gamma(1+\ep)/\Gamma(1-2\ep)\right]^4$ compared to the corresponding basis integrals $M'_{p,r}$ in the $G$-scheme.

%\appendix
\section{\label{app:simprels}Four- and five-loop constraints on the RG functions}
\def\theequation{C\arabic{equation}}
\setcounter{equation}{0}

As an example, we present the explicit results for $a_{n,m}$
for $m \leqslant 5$.

For $a_{2,m}$, we have
\begin{equation}
a_{2,m}=\frac{1}{m}\,\sum_{l=1}^{m-1} \, (l+1)a_{1,l}\beta_{m-l}=\frac{1}{m}\,\sum_{l=1}^{m-1} \, \frac{l+1}{l}\,\beta_{l}\beta_{m-l}\,,
\label{poln.2}
\end{equation}
which yields, after some algebra,
\begin{equation}
a_{2,2}=\beta^2_{1}\,,~~a_{2,3}=\frac{7}{6}\beta_{1}\beta_{2}\,,~~a_{2,4}=\frac{5}{6}\beta_{1}\beta_{3}+\frac{3}{8}\beta^2_{2}\,,~~
a_{2,5}=\frac{13}{20}\beta_{1}\beta_{4}+\frac{17}{30}\beta_{2}\beta_{3}\,.
\label{a2n}
\end{equation}

Using eq.~\eqref{poln.1}, $a_{n,m}$ ($n \geqslant 3$) can be expressed as
\begin{eqnarray}
\z a_{3,3}=\beta^3_{1}\,,~~a_{3,4}=\frac{23}{12}\beta^2_{1}\beta_{2}\,,~~a_{3,5}=\frac{43}{30}\beta^2_{1}\beta_{3}+\frac{157}{120}\beta_{1}\beta^2_{2}\,,~~\nonumber \\
\z a_{4,4}=\beta^4_{1}\,,~~a_{4,5}=\frac{163}{60}\beta^23_{1}\beta_{2}\,,~~a_{5,5}=\beta^3_{1}\,
\label{a3n}
\end{eqnarray}
and, in general, $a_{n,n}=\beta^n_{1}$.

It is convenient to separate out the terms containing $\zeta_i$ with $i\geqslant 3$ by writing
\begin{equation}
\beta_m(\delta)=\overline{\beta}_m+\delta_m, \qquad \delta_m=\sum_{i=3}^{2m-3} \beta^{\zeta_i}_{m}\zeta_i,
\label{bm}
\end{equation}
where $\overline{\beta}_m$ contain only rational numbers.

Then, we obtain
\begin{equation}
a_{2,4}=\overline{a}_{2,4}+\frac{5}{6}\beta_1\delta_3,~~a_{2,5}=\overline{a}_{2,5}+\frac{13}{20}\beta_1\delta_4+\frac{17}{20}\beta_2\delta_4,~
a_{3,5}=\overline{a}_{3,5}+\frac{43}{30}\beta_1^2\delta_3\,,
\label{bm1}
\end{equation}
where $\overline{a}_{n,m}$ contain only rational numbers.

Now we return to eq.~\eqref{gg} and rewrite the renormalization factor $Z_g$ as
\begin{equation}
Z_g=1+ \sum_{m=1}^{\infty}Z_g^{(m)}g^{m}, \qquad Z_g^{(m)}=\sum_{n=1}^{m}\frac{a_{n,m}}{\varepsilon^n}.
\label{gg1}
\end{equation}

Up to $m=5$, we have
\begin{equation}
Z_g^{(1)}=\frac{a_{11}}{\ep}=\frac{\beta_{1}}{\ep},
\qquad
Z_g^{(2)}=\frac{a_{12}}{\ep}+\frac{a_{22}}{\ep^2}=\frac{\beta_{2}}{2\ep}+\frac{\beta^2_{2}}{\ep^2},
\label{Zg12}
\end{equation}
and, for $m\geqslant 3$,
\begin{equation}
Z^{m}_g=\overline{Z}^{m}_g+\tilde{Z}^{m}_g\,,
\label{Zg3}
\end{equation}
where
\begin{equation}
\overline{Z}^{m}_g=\sum_{n=1}^{m-2}\frac{\overline{a}_{n,m}}{\varepsilon^n}+\frac{a_{m-1,m}}{\varepsilon^{m-1}}+\frac{a_{m,m}}{\varepsilon^m}\,
\label{oZgm}
\end{equation}
contains only rational numbers.

The factors $\tilde{Z}^{m}_g$ take the form
 \begin{equation}
 \tilde{Z}^{3}_g=\frac{\delta_3}{3\ep},~~\tilde{Z}^{4}_g=\frac{\delta_4}{4\ep}+\frac{5}{6\ep^2}\beta_1\delta_3,~~
 \tilde{Z}^{5}_g=\frac{\delta_5}{5\ep}+\left(\frac{13}{20\ep^2}\beta_1\delta_{4}+\frac{17}{30\ep^2}\beta_2\delta_{3}\right)+\frac{43}{30\ep^3}\beta^2_1\delta_3,
 \label{tZgm}
\end{equation}
where $\delta_i$ are defined in eq.~(\ref{bm}).

Now we can compare the results obtained in the standard $\overline{\text{MS}}$ scheme with those in the scheme where $\zeta_{2m} \to 0$ and $\zeta_{2l+1} \to \hat{\zeta}_{2l+1}$. In such a scheme, $\delta_m$ in eq.~\eqref{bm} is replaced by $\hat{\delta}_m$, defined as
\begin{equation}
\hat{\delta}_m=\sum_{l=1}^{m-1} \beta_m^{\zeta_{2l+1}}\hat{\zeta}_{2l+1}\,,
\label{hd}
\end{equation}
Moreover, the coupling $\hat{g}$ in this scheme is related to the coupling $g$ in the $\overline{\text{MS}}$ scheme via the finite renormalization
\begin{equation}
\hat{g}=g+\sum_{i \geqslant 1}c_i g^{i+1}\,,
\label{hg}
\end{equation}
where the coefficients $c_i$ remain finite in the limit $\ep \to 0$.

Consequently, we obtain
\begin{equation}
gZ_g\equiv g(1+\overline{Z}_g+\tilde{Z}_g)=\hat{g}(1+\overline{Z}_{\hat g}+\hat{Z}_{\hat g})\,,
\label{ZgZg}
\end{equation}
where
\begin{equation}
\hat{Z}_{g}=\tilde{Z}_{g}~\mbox{with}~\zeta_{2m}\to 0, \zeta_{2l+1}\to \hat{\zeta}_{2l+1}\,.
\label{hZ}
\end{equation}

Equating the coefficients of equal powers of $g$ in eq.~\eqref{ZgZg}, one finds
\begin{equation}
c_1 = 0 \quad \text{at } a^2 \qquad \text{and} \qquad c_2 = 0 \quad \text{at } a^3\,.
\label{c12}
\end{equation}

At $\mathcal{O}(a^4)$, we have
\begin{equation}
c_3 = \tilde{Z}^{(3)}_{g} - \hat{Z}^{(3)}_{g} = \frac{1}{3\ep}\left(\delta_3 - \hat{\delta}_3\right) = \frac{\beta_{3}^{\zeta_3}}{3\ep}\left(\zeta_3 - \hat{\zeta}_3\right) = -\frac{1}{2}\beta_{3}^{\zeta_3}\zeta_4 + \mathcal{O}(\ep)\,,
\label{c3}
\end{equation}
where we have used eq.~\eqref{hatZe} for $\hat{\zeta}_3$.

At $\mathcal{O}(a^5)$, after some algebra, we obtain
\begin{equation}
c_4 = \tilde{Z}^{(4)}_{g} - \hat{Z}^{(4)}_{g} - 2c_3 Z^{(1)}_{g} = \frac{1}{4\ep}\left(\delta_4 - \hat{\delta}_4\right) + \frac{\beta_1}{6\ep^2}\left(\delta_3 - \hat{\delta}_3\right)\,.
\label{c4}
\end{equation}

Substituting eqs.~\eqref{bm} and~\eqref{hd} for $\delta_m$ and $\hat{\delta}_m$, respectively, we find
\begin{equation}
c_4 = \frac{1}{4\ep}\Bigl[\beta_4^{\zeta_4}\zeta_4 + \beta_4^{\zeta_3}\left(\zeta_3 - \hat{\zeta}_3\right)\Bigr] + \frac{\beta_1}{6\ep^2}\beta_3^{\zeta_3}\left(\zeta_3 - \hat{\zeta}_3\right)\,.
\label{c4.1}
\end{equation}

Using the expression for $\hat{\zeta}_3$ again, this simplifies to
\begin{equation}
c_4 = \frac{1}{4\ep}\left(\beta_4^{\zeta_4} - \beta_1\beta_3^{\zeta_3}\right)\zeta_4 - \frac{3}{8}\beta_4^{\zeta_3}\zeta_4 + \mathcal{O}(\ep)\,.
\label{c4.2}
\end{equation}

Requiring $c_4$ to be finite in the limit $\ep \to 0$ yields the relation
\begin{equation}
\beta_4^{\zeta_4} = \beta_1\beta_3^{\zeta_3}\,,
\label{b44}
\end{equation}
which relates the coefficients of $\zeta_4$ and $\zeta_3$ in the four-loop and three-loop contributions to the $\beta$-function, respectively.

Applying eq.~\eqref{b44}, the coefficient $c_4$ reduces to the finite expression
\begin{equation}
c_4 = -\frac{3}{8}\beta_4^{\zeta_3}\zeta_4 + \mathcal{O}(\ep)\,.
\label{c4.3}
\end{equation}

At $\mathcal{O}(a^6)$, we have
\begin{align}
c_5 &{}= \tilde{Z}^{(5)}_{g} - \hat{Z}^{(5)}_{g} - 2c_4 Z^{(1)}_{g} - 3c_3 Z^{(2)}_{g} 
\nonumber \\
&{}= \frac{1}{5\ep}(\delta_5 - \hat{\delta}_5)
+ \frac{3\beta_1}{20\ep^2}(\delta_4 - \hat{\delta}_4) + \left(\frac{\beta_2}{15\ep^2} + \frac{\beta_1^2}{10\ep^3}\right)(\delta_3 - \hat{\delta}_3)\,.
\label{c5}
\end{align}

Substituting eqs.~\eqref{bm} and~\eqref{hd} for $\delta_m$ and $\hat{\delta}_m$, we obtain
\begin{eqnarray}
\z c_5 = \frac{1}{5\ep}\Bigl(\beta_5^{\zeta_4}\zeta_4 + \beta_5^{\zeta_6}\zeta_6 + \beta_5^{\zeta_3\zeta_4}\zeta_3\zeta_4\Bigr)
+ \frac{3\beta_1}{20\ep^2}\Bigl[\beta_4^{\zeta_4}\zeta_4 + \beta_4^{\zeta_3}(\zeta_3 - \hat{\zeta}_3) + \beta_4^{\zeta_5}(\zeta_5 - \hat{\zeta}_5)\Bigr] \nonumber \\
&& + \left(\frac{\beta_2}{15\ep^2} + \frac{\beta_1^2}{10\ep^3}\right)\beta_3^{\zeta_3}(\zeta_3 - \hat{\zeta}_3) + \mathcal{O}(\ep^0)\,.
\label{c5.1}
\end{eqnarray}

Using the expressions for $\hat{\zeta}_3$ and $\hat{\zeta}_5$, this simplifies to
\begin{eqnarray}
\z c_5 = \frac{3\beta_1}{20\ep^2}\left(\beta_4^{\zeta_4} - \beta_1\beta_3^{\zeta_3}\right)\zeta_4
+ \frac{1}{5\ep}\Biggl\{\left[\beta_5^{\zeta_4} - \frac{9}{8}\beta_1\beta_4^{\zeta_3} - \frac{1}{2}\beta_2\beta_3^{\zeta_3}\right]\zeta_4 \nonumber \\
&& + \left[\beta_5^{\zeta_6} - \frac{15}{8}\beta_1\beta_4^{\zeta_5}\right]\zeta_6 + \beta_5^{\zeta_3\zeta_4}\zeta_3\zeta_4\Biggr\}
+ \mathcal{O}(\ep^0)\,.
\label{c5.2}
\end{eqnarray}

Requiring $c_5$ to be finite in the limit $\ep \to 0$, the $\mathcal{O}(\ep^{-2})$ pole vanishes identically in agreement with eq.~\eqref{b44}. Requiring the $\mathcal{O}(\ep^{-1})$ pole to vanish as well yields the new relations
\begin{equation}
\beta_5^{\zeta_4} = \frac{9}{8}\beta_1\beta_4^{\zeta_3} + \frac{1}{2}\beta_2\beta_3^{\zeta_3}\,, \qquad 
\beta_5^{\zeta_6} = \frac{15}{8}\beta_1\beta_4^{\zeta_5}\,, \qquad 
\beta_5^{\zeta_3\zeta_4} = 0\,,
\label{b54}
\end{equation}
which relate the coefficients of $\zeta_4$, $\zeta_6$, and $\zeta_3\zeta_4$ in the five-loop contribution to the $\beta$-function to the coefficients of $\zeta_3$ and $\zeta_5$ in its three- and four-loop contributions.

Once eqs.~\eqref{b44} and~\eqref{b54} are imposed, the coefficient $c_5$ becomes finite.

%===================================================================================
%===================================================================================
%===================================================================================

\section{\label{app:9Lrels}Constraints on nine-loop AD and $\beta$-function}
\def\theequation{D\arabic{equation}}
\setcounter{equation}{0}

\subsection{The $\beta$-function}

\begin{align}
\beta_9^{\zeta_4} &= \frac{21}{16}\beta_1\beta_8^{\zeta_3} + \frac{15}{14}\beta_2\beta_7^{\zeta_3} + \frac{3}{4}\beta^{\text{R}}_{3}\beta_6^{\zeta_3} + \frac{3}{10}\beta^{\text{R}}_{4}\beta_5^{\zeta_3} - \frac{3}{8}\beta^{\text{R}}_{5}\beta_4^{\zeta_3} - \frac{3}{2}\beta^{\text{R}}_{6}\beta_3^{\zeta_3}\,, \label{b9z4} \\
\beta_9^{\zeta_3\zeta_4} &= \frac{21}{8}\beta_1\beta_8^{\zeta_3^2} + \frac{15}{7}\beta_2\beta_7^{\zeta_3^2} + \frac{3}{2}\beta^{\text{R}}_{3}\beta_6^{\zeta_3^2} + \frac{3}{5}\beta^{\text{R}}_{4}\beta_5^{\zeta_3^2} - \frac{3}{4}\beta_3^{\zeta_3}\beta_6^{\zeta_3} - \frac{3}{40}\beta_4^{\zeta_3}\beta_5^{\zeta_3}\,, \label{b9z3z4} \\
\beta_9^{\zeta_3^2\zeta_4} &= \frac{63}{16}\beta_1\beta_8^{\zeta_3^3} + \frac{45}{14}\beta_2\beta_7^{\zeta_3^3} + \frac{9}{4}\beta^{\text{R}}_{3}\beta_6^{\zeta_3^3} + \frac{9}{40}\beta_4^{\zeta_3}\beta_5^{\zeta_3^2}\,, \label{b9z32z4} \\
\beta_9^{\zeta_3^3\zeta_4} &= \frac{21}{4}\beta_1\beta_8^{\zeta_3^4} + \frac{3}{4}\beta_3^{\zeta_3}\beta_6^{\zeta_3^3} + \frac{7}{16}\beta_1\beta_8^{\zeta_{6,4,1,1}}\,, \label{b9z33z4} \\
\beta_9^{\zeta_4\zeta_5} &= \frac{21}{16}\beta_1\beta_8^{\zeta_3\zeta_5} - \frac{105}{16}\beta_1\beta_8^{\zeta_{5,3}} + \frac{15}{14}\beta_2\beta_7^{\zeta_3\zeta_5} - \frac{75}{14}\beta_2\beta_7^{\zeta_{5,3}} + \frac{3}{4}\beta^{\text{R}}_{3}\beta_6^{\zeta_3\zeta_5} \nonumber \\
&\hphantom{={}} - \frac{15}{4}\beta^{\text{R}}_{3}\beta_6^{\zeta_{5,3}} - \frac{3}{2}\beta_3^{\zeta_3}\beta_6^{\zeta_5} - \frac{3}{8}\beta_4^{\zeta_3}\beta_5^{\zeta_5} + \frac{3}{10}\beta_4^{\zeta_5}\beta_5^{\zeta_3}\,, \label{b9z4z5} 
\end{align}
\begin{align}
\beta_9^{\zeta_3\zeta_4\zeta_5} &= \frac{21}{8}\beta_1\beta_8^{\zeta_3^2\zeta_5} - \frac{105}{16}\beta_1\beta_8^{\zeta_3\zeta_{5,3}} + \frac{15}{7}\beta_2\beta_7^{\zeta_3^2\zeta_5} - \frac{75}{14}\beta_2\beta_7^{\zeta_3\zeta_{5,3}} - \frac{3}{4}\beta_3^{\zeta_3}\beta_6^{\zeta_3\zeta_5} \nonumber \\
&\hphantom{={}} - \frac{15}{4}\beta_3^{\zeta_3}\beta_6^{\zeta_{5,3}} + \frac{3}{5}\beta_4^{\zeta_5}\beta_5^{\zeta_3^2} + \frac{1}{2}\beta_9^{\zeta_{6,4,1,1}}\,, \label{b9z3z4z5} \\
\beta_9^{\zeta_3^2\zeta_4\zeta_5} &= \frac{1}{2}\beta_9^{\zeta_3\zeta_{6,4,1,1}} - \frac{25}{4}\beta_9^{\zeta_{6,4,3,1,1}}\,, \label{b9z32z4z5} \\
\beta_9^{\zeta_2\zeta_5^2} &= -\frac{7}{2}\beta_9^{\zeta_{6,4,1,1}}\,, \label{b9z2z52} \\
\beta_9^{\zeta_2\zeta_3\zeta_5^2} &= -\frac{7}{2}\beta_9^{\zeta_3\zeta_{6,4,1,1}} - \frac{9}{2}\beta_9^{\zeta_{6,4,3,1,1}}\,, \label{b9z2z3z52} \\
\beta_9^{\zeta_4\zeta_5^2} &= \frac{21}{16}\beta_1\beta_8^{\zeta_3\zeta_5^2} - \frac{105}{16}\beta_1\beta_8^{\zeta_5\zeta_{5,3}} + \frac{525}{32}\beta_1\beta_8^{\zeta_{5,5,3}} + \frac{189}{8}\beta_1\beta_8^{\zeta_{7,3,3}}\,, \label{b9z4z52} \\
\beta_9^{\zeta_6} &= -\frac{25}{16}\beta_1^3\beta_6^{\zeta_3} - \frac{195}{56}\beta_1^2\beta_2\beta_5^{\zeta_3} - \frac{15}{8}\beta_1^2\beta^{\text{R}}_{3}\beta_4^{\zeta_3} + \frac{3}{8}\beta_1^2\beta^{\text{R}}_{4}\beta_3^{\zeta_3} - \frac{255}{112}\beta_1\beta_2^2\beta_4^{\zeta_3} \nonumber \\
&\hphantom{={}} - \frac{75}{56}\beta_1\beta_2\beta^{\text{R}}_{3}\beta_3^{\zeta_3} + \frac{35}{16}\beta_1\beta_8^{\zeta_5} - \frac{5}{14}\beta_2^3\beta_3^{\zeta_3} + \frac{25}{14}\beta_2\beta_7^{\zeta_5} + \frac{5}{4}\beta^{\text{R}}_{3}\beta_6^{\zeta_5} + \frac{1}{2}\beta^{\text{R}}_{4}\beta_5^{\zeta_5} \nonumber \\
&\hphantom{={}} - \frac{5}{8}\beta^{\text{R}}_{5}\beta_4^{\zeta_5}\,, \label{b9z6} \\
\beta_9^{\zeta_3\zeta_6} &= -\frac{25}{8}\beta_1^3\beta_6^{\zeta_3^2} - \frac{195}{28}\beta_1^2\beta_2\beta_5^{\zeta_3^2} - \frac{3}{2}\beta_1^2\beta_3^{\zeta_3}\beta_4^{\zeta_3} - \frac{75}{56}\beta_1\beta_2\left(\beta_3^{\zeta_3}\right)^2 + \frac{35}{16}\beta_1\beta_8^{\zeta_3\zeta_5} \nonumber \\
&\hphantom{={}} + \frac{25}{14}\beta_2\beta_7^{\zeta_3\zeta_5} + \frac{5}{4}\beta^{\text{R}}_{3}\beta_6^{\zeta_3\zeta_5} + \frac{5}{4}\beta_3^{\zeta_3}\beta_6^{\zeta_5} + \frac{1}{2}\beta_4^{\zeta_3}\beta_5^{\zeta_5} - \frac{5}{8}\beta_4^{\zeta_5}\beta_5^{\zeta_3}\,, \label{b9z3z6} \\
\beta_9^{\zeta_3^2\zeta_6} &= -\frac{75}{16}\beta_1^3\beta_6^{\zeta_3^3} + \frac{35}{16}\beta_1\beta_8^{\zeta_3^2\zeta_5} + \frac{25}{14}\beta_2\beta_7^{\zeta_3^2\zeta_5} + \frac{5}{4}\beta_3^{\zeta_3}\beta_6^{\zeta_3\zeta_5} - \frac{5}{8}\beta_4^{\zeta_5}\beta_5^{\zeta_3^2} + \frac{9}{4}\beta_9^{\zeta_{6,4,1,1}}\,, \label{b9z32z6} \\
\beta_9^{\zeta_3^3\zeta_6} &= \frac{9}{4}\beta_9^{\zeta_3\zeta_{6,4,1,1}} - \frac{3}{4}\beta_9^{\zeta_{6,4,3,1,1}}\,, \label{b9z33z6}
\\
\beta_9^{\zeta_5\zeta_6} &= -\frac{25}{16}\beta_1^3\beta_6^{\zeta_3\zeta_5} + \frac{125}{16}\beta_1^3\beta_6^{\zeta_{5,3}} + \frac{3}{8}\beta_1^2\beta_3^{\zeta_3}\beta_4^{\zeta_5} + \frac{35}{8}\beta_1\beta_8^{\zeta_5^2} - \frac{105}{8}\beta_1\beta_8^{\zeta_{7,3}} \nonumber \\
&\hphantom{={}} + \frac{25}{7}\beta_2\beta_7^{\zeta_5^2} - \frac{75}{7}\beta_2\beta_7^{\zeta_{7,3}} - \frac{1}{8}\beta_4^{\zeta_5}\beta_5^{\zeta_5} - \frac{5}{2}\beta_9^{\zeta_{5,3,3}}\,, \label{b9z5z6} \\
\beta_9^{\zeta_3\zeta_5\zeta_6} &= \frac{35}{8}\beta_1\beta_8^{\zeta_3\zeta_5^2} - \frac{105}{8}\beta_1\beta_8^{\zeta_3\zeta_{7,3}} - \frac{5}{2}\beta_9^{\zeta_3\zeta_{5,3,3}}\,, \label{b9z3z5z6} \\
\beta_9^{\zeta_2\zeta_3\zeta_7} &= -10\beta_9^{\zeta_{6,4,1,1}}\,, \label{b9z2z3z7} \\
\beta_9^{\zeta_2\zeta_3^2\zeta_7} &= -10\beta_9^{\zeta_3\zeta_{6,4,1,1}} - 16\beta_9^{\zeta_{6,4,3,1,1}}\,, \label{b9z2z32z7}
\\
\beta_9^{\zeta_4\zeta_7} &= \frac{21}{16}\beta_1\beta_8^{\zeta_3\zeta_7} - \frac{147}{8}\beta_1\beta_8^{\zeta_{7,3}} + \frac{15}{14}\beta_2\beta_7^{\zeta_3\zeta_7} - 15\beta_2\beta_7^{\zeta_{7,3}} \nonumber \\
&\hphantom{={}} - \frac{3}{2}\beta_3^{\zeta_3}\beta_6^{\zeta_7} - \frac{3}{8}\beta_4^{\zeta_3}\beta_5^{\zeta_7} + 3\beta_9^{\zeta_{5,3,3}}\,, \label{b9z4z7} \\
\beta_9^{\zeta_3\zeta_4\zeta_7} &= \frac{21}{8}\beta_1\beta_8^{\zeta_3^2\zeta_7} - \frac{147}{8}\beta_1\beta_8^{\zeta_3\zeta_{7,3}} + 3\beta_9^{\zeta_3\zeta_{5,3,3}}\,, \label{b9z3z4z7} \\
\beta_9^{\zeta_2\zeta_5\zeta_7} &= 20\beta_9^{\zeta_{5,3,3,3}}\,, \label{b9z2z5z7} \\
\beta_9^{\zeta_6\zeta_7} &= \frac{35}{16}\beta_1\beta_8^{\zeta_5\zeta_7} - \frac{525}{16}\beta_1\beta_8^{\zeta_{9,3}} + \frac{39655}{256}\beta_1\beta_8^{\zeta_{6,4,1,1}} - \frac{5}{8}\beta_4^{\zeta_5}\beta_5^{\zeta_7} - 4\beta_9^{\zeta_{7,3,3}}\,, \label{b9z6z7} 
\end{align}
\begin{align}
\beta_9^{\zeta_8} &= \frac{63}{16}\beta_1^5\beta_4^{\zeta_3} + \frac{45}{4}\beta_1^4\beta_2\beta_3^{\zeta_3} - \frac{175}{32}\beta_1^3\beta_6^{\zeta_5} - \frac{195}{16}\beta_1^2\beta_2\beta_5^{\zeta_5} - \frac{105}{16}\beta_1^2\beta^{\text{R}}_{3}\beta_4^{\zeta_5} + \frac{63}{32}\beta_1^2\beta_7^{\zeta_3^2} \nonumber \\
&\hphantom{={}} - \frac{255}{32}\beta_1\beta_2^2\beta_4^{\zeta_5} - \frac{63}{256}\beta_1\left(\beta_4^{\zeta_3}\right)^2 - \frac{189}{160}\beta_1\beta_3^{\zeta_3}\beta_5^{\zeta_3} + \frac{99}{32}\beta_1\beta_2\beta_6^{\zeta_3^2} + \frac{63}{32}\beta_1\beta^{\text{R}}_{3}\beta_5^{\zeta_3^2} \nonumber \\
&\hphantom{={}} + \frac{49}{16}\beta_1\beta_8^{\zeta_7} + \frac{9}{8}\beta_2^2\beta_5^{\zeta_3^2} - \frac{27}{32}\beta_2\beta_3^{\zeta_3}\beta_4^{\zeta_3} + \frac{5}{2}\beta_2\beta_7^{\zeta_7} + \frac{7}{4}\beta^{\text{R}}_{3}\beta_6^{\zeta_7} + \frac{7}{10}\beta^{\text{R}}_{4}\beta_5^{\zeta_7} \nonumber \\
&\hphantom{={}} - \frac{29}{12}\beta_9^{\zeta_{5,3}}\,, \label{b9z8} \\
\beta_9^{\zeta_3\zeta_8} &= -\frac{175}{32}\beta_1^3\beta_6^{\zeta_3\zeta_5} - \frac{105}{16}\beta_1^2\beta_3^{\zeta_3}\beta_4^{\zeta_5} + \frac{189}{32}\beta_1^2\beta_7^{\zeta_3^3} + \frac{297}{32}\beta_1\beta_2\beta_6^{\zeta_3^3} - \frac{63}{160}\beta_1\beta_3^{\zeta_3}\beta_5^{\zeta_3^2} \nonumber \\
&\hphantom{={}} + \frac{49}{16}\beta_1\beta_8^{\zeta_3\zeta_7} + \frac{5}{2}\beta_2\beta_7^{\zeta_3\zeta_7} + \frac{7}{4}\beta_3^{\zeta_3}\beta_6^{\zeta_7} + \frac{7}{10}\beta_4^{\zeta_3}\beta_5^{\zeta_7} - \frac{29}{12}\beta_9^{\zeta_3\zeta_{5,3}}\,, \label{b9z3z8} \\
\beta_9^{\zeta_3^2\zeta_8} &= \frac{49}{16}\beta_1\beta_8^{\zeta_3^2\zeta_7} - \frac{29}{12}\beta_9^{\zeta_3^2\zeta_{5,3}}\,, \label{b9z32z8} \\
\beta_9^{\zeta_5\zeta_8} &= \frac{63}{32}\beta_1^2\beta_7^{\zeta_3^2\zeta_5} - \frac{315}{32}\beta_1^2\beta_7^{\zeta_3\zeta_{5,3}} - \frac{315}{64}\beta_1^2\beta_7^{\zeta_{5,3,3}} + \frac{49}{16}\beta_1\beta_8^{\zeta_5\zeta_7} - \frac{147}{8}\beta_1\beta_8^{\zeta_{9,3}} \nonumber \\
&\hphantom{={}} + \frac{1421}{16}\beta_1\beta_8^{\zeta_{6,4,1,1}} + \frac{7}{10}\beta_4^{\zeta_5}\beta_5^{\zeta_7} - \frac{29}{12}\beta_9^{\zeta_5\zeta_{5,3}} + \frac{145}{12}\beta_9^{\zeta_{5,5,3}} + \frac{29}{2}\beta_9^{\zeta_{7,3,3}}\,, \label{b9z5z8} \\
\beta_9^{\zeta_7\zeta_8} &= -\frac{29}{12}\beta_9^{\zeta_7\zeta_{5,3}} - \frac{287}{6}\beta_9^{\zeta_{7,3,5}} + \frac{157}{4}\beta_9^{\zeta_{9,3,3}} - \frac{7811}{144}\beta_9^{\zeta_{6,4,3,1,1}}\,, \label{b9z7z8}
\end{align}
\begin{align}
\beta_9^{\zeta_2\zeta_9} &= 45\beta_9^{\zeta_{5,3,3}}\,, \label{b9z2z9} \\
\beta_9^{\zeta_2\zeta_3\zeta_9} &= 45\beta_9^{\zeta_3\zeta_{5,3,3}}\,, \label{b9z2z3z9} \\
\beta_9^{\zeta_4\zeta_9} &= \frac{21}{16}\beta_1\beta_8^{\zeta_3\zeta_9} - \frac{567}{16}\beta_1\beta_8^{\zeta_{9,3}} + \frac{12593}{96}\beta_1\beta_8^{\zeta_{6,4,1,1}} - \frac{3}{2}\beta_3^{\zeta_3}\beta_6^{\zeta_9} + 25\beta_9^{\zeta_{5,5,3}} + 28\beta_9^{\zeta_{7,3,3}}\,, \label{b9z4z9} \\
\beta_9^{\zeta_6\zeta_9} &= -17\beta_9^{\zeta_{7,3,5}} + \frac{29}{2}\beta_9^{\zeta_{9,3,3}} + \frac{84185}{144}\beta_9^{\zeta_{6,4,3,1,1}}\,, \label{b9z6z9} \\
\beta_9^{\zeta_{10}} &= \frac{189}{8}\beta_1^5\beta_4^{\zeta_5} - \frac{33}{8}\beta_1^4\beta_5^{\zeta_3^2} + \frac{33}{80}\beta_1^3\left(\beta_3^{\zeta_3}\right)^2 - \frac{105}{8}\beta_1^3\beta_6^{\zeta_7} - \frac{117}{4}\beta_1^2\beta_2\beta_5^{\zeta_7} - \frac{793}{94}\beta_9^{\zeta_{7,3}} \nonumber \\
&\hphantom{={}} - \frac{8715}{1504}\beta_1^2\beta_7^{\zeta_{5,3}} + \frac{1089}{224}\beta_1\beta_2\beta_6^{\zeta_3\zeta_5} - \frac{13695}{1504}\beta_1\beta_2\beta_6^{\zeta_{5,3}} - \frac{99}{128}\beta_1\beta_4^{\zeta_3}\beta_4^{\zeta_5} + \frac{9}{4}\beta^{\text{R}}_{3}\beta_6^{\zeta_9} \nonumber \\
&\hphantom{={}} + \frac{63}{16}\beta_1\beta_8^{\zeta_9} - \frac{297}{224}\beta_2\beta_3^{\zeta_3}\beta_4^{\zeta_5} + \frac{45}{14}\beta_2\beta_7^{\zeta_9} + \frac{99}{32}\beta_1^2\beta_7^{\zeta_3\zeta_5} - \frac{297}{160}\beta_1\beta_3^{\zeta_3}\beta_5^{\zeta_5}\,, \label{b9z10} \\
\beta_9^{\zeta_3\zeta_{10}} &= \frac{99}{16}\beta_1^2\beta_7^{\zeta_3^2\zeta_5} - \frac{8715}{1504}\beta_1^2\beta_7^{\zeta_3\zeta_{5,3}} + \frac{63}{16}\beta_1\beta_8^{\zeta_3\zeta_9} - \frac{5593}{128}\beta_1\beta_8^{\zeta_{6,4,1,1}} + \frac{9}{4}\beta_3^{\zeta_3}\beta_6^{\zeta_9} \nonumber \\
&\hphantom{={}}  - \frac{793}{94}\beta_9^{\zeta_3\zeta_{7,3}}\,, \label{b9z3z10} \\
\beta_9^{\zeta_5\zeta_{10}} &= -\frac{793}{94}\beta_9^{\zeta_5\zeta_{7,3}} - \frac{2520}{47}\beta_9^{\zeta_{7,3,5}} + \frac{2379}{47}\beta_9^{\zeta_{9,3,3}} - \frac{210685}{1128}\beta_9^{\zeta_{6,4,3,1,1}}\,, \label{b9z5z10} 
\\
\beta_9^{\zeta_2\zeta_{11}} &= \frac{275}{2}\beta_9^{\zeta_{5,5,3}} + \frac{407}{2}\beta_9^{\zeta_{7,3,3}}\,, \label{b9z2z11} \\
\beta_9^{\zeta_4\zeta_{11}} &= 55\beta_9^{\zeta_{7,3,5}} + 126\beta_9^{\zeta_{9,3,3}} - \frac{27199}{24}\beta_9^{\zeta_{6,4,3,1,1}}\,, \label{b9z4z11} 
\end{align}
\begin{align}
\beta_9^{\zeta_{12}} &= \frac{945945}{353792}\beta_1^3\beta_6^{\zeta_3^3} - \frac{825}{32}\beta_1^3\beta_6^{\zeta_9} + \frac{53625}{11056}\beta_1^2\beta_7^{\zeta_5^2} + \frac{189189}{44224}\beta_1^2\beta_7^{\zeta_3\zeta_7} + \frac{3}{4}\hat{\zeta}_{7,3(2)}^{12}\beta_1^2\beta_7^{\zeta_{7,3}} \nonumber \\
&\hphantom{={}} - \frac{53625}{88448}\beta_1\left(\beta_4^{\zeta_5}\right)^2 - \frac{567567}{221120}\beta_1\beta_3^{\zeta_3}\beta_5^{\zeta_7} + \frac{77}{16}\beta_1\beta_8^{\zeta_{11}} - \frac{609609}{176896}\beta_1\beta_8^{\zeta_3\zeta_{5,3}} \nonumber \\
&\hphantom{={}} + \frac{87087}{22112}\beta_3^{\zeta_3}\beta_6^{\zeta_{5,3}} + \frac{55}{14}\beta_2\beta_7^{\zeta_{11}} - \frac{62205}{22112}\beta_2\beta_7^{\zeta_3\zeta_{5,3}} \nonumber \\
&\hphantom{={}}  + \frac{125125}{353792}\hat{\zeta}_{5,3,3(1)}^{12}\left(49\beta_1\beta_8^{\zeta_{5,3,3}} + 40\beta_2\beta_7^{\zeta_{5,3,3}}\right) + \hat{\zeta}_{9,3(0)}^{12}\beta_9^{\zeta_{9,3}} + \hat{\zeta}_{6,4,1,1(0)}^{12}\beta_9^{\zeta_{6,4,1,1}}\,, \label{b9z12} \\
\beta_9^{\zeta_3\zeta_{12}} &= \hat{\zeta}_{9,3(0)}^{12}\beta_9^{\zeta_3\zeta_{9,3}} + \hat{\zeta}_{7,3,5(0)}^{3,12}\beta_9^{\zeta_{7,3,5}} + \hat{\zeta}_{9,3,3(0)}^{3,12}\beta_9^{\zeta_{9,3,3}} + \hat{\zeta}_{6,4,1,1(0)}^{12}\beta_9^{\zeta_3\zeta_{6,4,1,1}} \nonumber \\
&\hphantom{={}} + \hat{\zeta}_{6,4,3,1,1(0)}^{3,12}\beta_9^{\zeta_{6,4,3,1,1}}\,, \label{b9z3z12} \\
\beta_9^{\zeta_2\zeta_{13}} &= \frac{1001}{2}\beta_9^{\zeta_{7,3,5}} + \frac{1209}{2}\beta_9^{\zeta_{9,3,3}} - \frac{56717}{120}\beta_9^{\zeta_{6,4,3,1,1}}\,, \label{b9z2z13}
\end{align}
\begin{align}
\beta_9^{\zeta_{14}} &= \frac{91}{16}\beta_1\beta_8^{\zeta_{13}} - \frac{4147}{768}\beta_1\beta_8^{\zeta_5\zeta_{5,3}} - \frac{72163}{6016}\beta_1\beta_8^{\zeta_3\zeta_{7,3}} + \frac{7}{8}\hat{\zeta}_{5,5,3(1)}^{14}\beta_1\beta_8^{\zeta_{5,5,3}} \nonumber \\
&\hphantom{={}} + \frac{7}{8}\hat{\zeta}_{7,3,3(1)}^{14}\beta_1\beta_8^{\zeta_{7,3,3}} + \hat{\zeta}_{9,5(0)}^{14}\beta_9^{\zeta_{9,5}} + \hat{\zeta}_{11,3(0)}^{14}\beta_9^{\zeta_{11,3}} + \hat{\zeta}_{5,3,3,3(0)}^{14}\beta_9^{\zeta_{5,3,3,3}}\,, \label{b9z14} \\
\beta_9^{\zeta_4\zeta_{5,3}} &= \frac{21}{16}\beta_1\beta_8^{\zeta_3\zeta_{5,3}} + \frac{21}{16}\beta_1\beta_8^{\zeta_{5,3,3}} + \frac{15}{14}\beta_2\beta_7^{\zeta_3\zeta_{5,3}} + \frac{15}{14}\beta_2\beta_7^{\zeta_{5,3,3}} - \frac{3}{2}\beta_3^{\zeta_3}\beta_6^{\zeta_{5,3}} \nonumber \\
&\hphantom{={}} - \frac{3}{2}\beta_9^{\zeta_{6,4,1,1}}\,, \label{b9z4z53} \\
\beta_9^{\zeta_3\zeta_4\zeta_{5,3}} &= -\frac{3}{2}\beta_9^{\zeta_3\zeta_{6,4,1,1}} - \frac{5}{2}\beta_9^{\zeta_{6,4,3,1,1}}\,, \label{b9z3z4z53} \\
\beta_9^{\zeta_2\zeta_5\zeta_{5,3}} &= \frac{21}{2}\beta_9^{\zeta_{6,4,3,1,1}}\,, \label{b9z2z5z53} \\
\beta_9^{\zeta_6\zeta_{5,3}} &= \frac{35}{16}\beta_1\beta_8^{\zeta_5\zeta_{5,3}} + \frac{1}{2}\beta_9^{\zeta_{5,3,3,3}}\,, \label{b9z6z53} \\
\beta_9^{\zeta_2\zeta_{7,3}} &= -3\beta_9^{\zeta_{6,4,1,1}}\,, \label{b9z2z73} \\
\beta_9^{\zeta_2\zeta_3\zeta_{7,3}} &= -3\beta_9^{\zeta_3\zeta_{6,4,1,1}} - 3\beta_9^{\zeta_{6,4,3,1,1}}\,, \label{b9z2z3z73} \\
\beta_9^{\zeta_4\zeta_{7,3}} &= \frac{21}{16}\beta_1\beta_8^{\zeta_3\zeta_{7,3}} + \frac{21}{16}\beta_1\beta_8^{\zeta_{7,3,3}}\,, \label{b9z4z73} \\
\beta_9^{\zeta_2\zeta_{9,3}} &= \frac{10}{3}\beta_9^{\zeta_{5,3,3,3}}\,, \label{b9z2z93} \\
\beta_9^{\zeta_4\zeta_{5,3,3}} &= 10\beta_9^{\zeta_{6,4,3,1,1}}\,, \label{b9z4z533} \\
\beta_9^{\zeta_2\zeta_{5,5,3}} &= -\frac{14}{5}\beta_9^{\zeta_{6,4,3,1,1}}\,, \label{b9z2z553} \\
\beta_9^{\zeta_2\zeta_{7,3,3}} &= 2\beta_9^{\zeta_{6,4,3,1,1}}\,. \label{b9z2z733}
\end{align}
Here, the coefficients $\hat{\zeta}_{6,4,1,1(0)}^{12}$, $\hat{\zeta}_{5,3,3(1)}^{12}$, and $\hat{\zeta}_{9,3(0)}^{12}$ are defined in eqs.~\eqref{eq:hz_6411_12}--\eqref{eq:hz_533_12}.

\subsection{The anomalous dimension}

\begin{align}
\gamma_9^{\zeta_4} &= \frac{3}{2} \sum_{i=3}^8 \gamma^{\zeta_3}_i \beta^\text{R}_{9-i} - \frac{3}{16}\gamma_1\beta_8^{\zeta_3} - \frac{3}{7}\gamma_2\beta_7^{\zeta_3} - \frac{3}{4}\gamma^{\text{R}}_{3}\beta_6^{\zeta_3} - \frac{6}{5}\gamma^{\text{R}}_{4}\beta_5^{\zeta_3} - \frac{15}{8}\gamma^{\text{R}}_{5}\beta_4^{\zeta_3} \nonumber \\
&\hphantom{={}}  - 3\gamma^{\text{R}}_{6}\beta_3^{\zeta_3}\,, \label{g9z4} \\
\gamma_9^{\zeta_3\zeta_4} &= 3\beta_1\gamma_8^{\zeta_3^2} + 3\beta_2\gamma_7^{\zeta_3^2} + 3\beta^{\text{R}}_{3}\gamma_6^{\zeta_3^2} + 3\beta^{\text{R}}_{4}\gamma_5^{\zeta_3^2} + \frac{3}{4}\beta_6^{\zeta_3}\gamma_3^{\zeta_3} + \frac{3}{10}\beta_5^{\zeta_3}\gamma_4^{\zeta_3} \nonumber \\
&\hphantom{={}} - \frac{3}{8}\beta_4^{\zeta_3}\gamma_5^{\zeta_3} - \frac{3}{2}\beta_3^{\zeta_3}\gamma_6^{\zeta_3} - \frac{3}{8}\gamma_1\beta_8^{\zeta_3^2} - \frac{6}{7}\gamma_2\beta_7^{\zeta_3^2} - \frac{3}{2}\gamma^{\text{R}}_{3}\beta_6^{\zeta_3^2} - \frac{12}{5}\gamma^{\text{R}}_{4}\beta_5^{\zeta_3^2}\,, \label{g9z3z4} \\
\gamma_9^{\zeta_3^2\zeta_4} &= \frac{9}{2} \sum_{i=6}^8 \gamma^{\zeta_3^3}_i \beta^\text{R}_{9-i} + \frac{9}{8}\beta_4^{\zeta_3}\gamma_5^{\zeta_3^2} - \frac{9}{10}\beta_5^{\zeta_3^2}\gamma_4^{\zeta_3} - \frac{9}{7}\gamma_2\beta_7^{\zeta_3^3} - \frac{9}{4}\gamma^{\text{R}}_{3}\beta_6^{\zeta_3^3}\,, \label{g9z32z4} \\
\gamma_9^{\zeta_3^3\zeta_4} &= 6\beta_1\gamma_8^{\zeta_3^4} + \frac{1}{2}\beta_1\gamma_8^{\zeta_{6,4,1,1}} + \frac{3}{2}\beta_3^{\zeta_3}\gamma_6^{\zeta_3^3} - \frac{3}{4}\beta_6^{\zeta_3^3}\gamma_3^{\zeta_3} - \frac{3}{4}\gamma_1\beta_8^{\zeta_3^4} - \frac{1}{16}\gamma_1\beta_8^{\zeta_{6,4,1,1}}\,, \label{g9z33z4} \\
\gamma_9^{\zeta_4\zeta_5} &= \frac{3}{2}\beta_1\gamma_8^{\zeta_3\zeta_5} - \frac{15}{2}\beta_1\gamma_8^{\zeta_{5,3}} + \frac{3}{2}\beta_2\gamma_7^{\zeta_3\zeta_5} - \frac{15}{2}\beta_2\gamma_7^{\zeta_{5,3}} + \frac{3}{2}\beta^{\text{R}}_{3}\gamma_6^{\zeta_3\zeta_5} - \frac{15}{2}\beta^{\text{R}}_{3}\gamma_6^{\zeta_{5,3}} \nonumber \\
&\hphantom{={}} + \frac{3}{2}\beta_6^{\zeta_5}\gamma_3^{\zeta_3} + \frac{3}{2}\beta_5^{\zeta_5}\gamma_4^{\zeta_3} - \frac{6}{5}\beta_5^{\zeta_3}\gamma_4^{\zeta_5} + \frac{3}{2}\beta_4^{\zeta_5}\gamma_5^{\zeta_3} - \frac{15}{8}\beta_4^{\zeta_3}\gamma_5^{\zeta_5} - 3\beta_3^{\zeta_3}\gamma_6^{\zeta_5} \nonumber \\
&\hphantom{={}} - \frac{3}{16}\gamma_1\beta_8^{\zeta_3\zeta_5} + \frac{15}{16}\gamma_1\beta_8^{\zeta_{5,3}} - \frac{3}{7}\gamma_2\beta_7^{\zeta_3\zeta_5} + \frac{15}{7}\gamma_2\beta_7^{\zeta_{5,3}} - \frac{3}{4}\gamma^{\text{R}}_{3}\beta_6^{\zeta_3\zeta_5} + \frac{15}{4}\gamma^{\text{R}}_{3}\beta_6^{\zeta_{5,3}}\,, \label{g9z4z5} \\
\gamma_9^{\zeta_3\zeta_4\zeta_5} &= 3\beta_1\gamma_8^{\zeta_3^2\zeta_5} - \frac{15}{2}\beta_1\gamma_8^{\zeta_3\zeta_{5,3}} + 3\beta_2\gamma_7^{\zeta_3^2\zeta_5} - \frac{15}{2}\beta_2\gamma_7^{\zeta_3\zeta_{5,3}} + \frac{3}{4}\beta_6^{\zeta_3\zeta_5}\gamma_3^{\zeta_3} + \frac{15}{4}\beta_6^{\zeta_{5,3}}\gamma_3^{\zeta_3} \nonumber \\
&\hphantom{={}} - \frac{12}{5}\beta_5^{\zeta_3^2}\gamma_4^{\zeta_5} + 3\beta_4^{\zeta_5}\gamma_5^{\zeta_3^2} - \frac{3}{2}\beta_3^{\zeta_3}\gamma_6^{\zeta_3\zeta_5} - \frac{15}{2}\beta_3^{\zeta_3}\gamma_6^{\zeta_{5,3}} - \frac{3}{8}\gamma_1\beta_8^{\zeta_3^2\zeta_5} + \frac{15}{16}\gamma_1\beta_8^{\zeta_3\zeta_{5,3}} \nonumber \\
&\hphantom{={}} - \frac{6}{7}\gamma_2\beta_7^{\zeta_3^2\zeta_5} + \frac{15}{7}\gamma_2\beta_7^{\zeta_3\zeta_{5,3}} + \frac{1}{2}\gamma_9^{\zeta_{6,4,1,1}}\,, \label{g9z3z4z5} \\
\gamma_9^{\zeta_3^2\zeta_4\zeta_5} &= \frac{1}{2}\gamma_9^{\zeta_3\zeta_{6,4,1,1}} - \frac{25}{4}\gamma_9^{\zeta_{6,4,3,1,1}}\,, \label{g9z32z4z5} \\
\gamma_9^{\zeta_2\zeta_5^2} &= -\frac{7}{2}\gamma_9^{\zeta_{6,4,1,1}}\,, \label{g9z2z52} \\
\gamma_9^{\zeta_2\zeta_3\zeta_5^2} &= -\frac{7}{2}\gamma_9^{\zeta_3\zeta_{6,4,1,1}} - \frac{9}{2}\gamma_9^{\zeta_{6,4,3,1,1}}\,, \label{g9z2z3z52} \\
\gamma_9^{\zeta_4\zeta_5^2} &= \frac{3}{2}\beta_1\gamma_8^{\zeta_3\zeta_5^2} - \frac{15}{2}\beta_1\gamma_8^{\zeta_5\zeta_{5,3}} + \frac{75}{4}\beta_1\gamma_8^{\zeta_{5,5,3}} + 27\beta_1\gamma_8^{\zeta_{7,3,3}} \nonumber \\
&\hphantom{={}} - \frac{3}{16}\gamma_1\beta_8^{\zeta_3\zeta_5^2} + \frac{15}{16}\gamma_1\beta_8^{\zeta_5\zeta_{5,3}} - \frac{75}{32}\gamma_1\beta_8^{\zeta_{5,5,3}} - \frac{27}{8}\gamma_1\beta_8^{\zeta_{7,3,3}}\,, \label{g9z4z52}
\end{align}
\begin{align}
\gamma_9^{\zeta_6} &= \frac{5}{2} \sum_{i=4}^8 \gamma^{\zeta_5}_i \beta^\text{R}_{9-i} -\frac{15}{2} \sum_{i=3}^5 \gamma^{\zeta_3}_i \beta^\text{R}_{7-i}\beta^2_1  -\frac{5}{2}\beta_1^3\gamma_6^{\zeta_3} + \frac{15}{16}\beta_1^2\gamma_1\beta_6^{\zeta_3} + \frac{15}{7}\beta_1^2\gamma_2\beta_5^{\zeta_3} \nonumber \\
&\hphantom{={}}  + \frac{15}{4}\beta_1^2\gamma^{\text{R}}_{3}\beta_4^{\zeta_3} + 6\beta_1^2\gamma^{\text{R}}_{4}\beta_3^{\zeta_3} - \frac{15}{2}\beta_1\beta_2^2\gamma_4^{\zeta_3} + \frac{15}{8}\beta_1\beta_2\gamma_1\beta_5^{\zeta_3} + \frac{30}{7}\beta_1\beta_2\gamma_2\beta_4^{\zeta_3} \nonumber \\
&\hphantom{={}}   + \frac{15}{8}\beta_1\beta^{\text{R}}_{3}\gamma_1\beta_4^{\zeta_3} - 15\beta_1\beta_2\beta^{\text{R}}_{3}\gamma_3^{\zeta_3} + \frac{15}{2}\beta_1\beta_2\gamma^{\text{R}}_{3}\beta_3^{\zeta_3} + \frac{30}{7}\beta_1\beta^{\text{R}}_{3}\gamma_2\beta_3^{\zeta_3} - \frac{5}{2}\beta_2^3\gamma_3^{\zeta_3} \nonumber \\
&\hphantom{={}} + \frac{15}{8}\beta_1\beta^{\text{R}}_{4}\gamma_1\beta_3^{\zeta_3} + \frac{15}{16}\beta_2^2\gamma_1\beta_4^{\zeta_3} + \frac{15}{7}\beta_2^2\gamma_2\beta_3^{\zeta_3} + \frac{15}{8}\beta_2\beta^{\text{R}}_{3}\gamma_1\beta_3^{\zeta_3} \nonumber \\
&\hphantom{={}} - \frac{5}{16}\gamma_1\beta_8^{\zeta_5} - \frac{5}{7}\gamma_2\beta_7^{\zeta_5} - \frac{5}{4}\gamma^{\text{R}}_{3}\beta_6^{\zeta_5} - 2\gamma^{\text{R}}_{4}\beta_5^{\zeta_5} - \frac{25}{8}\gamma^{\text{R}}_{5}\beta_4^{\zeta_5}\,, \label{g9z6} \\
\gamma_9^{\zeta_3\zeta_6} &= -5\beta_1^3\gamma_6^{\zeta_3^2} + \frac{15}{8}\beta_1^2\gamma_1\beta_6^{\zeta_3^2} - 15\beta_1^2\beta_2\gamma_5^{\zeta_3^2} + \frac{30}{7}\beta_1^2\gamma_2\beta_5^{\zeta_3^2} - \frac{3}{2}\beta_1^2\beta_3^{\zeta_3}\gamma_4^{\zeta_3} - \frac{15}{4}\beta_1^2\beta_4^{\zeta_3}\gamma_3^{\zeta_3} \nonumber \\
&\hphantom{={}} + \frac{15}{4}\beta_1\beta_2\gamma_1\beta_5^{\zeta_3^2} - \frac{15}{2}\beta_1\beta_2\beta_3^{\zeta_3}\gamma_3^{\zeta_3} + \frac{15}{4}\beta_1\gamma_1\beta_3^{\zeta_3}\beta_4^{\zeta_3} + \frac{30}{7}\beta_1\gamma_2\left(\beta_3^{\zeta_3}\right)^2 \nonumber \\
&\hphantom{={}} + \frac{15}{8}\beta_2\gamma_1\left(\beta_3^{\zeta_3}\right)^2 + \frac{5}{2} \sum_{i=6}^8 \gamma^{\zeta_3 \zeta_5}_i \beta^{\text{R}}_{9-i} + \frac{5}{2} \sum_{i=4}^6 \gamma^{\zeta_{5}}_i \beta^{\zeta_{3}}_{9-i} - \frac{25}{8}\beta_4^{\zeta_5}\gamma_5^{\zeta_3} - 2\beta_5^{\zeta_5}\gamma_4^{\zeta_3} \nonumber \\
&\hphantom{={}} - \frac{5}{4}\beta_6^{\zeta_5}\gamma_3^{\zeta_3} - \frac{5}{16}\gamma_1\beta_8^{\zeta_3\zeta_5} - \frac{5}{7}\gamma_2\beta_7^{\zeta_3\zeta_5} - \frac{5}{4}\gamma^{\text{R}}_{3}\beta_6^{\zeta_3\zeta_5}\,, \label{g9z3z6} \\
\gamma_9^{\zeta_3^2\zeta_6} &= -\frac{15}{2}\beta_1^3\gamma_6^{\zeta_3^3} + \frac{45}{16}\beta_1^2\gamma_1\beta_6^{\zeta_3^3} + \frac{5}{2}\beta_1\gamma_8^{\zeta_3^2\zeta_5} + \frac{5}{2}\beta_2\gamma_7^{\zeta_3^2\zeta_5} + \frac{5}{2}\beta_3^{\zeta_3}\gamma_6^{\zeta_3\zeta_5} + \frac{5}{2}\beta_5^{\zeta_3^2}\gamma_4^{\zeta_5} \nonumber \\
&\hphantom{={}} - \frac{25}{8}\beta_4^{\zeta_5}\gamma_5^{\zeta_3^2} - \frac{5}{4}\beta_6^{\zeta_3\zeta_5}\gamma_3^{\zeta_3} - \frac{5}{16}\gamma_1\beta_8^{\zeta_3^2\zeta_5} - \frac{5}{7}\gamma_2\beta_7^{\zeta_3^2\zeta_5} + \frac{9}{4}\gamma_9^{\zeta_{6,4,1,1}}\,, \label{g9z32z6} \\
\gamma_9^{\zeta_3^3\zeta_6} &= \frac{9}{4}\gamma_9^{\zeta_3\zeta_{6,4,1,1}} - \frac{3}{4}\gamma_9^{\zeta_{6,4,3,1,1}}\,, \label{g9z33z6} \\
\gamma_9^{\zeta_5\zeta_6} &= -\frac{5}{2}\beta_1^3\gamma_6^{\zeta_3\zeta_5} + \frac{25}{2}\beta_1^3\gamma_6^{\zeta_{5,3}} + \frac{15}{16}\beta_1^2\gamma_1\beta_6^{\zeta_3\zeta_5} - \frac{75}{16}\beta_1^2\gamma_1\beta_6^{\zeta_{5,3}} + 6\beta_1^2\beta_3^{\zeta_3}\gamma_4^{\zeta_5} + \frac{1}{2}\beta_5^{\zeta_5}\gamma_4^{\zeta_5} \nonumber \\
&\hphantom{={}} + \frac{15}{8}\beta_1\gamma_1\beta_3^{\zeta_3}\beta_4^{\zeta_5} + 5\beta_1\gamma_8^{\zeta_5^2} - 15\beta_1\gamma_8^{\zeta_{7,3}} + 5\beta_2\gamma_7^{\zeta_5^2} - 15\beta_2\gamma_7^{\zeta_{7,3}} - \frac{15}{2}\beta_1^2\beta_4^{\zeta_5}\gamma_3^{\zeta_3} \nonumber \\
&\hphantom{={}} - \frac{5}{8}\beta_4^{\zeta_5}\gamma_5^{\zeta_5} - \frac{5}{8}\gamma_1\beta_8^{\zeta_5^2} + \frac{15}{8}\gamma_1\beta_8^{\zeta_{7,3}} - \frac{10}{7}\gamma_2\beta_7^{\zeta_5^2} + \frac{30}{7}\gamma_2\beta_7^{\zeta_{7,3}} - \frac{5}{2}\gamma_9^{\zeta_{5,3,3}}\,, \label{g9z5z6}
\\
\gamma_9^{\zeta_3\zeta_5\zeta_6} &= 5\beta_1\gamma_8^{\zeta_3\zeta_5^2} - 15\beta_1\gamma_8^{\zeta_3\zeta_{7,3}} - \frac{5}{8}\gamma_1\beta_8^{\zeta_3\zeta_5^2} + \frac{15}{8}\gamma_1\beta_8^{\zeta_3\zeta_{7,3}} - \frac{5}{2}\gamma_9^{\zeta_3\zeta_{5,3,3}}\,, \label{g9z3z5z6} \\
\gamma_9^{\zeta_2\zeta_3\zeta_7} &= -10\gamma_9^{\zeta_{6,4,1,1}}\,, \label{g9z2z3z7} \\
\gamma_9^{\zeta_2\zeta_3^2\zeta_7} &= -10\gamma_9^{\zeta_3\zeta_{6,4,1,1}} - 16\gamma_9^{\zeta_{6,4,3,1,1}}\,, \label{g9z2z32z7} \\
\gamma_9^{\zeta_4\zeta_7} &= \frac{3}{2}\,\sum_{i=7}^8\gamma^{\zeta_{3}\zeta_{7}}_i\beta_{9-i}-21\sum_{i=7}^8\gamma^{\zeta_{7,3}}_i\beta_{9-i} + \frac{3}{2}\beta_6^{\zeta_7}\gamma_3^{\zeta_3} + \frac{3}{2}\beta_5^{\zeta_7}\gamma_4^{\zeta_3} - \frac{15}{8}\beta_4^{\zeta_3}\gamma_5^{\zeta_7} \nonumber \\
&\hphantom{={}} - 3\beta_3^{\zeta_3}\gamma_6^{\zeta_7} - \frac{3}{16}\gamma_1\beta_8^{\zeta_3\zeta_7} + \frac{21}{8}\gamma_1\beta_8^{\zeta_{7,3}} - \frac{3}{7}\gamma_2\beta_7^{\zeta_3\zeta_7} + 6\gamma_2\beta_7^{\zeta_{7,3}} + 3\gamma_9^{\zeta_{5,3,3}}\,, \label{g9z4z7} \\
\gamma_9^{\zeta_3\zeta_4\zeta_7} &= 3\beta_1\gamma_8^{\zeta_3^2\zeta_7} - 21\beta_1\gamma_8^{\zeta_3\zeta_{7,3}} - \frac{3}{8}\gamma_1\beta_8^{\zeta_3^2\zeta_7} + \frac{21}{8}\gamma_1\beta_8^{\zeta_3\zeta_{7,3}} + 3\gamma_9^{\zeta_3\zeta_{5,3,3}}\,, \label{g9z3z4z7} \\
\gamma_9^{\zeta_2\zeta_5\zeta_7} &= 20\gamma_9^{\zeta_{5,3,3,3}}\,, \label{g9z2z5z7} 
\end{align}
\begin{align}
\gamma_9^{\zeta_6\zeta_7} &= \frac{5}{2}\beta_1\gamma_8^{\zeta_5\zeta_7} - \frac{75}{2}\beta_1\gamma_8^{\zeta_{9,3}} + \frac{5665}{32}\beta_1\gamma_8^{\zeta_{6,4,1,1}} + \frac{5}{2}\beta_5^{\zeta_7}\gamma_4^{\zeta_5} - \frac{25}{8}\beta_4^{\zeta_5}\gamma_5^{\zeta_7} \nonumber \\
&\hphantom{={}} - \frac{5}{16}\gamma_1\beta_8^{\zeta_5\zeta_7} + \frac{75}{16}\gamma_1\beta_8^{\zeta_{9,3}} - \frac{5665}{256}\gamma_1\beta_8^{\zeta_{6,4,1,1}} - 4\gamma_9^{\zeta_{7,3,3}}\,, \label{g9z6z7} \\
\gamma_9^{\zeta_8} &= \frac{21}{2}\beta_1^5\gamma_4^{\zeta_3} - \frac{105}{16}\beta_1^4\gamma_1\beta_4^{\zeta_3} + \frac{105}{2}\beta_1^4\beta_2\gamma_3^{\zeta_3} - 15\beta_1^4\gamma_2\beta_3^{\zeta_3} - \frac{105}{4}\beta_1^3\beta_2\gamma_1\beta_3^{\zeta_3} \nonumber \\
&\hphantom{={}} - \frac{35}{4}\beta_1^3\gamma_6^{\zeta_5} + \frac{105}{32}\beta_1^2\gamma_1\beta_6^{\zeta_5} - \frac{105}{4}\beta_1^2\beta_2\gamma_5^{\zeta_5} + \frac{15}{2}\beta_1^2\gamma_2\beta_5^{\zeta_5} - \frac{105}{4}\beta_1^2\beta^{\text{R}}_{3}\gamma_4^{\zeta_5} \nonumber \\
&\hphantom{={}} + \frac{105}{8}\beta_1^2\gamma^{\text{R}}_{3}\beta_4^{\zeta_5} + \frac{21}{8}\beta_1^2\gamma_7^{\zeta_3^2} - \frac{105}{4}\beta_1\beta_2^2\gamma_4^{\zeta_5} + \frac{105}{16}\beta_1\beta_2\gamma_1\beta_5^{\zeta_5} + 15\beta_1\beta_2\gamma_2\beta_4^{\zeta_5} \nonumber \\
&\hphantom{={}} + \frac{105}{16}\beta_1\beta^{\text{R}}_{3}\gamma_1\beta_4^{\zeta_5} + \frac{21}{4}\beta_1\beta_2\gamma_6^{\zeta_3^2} + \frac{21}{4}\beta_1\beta^{\text{R}}_{3}\gamma_5^{\zeta_3^2} - \frac{21}{8}\beta_1\beta_3^{\zeta_3}\gamma_5^{\zeta_3} - \frac{21}{32}\beta_1\beta_4^{\zeta_3}\gamma_4^{\zeta_3} \nonumber \\
&\hphantom{={}} + \frac{21}{40}\beta_1\beta_5^{\zeta_3}\gamma_3^{\zeta_3} - \frac{21}{32}\beta_1\gamma_1\beta_7^{\zeta_3^2} - \frac{3}{2}\beta_1\gamma_2\beta_6^{\zeta_3^2} - \frac{21}{8}\beta_1\gamma^{\text{R}}_{3}\beta_5^{\zeta_3^2} + \frac{7}{2}\beta_1\gamma_8^{\zeta_7} \nonumber \\
&\hphantom{={}} + \frac{105}{32}\beta_2^2\gamma_1\beta_4^{\zeta_5} + \frac{21}{8}\beta_2^2\gamma_5^{\zeta_3^2} - \frac{21}{32}\beta_2\gamma_1\beta_6^{\zeta_3^2} - \frac{3}{2}\beta_2\gamma_2\beta_5^{\zeta_3^2} - \frac{21}{8}\beta_2\beta_3^{\zeta_3}\gamma_4^{\zeta_3} \nonumber \\
&\hphantom{={}} - \frac{21}{32}\beta_2\beta_4^{\zeta_3}\gamma_3^{\zeta_3} + \frac{39}{16}\gamma_2\beta_3^{\zeta_3}\beta_4^{\zeta_3} - \frac{21}{32}\beta^{\text{R}}_{3}\gamma_1\beta_5^{\zeta_3^2} - \frac{21}{8}\beta^{\text{R}}_{3}\beta_3^{\zeta_3}\gamma_3^{\zeta_3} + \frac{21}{8}\gamma^{\text{R}}_{3}\left(\beta_3^{\zeta_3}\right)^2 \nonumber \\
&\hphantom{={}} + \frac{105}{256}\gamma_1\left(\beta_4^{\zeta_3}\right)^2 + \frac{147}{160}\gamma_1\beta_3^{\zeta_3}\beta_5^{\zeta_3} + \frac{7}{2}\beta_2\gamma_7^{\zeta_7} + \frac{7}{2}\beta^{\text{R}}_{3}\gamma_6^{\zeta_7} + \frac{7}{2}\beta^{\text{R}}_{4}\gamma_5^{\zeta_7} \nonumber \\
&\hphantom{={}} - \frac{7}{16}\gamma_1\beta_8^{\zeta_7} - \gamma_2\beta_7^{\zeta_7} - \frac{7}{4}\gamma^{\text{R}}_{3}\beta_6^{\zeta_7} - \frac{14}{5}\gamma^{\text{R}}_{4}\beta_5^{\zeta_7} - \frac{29}{12}\gamma_9^{\zeta_{5,3}}\,, \label{g9z8}
\\
\gamma_9^{\zeta_3\zeta_8} &= -\frac{35}{4}\beta_1^3\gamma_6^{\zeta_3\zeta_5} + \frac{105}{32}\beta_1^2\gamma_1\beta_6^{\zeta_3\zeta_5} - \frac{105}{4}\beta_1^2\beta_3^{\zeta_3}\gamma_4^{\zeta_5} + \frac{105}{8}\beta_1^2\beta_4^{\zeta_5}\gamma_3^{\zeta_3} + \frac{63}{8}\beta_1^2\gamma_7^{\zeta_3^3} \nonumber \\
&\hphantom{={}} + \frac{105}{16}\beta_1\gamma_1\beta_3^{\zeta_3}\beta_4^{\zeta_5} + \frac{63}{4}\beta_1\beta_2\gamma_6^{\zeta_3^3} - \frac{9}{2}\beta_1\gamma_2\beta_6^{\zeta_3^3} - \frac{63}{32}\beta_1\gamma_1\beta_7^{\zeta_3^3} - \frac{63}{40}\beta_1\beta_5^{\zeta_3^2}\gamma_3^{\zeta_3} \nonumber \\
&\hphantom{={}} + \frac{7}{2}\beta_1\gamma_8^{\zeta_3\zeta_7} - \frac{63}{32}\beta_2\gamma_1\beta_6^{\zeta_3^3} + \frac{189}{160}\gamma_1\beta_3^{\zeta_3}\beta_5^{\zeta_3^2} + \frac{7}{2}\beta_2\gamma_7^{\zeta_3\zeta_7} + \frac{7}{2}\beta_3^{\zeta_3}\gamma_6^{\zeta_7} + \frac{7}{2}\beta_4^{\zeta_3}\gamma_5^{\zeta_7} \nonumber \\
&\hphantom{={}} - \frac{14}{5}\beta_5^{\zeta_7}\gamma_4^{\zeta_3} - \frac{7}{4}\beta_6^{\zeta_7}\gamma_3^{\zeta_3} - \frac{7}{16}\gamma_1\beta_8^{\zeta_3\zeta_7} - \gamma_2\beta_7^{\zeta_3\zeta_7} - \frac{29}{12}\gamma_9^{\zeta_3\zeta_{5,3}}\,, \label{g9z3z8} \\
\gamma_9^{\zeta_3^2\zeta_8} &= \frac{7}{2}\beta_1\gamma_8^{\zeta_3^2\zeta_7} - \frac{7}{16}\gamma_1\beta_8^{\zeta_3^2\zeta_7} - \frac{29}{12}\gamma_9^{\zeta_3^2\zeta_{5,3}}\,, \label{g9z32z8} \\
\gamma_9^{\zeta_5\zeta_8} &= \frac{21}{8}\beta_1^2\gamma_7^{\zeta_3^2\zeta_5} - \frac{105}{8}\beta_1^2\gamma_7^{\zeta_3\zeta_{5,3}} - \frac{105}{16}\beta_1^2\gamma_7^{\zeta_{5,3,3}} - \frac{21}{32}\beta_1\gamma_1\beta_7^{\zeta_3^2\zeta_5} + \frac{105}{32}\beta_1\gamma_1\beta_7^{\zeta_3\zeta_{5,3}} \nonumber \\
&\hphantom{={}} + \frac{105}{64}\beta_1\gamma_1\beta_7^{\zeta_{5,3,3}} + \frac{7}{2}\beta_1\gamma_8^{\zeta_5\zeta_7} - 21\beta_1\gamma_8^{\zeta_{9,3}} + \frac{203}{2}\beta_1\gamma_8^{\zeta_{6,4,1,1}} + \frac{7}{2}\beta_4^{\zeta_5}\gamma_5^{\zeta_7} \nonumber \\
&\hphantom{={}}  - \frac{14}{5}\beta_5^{\zeta_7}\gamma_4^{\zeta_5} - \frac{7}{16}\gamma_1\beta_8^{\zeta_5\zeta_7} + \frac{21}{8}\gamma_1\beta_8^{\zeta_{9,3}} - \frac{203}{16}\gamma_1\beta_8^{\zeta_{6,4,1,1}} - \frac{29}{12}\gamma_9^{\zeta_5\zeta_{5,3}} + \frac{145}{12}\gamma_9^{\zeta_{5,5,3}}  \nonumber \\
&\hphantom{={}}  + \frac{29}{2}\gamma_9^{\zeta_{7,3,3}}\,, \label{g9z5z8} \\
\gamma_9^{\zeta_7\zeta_8} &= -\frac{29}{12}\gamma_9^{\zeta_7\zeta_{5,3}} - \frac{287}{6}\gamma_9^{\zeta_{7,3,5}} + \frac{157}{4}\gamma_9^{\zeta_{9,3,3}} - \frac{7811}{144}\gamma_9^{\zeta_{6,4,3,1,1}}\,, \label{g9z7z8} \\
\gamma_9^{\zeta_2\zeta_9} &= 45\gamma_9^{\zeta_{5,3,3}}\,, \label{g9z2z9} \\
\gamma_9^{\zeta_2\zeta_3\zeta_9} &= 45\gamma_9^{\zeta_3\zeta_{5,3,3}}\,, \label{g9z2z3z9} 
\end{align}
\begin{align}
\gamma_9^{\zeta_4\zeta_9} &= \frac{3}{2}\beta_1\gamma_8^{\zeta_3\zeta_9} - \frac{81}{2}\beta_1\gamma_8^{\zeta_{9,3}} + \frac{1799}{12}\beta_1\gamma_8^{\zeta_{6,4,1,1}} + \frac{3}{2}\beta_6^{\zeta_9}\gamma_3^{\zeta_3} - 3\beta_3^{\zeta_3}\gamma_6^{\zeta_9} \nonumber \\
&\hphantom{={}} - \frac{3}{16}\gamma_1\beta_8^{\zeta_3\zeta_9} + \frac{81}{16}\gamma_1\beta_8^{\zeta_{9,3}} - \frac{1799}{96}\gamma_1\beta_8^{\zeta_{6,4,1,1}} + 25\gamma_9^{\zeta_{5,5,3}} + 28\gamma_9^{\zeta_{7,3,3}}\,, \label{g9z4z9} \\
\gamma_9^{\zeta_6\zeta_9} &= -17\gamma_9^{\zeta_{7,3,5}} + \frac{29}{2}\gamma_9^{\zeta_{9,3,3}} + \frac{84185}{144}\gamma_9^{\zeta_{6,4,3,1,1}}\,, \label{g9z6z9} \\
\gamma_9^{\zeta_{10}} &= 63\beta_1^5\gamma_4^{\zeta_5} - \frac{315}{8}\beta_1^4\gamma_1\beta_4^{\zeta_5} - \frac{33}{4}\beta_1^4\gamma_5^{\zeta_3^2} + \frac{33}{8}\beta_1^3\gamma_1\beta_5^{\zeta_3^2} + \frac{33}{20}\beta_1^3\beta_3^{\zeta_3}\gamma_3^{\zeta_3} \nonumber \\
&\hphantom{={}} - 21\beta_1^3\gamma_6^{\zeta_7} + \frac{63}{8}\beta_1^2\gamma_1\beta_6^{\zeta_7} - 63\beta_1^2\beta_2\gamma_5^{\zeta_7} + 18\beta_1^2\gamma_2\beta_5^{\zeta_7} - \frac{99}{80}\beta_1^2\gamma_1\left(\beta_3^{\zeta_3}\right)^2 \nonumber \\
&\hphantom{={}} + \frac{33}{8}\beta_1^2\gamma_7^{\zeta_3\zeta_5} - \frac{2905}{376}\beta_1^2\gamma_7^{\zeta_{5,3}} + \frac{63}{4}\beta_1\beta_2\gamma_1\beta_5^{\zeta_7} + \frac{33}{4}\beta_1\beta_2\gamma_6^{\zeta_3\zeta_5} - \frac{2905}{188}\beta_1\beta_2\gamma_6^{\zeta_{5,3}} \nonumber \\
&\hphantom{={}} - \frac{33}{14}\beta_1\gamma_2\beta_6^{\zeta_3\zeta_5} + \frac{415}{94}\beta_1\gamma_2\beta_6^{\zeta_{5,3}} - \frac{33}{32}\beta_1\gamma_1\beta_7^{\zeta_3\zeta_5} + \frac{2905}{1504}\beta_1\gamma_1\beta_7^{\zeta_{5,3}} + \frac{33}{40}\beta_1\beta_5^{\zeta_5}\gamma_3^{\zeta_3} \nonumber \\
&\hphantom{={}} - \frac{33}{32}\beta_1\beta_4^{\zeta_5}\gamma_4^{\zeta_3} - \frac{33}{32}\beta_1\beta_4^{\zeta_3}\gamma_4^{\zeta_5} - \frac{33}{8}\beta_1\beta_3^{\zeta_3}\gamma_5^{\zeta_5} + \frac{9}{2}\beta_1\gamma_8^{\zeta_9} - \frac{33}{32}\beta_2\gamma_1\beta_6^{\zeta_3\zeta_5} \nonumber \\
&\hphantom{={}} + \frac{2905}{1504}\beta_2\gamma_1\beta_6^{\zeta_{5,3}} - \frac{33}{32}\beta_2\beta_4^{\zeta_5}\gamma_3^{\zeta_3} - \frac{33}{8}\beta_2\beta_3^{\zeta_3}\gamma_4^{\zeta_5} + \frac{429}{112}\gamma_2\beta_3^{\zeta_3}\beta_4^{\zeta_5} \nonumber \\
&\hphantom{={}} + \frac{165}{128}\gamma_1\beta_4^{\zeta_3}\beta_4^{\zeta_5} + \frac{231}{160}\gamma_1\beta_3^{\zeta_3}\beta_5^{\zeta_5} + \frac{9}{2}\beta_2\gamma_7^{\zeta_9} + \frac{9}{2}\beta^{\text{R}}_{3}\gamma_6^{\zeta_9} \nonumber \\
&\hphantom{={}} - \frac{9}{16}\gamma_1\beta_8^{\zeta_9} - \frac{9}{7}\gamma_2\beta_7^{\zeta_9} - \frac{9}{4}\gamma^{\text{R}}_{3}\beta_6^{\zeta_9} - \frac{793}{94}\gamma_9^{\zeta_{7,3}}\,, \label{g9z10}
\\
\gamma_9^{\zeta_3\zeta_{10}} &= \frac{33}{4}\beta_1^2\gamma_7^{\zeta_3^2\zeta_5} - \frac{2905}{376}\beta_1^2\gamma_7^{\zeta_3\zeta_{5,3}} - \frac{33}{16}\beta_1\gamma_1\beta_7^{\zeta_3^2\zeta_5} + \frac{2905}{1504}\beta_1\gamma_1\beta_7^{\zeta_3\zeta_{5,3}} - \frac{799}{16}\beta_1\gamma_8^{\zeta_{6,4,1,1}} \nonumber \\
&\hphantom{={}}  - \frac{9}{16}\gamma_1\beta_8^{\zeta_3\zeta_9} + \frac{9}{2}\beta_1\gamma_8^{\zeta_3\zeta_9} + \frac{9}{2}\beta_3^{\zeta_3}\gamma_6^{\zeta_9} - \frac{9}{4}\beta_6^{\zeta_9}\gamma_3^{\zeta_3} + \frac{799}{128}\gamma_1\beta_8^{\zeta_{6,4,1,1}} - \frac{793}{94}\gamma_9^{\zeta_3\zeta_{7,3}}\,, \label{g9z3z10} \\
\gamma_9^{\zeta_5\zeta_{10}} &= -\frac{793}{94}\gamma_9^{\zeta_5\zeta_{7,3}} - \frac{2520}{47}\gamma_9^{\zeta_{7,3,5}} + \frac{2379}{47}\gamma_9^{\zeta_{9,3,3}} - \frac{210685}{1128}\gamma_9^{\zeta_{6,4,3,1,1}}\,, \label{g9z5z10} \\
\gamma_9^{\zeta_2\zeta_{11}} &= \frac{275}{2}\gamma_9^{\zeta_{5,5,3}} + \frac{407}{2}\gamma_9^{\zeta_{7,3,3}}\,, \label{g9z2z11} \\
\gamma_9^{\zeta_4\zeta_{11}} &= 55\gamma_9^{\zeta_{7,3,5}} + 126\gamma_9^{\zeta_{9,3,3}} - \frac{27199}{24}\gamma_9^{\zeta_{6,4,3,1,1}}\,, \label{g9z4z11} \\
\gamma_9^{\zeta_3\zeta_{12}} &= \hat{\zeta}_{9,3(0)}^{12}\gamma_9^{\zeta_3\zeta_{9,3}} + \hat{\zeta}_{7,3,5(0)}^{3,12}\gamma_9^{\zeta_{7,3,5}} + \hat{\zeta}_{9,3,3(0)}^{3,12}\gamma_9^{\zeta_{9,3,3}} + \hat{\zeta}_{6,4,1,1(0)}^{12}\gamma_9^{\zeta_3\zeta_{6,4,1,1}} \nonumber \\
&\hphantom{={}} + \hat{\zeta}_{6,4,3,1,1(0)}^{3,12}\gamma_9^{\zeta_{6,4,3,1,1}}\,, \label{g9z3z12_full} \\
\gamma_9^{\zeta_2\zeta_{13}} &= \frac{1001}{2}\gamma_9^{\zeta_{7,3,5}} + \frac{1209}{2}\gamma_9^{\zeta_{9,3,3}} - \frac{56717}{120}\gamma_9^{\zeta_{6,4,3,1,1}}\,, \label{g9z2z13} \\
\gamma_9^{\zeta_{14}} &= \frac{13}{2}\beta_1\gamma_8^{\zeta_{13}} - \frac{4147}{672}\beta_1\gamma_8^{\zeta_5\zeta_{5,3}} - \frac{10309}{752}\beta_1\gamma_8^{\zeta_3\zeta_{7,3}} - \frac{13}{16}\gamma_1\beta_8^{\zeta_{13}} + \frac{4147}{5376}\gamma_1\beta_8^{\zeta_5\zeta_{5,3}} \nonumber \\
&\hphantom{={}} + \frac{10309}{6016}\gamma_1\beta_8^{\zeta_3\zeta_{7,3}} + \hat{\zeta}_{9,5(0)}^{14}\gamma_9^{\zeta_{9,5}} + \hat{\zeta}_{11,3(0)}^{14}\gamma_9^{\zeta_{11,3}} + \hat{\zeta}_{5,3,3,3(0)}^{14}\gamma_9^{\zeta_{5,3,3,3}} \nonumber \\
&\hphantom{={}}  + \hat{\zeta}_{5,5,3(1)}^{14}\left(\beta_1\gamma_8^{\zeta_{5,5,3}} - \frac{1}{8}\gamma_1\beta_8^{\zeta_{5,5,3}}\right) + \hat{\zeta}_{7,3,3(1)}^{14}\left(\beta_1\gamma_8^{\zeta_{7,3,3}} - \frac{1}{8}\gamma_1\beta_8^{\zeta_{7,3,3}}\right)\,, \label{g9z14_full}
\end{align}
\begin{align}
\gamma_9^{\zeta_{12}} &= \frac{189189}{44224}\beta_1^3\gamma_6^{\zeta_3^3} - \frac{165}{4}\beta_1^3\gamma_6^{\zeta_9} - \frac{567567}{353792}\beta_1^2\gamma_1\beta_6^{\zeta_3^3} + \frac{495}{32}\beta_1^2\gamma_1\beta_6^{\zeta_9} \nonumber \\
&\hphantom{={}} + \frac{17875}{2764}\beta_1^2\gamma_7^{\zeta_5^2} + \frac{63063}{11056}\beta_1^2\gamma_7^{\zeta_3\zeta_7} + \hat{\zeta}_{7,3(2)}^{12}\beta_1^2\gamma_7^{\zeta_{7,3}} - \frac{17875}{11056}\beta_1\gamma_1\beta_7^{\zeta_5^2} \nonumber \\
&\hphantom{={}} - \frac{63063}{44224}\beta_1\gamma_1\beta_7^{\zeta_3\zeta_7} - \frac{63063}{11056}\beta_1\beta_3^{\zeta_3}\gamma_5^{\zeta_7} + \frac{63063}{55280}\beta_1\beta_5^{\zeta_7}\gamma_3^{\zeta_3} - \frac{17875}{11056}\beta_1\beta_4^{\zeta_5}\gamma_4^{\zeta_5} \nonumber \\
&\hphantom{={}} + \frac{11}{2}\beta_1\gamma_8^{\zeta_{11}} - \frac{87087}{22112}\beta_1\gamma_8^{\zeta_3\zeta_{5,3}} - \frac{1}{4}\hat{\zeta}_{7,3(2)}^{12}\beta_1\gamma_1\beta_7^{\zeta_{7,3}} + \frac{875875}{44224}\hat{\zeta}_{5,3,3(1)}^{12}\beta_1\gamma_8^{\zeta_{5,3,3}} \nonumber \\
&\hphantom{={}} + \frac{89375}{88448}\gamma_1\left(\beta_4^{\zeta_5}\right)^2 + \frac{441441}{221120}\gamma_1\beta_3^{\zeta_3}\beta_5^{\zeta_7} + \frac{87087}{11056}\beta_3^{\zeta_3}\gamma_6^{\zeta_{5,3}} - \frac{87087}{22112}\beta_6^{\zeta_{5,3}}\gamma_3^{\zeta_3} \nonumber \\
&\hphantom{={}} + \frac{11}{2}\beta_2\gamma_7^{\zeta_{11}} - \frac{87087}{22112}\beta_2\gamma_7^{\zeta_3\zeta_{5,3}} - \frac{11}{16}\gamma_1\beta_8^{\zeta_{11}} + \frac{87087}{176896}\gamma_1\beta_8^{\zeta_3\zeta_{5,3}} \nonumber \\
&\hphantom{={}} - \frac{11}{7}\gamma_2\beta_7^{\zeta_{11}} + \frac{12441}{11056}\gamma_2\beta_7^{\zeta_3\zeta_{5,3}}  + \hat{\zeta}_{9,3(0)}^{12}\gamma_9^{\zeta_{9,3}} + \hat{\zeta}_{6,4,1,1(0)}^{12}\gamma_9^{\zeta_{6,4,1,1}} \nonumber \\
&\hphantom{={}}+ \hat{\zeta}_{5,3,3(1)}^{12}\left(\frac{875875}{44224}\beta_2\gamma_7^{\zeta_{5,3,3}} - \frac{875875}{353792}\gamma_1\beta_8^{\zeta_{5,3,3}} - \frac{125125}{22112}\gamma_2\beta_7^{\zeta_{5,3,3}}\right)\,, \label{g9z12_full} 
\end{align}
\begin{align}
\gamma_9^{\zeta_4\zeta_{5,3}} &= \frac{3}{2}\beta_1\gamma_8^{\zeta_3\zeta_{5,3}} + \frac{3}{2}\beta_1\gamma_8^{\zeta_{5,3,3}} + \frac{3}{2}\beta_2\gamma_7^{\zeta_3\zeta_{5,3}} + \frac{3}{2}\beta_2\gamma_7^{\zeta_{5,3,3}} + \frac{3}{2}\beta_6^{\zeta_{5,3}}\gamma_3^{\zeta_3} - 3\beta_3^{\zeta_3}\gamma_6^{\zeta_{5,3}} \nonumber \\
&\hphantom{={}} - \frac{3}{16}\gamma_1\beta_8^{\zeta_3\zeta_{5,3}} - \frac{3}{16}\gamma_1\beta_8^{\zeta_{5,3,3}} - \frac{3}{7}\gamma_2\beta_7^{\zeta_3\zeta_{5,3}} - \frac{3}{7}\gamma_2\beta_7^{\zeta_{5,3,3}} - \frac{3}{2}\gamma_9^{\zeta_{6,4,1,1}}\,, \label{g9z4z53} \\
\gamma_9^{\zeta_3\zeta_4\zeta_{5,3}} &= -\frac{3}{2}\gamma_9^{\zeta_3\zeta_{6,4,1,1}} - \frac{5}{2}\gamma_9^{\zeta_{6,4,3,1,1}}\,, \label{g9z3z4z53} \\
\gamma_9^{\zeta_2\zeta_5\zeta_{5,3}} &= \frac{21}{2}\gamma_9^{\zeta_{6,4,3,1,1}}\,, \label{g9z2z5z53} \\
\gamma_9^{\zeta_6\zeta_{5,3}} &= \frac{5}{2}\beta_1\gamma_8^{\zeta_5\zeta_{5,3}} - \frac{5}{16}\gamma_1\beta_8^{\zeta_5\zeta_{5,3}} + \frac{1}{2}\gamma_9^{\zeta_{5,3,3,3}}\,, \label{g9z6z53} \\
\gamma_9^{\zeta_2\zeta_{7,3}} &= -3\gamma_9^{\zeta_{6,4,1,1}}\,, \label{g9z2z73} \\
\gamma_9^{\zeta_2\zeta_3\zeta_{7,3}} &= -3\gamma_9^{\zeta_3\zeta_{6,4,1,1}} - 3\gamma_9^{\zeta_{6,4,3,1,1}}\,, \label{g9z2z3z73} \\
\gamma_9^{\zeta_4\zeta_{7,3}} &= \frac{3}{2}\beta_1\gamma_8^{\zeta_3\zeta_{7,3}} + \frac{3}{2}\beta_1\gamma_8^{\zeta_{7,3,3}} - \frac{3}{16}\gamma_1\beta_8^{\zeta_3\zeta_{7,3}} - \frac{3}{16}\gamma_1\beta_8^{\zeta_{7,3,3}}\,, \label{g9z4z73} \\
\gamma_9^{\zeta_2\zeta_{9,3}} &= \frac{10}{3}\gamma_9^{\zeta_{5,3,3,3}}\,, \label{g9z2z93} \\
\gamma_9^{\zeta_4\zeta_{5,3,3}} &= 10\gamma_9^{\zeta_{6,4,3,1,1}}\,, \label{g9z4z533} \\
\gamma_9^{\zeta_2\zeta_{5,5,3}} &= -\frac{14}{5}\gamma_9^{\zeta_{6,4,3,1,1}}\,, \label{g9z2z553} \\
\gamma_9^{\zeta_2\zeta_{7,3,3}} &= 2\gamma_9^{\zeta_{6,4,3,1,1}}\,. \label{g9z2z733}
\end{align}
Here, the coefficients $\hat{\zeta}_{6,4,1,1(0)}^{12}$, $\hat{\zeta}_{5,3,3(1)}^{12}$, and $\hat{\zeta}_{9,3(0)}^{12}$ are defined in eqs.~\eqref{eq:hz_6411_12}--\eqref{eq:hz_533_12}.

\section{Closed-form expressions for the relations}
\def\theequation{E\arabic{equation}}
\setcounter{equation}{0}
\label{app:rels}

We should remind ourselves again that the first perturbative order containing the Riemann zeta function (namely, $\zeta_3$)  is the three-loop order, $\beta_3$ and $\gamma_3$, in our ansatz. Every subsequent order can also include any MZV of weight increased by 2 compared to the previous one, i.e.,
\begin{align}\label{eq:FZany=0}
F_{n}^{\zeta_{ \boldsymbol{\Lambda} }} = 0, 
\qquad n < \frac{ \left\lvert \boldsymbol{\Lambda} \right\rvert +3}{2}, 
\qquad F=\beta\text{ or }\gamma.
\end{align}
The multi--multi-index notation for MZV products was introduced in eqs.~\eqref{eq:zetaexpansatz} and \eqref{eq:zetaexpansatz2}. In addition, we will sometimes take the liberty of diverging from the notation of ref.~\cite{Baikov:2018wgs} and writing
\begin{align}
F_{n}^{2m; \boldsymbol{\Lambda}} = F_{n}^{\zeta_{2m}\zeta_{ \boldsymbol{\Lambda} }}
\end{align}
for brevity. Note also that, in this section, we assume that all Riemann zeta functions of even arguments are accumulated in a standalone prefactor $\zeta_{2m}$ in isolation from the product of other MZVs, $\zeta_{ \boldsymbol{\Lambda} }$.

Moreover, in our ansatz, we postulate that all MZV products containing a single factor of $\zeta_2$ (and not its higher powers) vanish because of the scheme choice,
\begin{align}\label{eq:FZ2Zanything}
F_{n}^{\zeta_2 \zeta_{ \boldsymbol{\Lambda} }} = 0, 
\qquad n \geqslant 2,
\end{align}
except for the fifteen sequences given below: 
\begin{align}
F_{n}^{\zeta_2\zeta_9} &= 45F_{n}^{\zeta_{5,3,3}}, \qquad n \geqslant 7, \label{eq:Fnz2z9} \\
F_{n}^{\zeta_2\zeta_5^2} &= -\frac{7}{2}F_{n}^{\zeta_{6,4,1,1}}, \qquad n \geqslant 8, \label{g10z2z52} \\
F_{n}^{\zeta_2\zeta_3\zeta_7} &= -10F_{n}^{\zeta_{6,4,1,1}}, \qquad n \geqslant 8, \label{g10z2z3z7} \\
F_{n}^{\zeta_2\zeta_{7,3}} &= -3F_{n}^{\zeta_{6,4,1,1}}, \qquad n \geqslant 8, \label{g10z2z73} \\
F_{n}^{\zeta_2\zeta_{11}} &= \frac{275}{2}F_{n}^{\zeta_{5,5,3}} + \frac{407}{2}F_{n}^{\zeta_{7,3,3}}, \qquad n \geqslant 8, \label{g10z2z11} \\
F_{n}^{\zeta_2\zeta_3\zeta_9} &= 45F_{n}^{\zeta_3\zeta_{5,3,3}}, \qquad n \geqslant 9, \label{g10z2z3z9} \\
F_{n}^{\zeta_2\zeta_{9,3}} &= \frac{10}{3}F_{n}^{\zeta_{5,3,3,3}}, \qquad n \geqslant 9, \label{g10z2z93} \\
F_{n}^{\zeta_2\zeta_5\zeta_7} &= 20F_{n}^{\zeta_{5,3,3,3}}, \qquad n \geqslant 9, \label{g10z2z5z7} \\
F_{n}^{\zeta_2\zeta_3\zeta_5^2} &= -\frac{7}{2}F_{n}^{\zeta_3\zeta_{6,4,1,1}} - \frac{9}{2}F_{n}^{\zeta_{6,4,3,1,1}}, \qquad n \geqslant 9, \label{g10z2z3z52} \\
F_{n}^{\zeta_2\zeta_3^2\zeta_7} &= -10F_{n}^{\zeta_3\zeta_{6,4,1,1}} - 16F_{n}^{\zeta_{6,4,3,1,1}} , \qquad n \geqslant 9, \label{g10z2z32z7} \\
F_{n}^{\zeta_2\zeta_5\zeta_{5,3}} &= \frac{21}{2}F_{n}^{\zeta_{6,4,3,1,1}}, \qquad n \geqslant 9, \label{g10z2z5z53} \\
F_{n}^{\zeta_2\zeta_3\zeta_{7,3}} &= -3F_{n}^{\zeta_3\zeta_{6,4,1,1}} - 3F_{n}^{\zeta_{6,4,3,1,1}}, \qquad n \geqslant 9, \label{g10z2z3z73} \\
F_{n}^{\zeta_2\zeta_{5,5,3}} &= -\frac{14}{5}F_{n}^{\zeta_{6,4,3,1,1}}, \qquad n \geqslant 9, \label{g10z2z553} \\
F_{n}^{\zeta_2\zeta_{7,3,3}} &= 2F_{n}^{\zeta_{6,4,3,1,1}}, \qquad n \geqslant 9, \label{g10z2z733} \\
F_{n}^{\zeta_2\zeta_{13}} &= \frac{1001}{2}F_{n}^{\zeta_{7,3,5}} + \frac{1209}{2}F_{n}^{\zeta_{9,3,3}} - \frac{56717}{120}F_{n}^{\zeta_{6,4,3,1,1}}, \qquad n \geqslant 9. \label{eq:Fnz2z13}
\end{align}

All other relations express $F_{n}^{\zeta_{2m} \zeta_{\boldsymbol{\Lambda}}}$, a rational coefficient in a term containing a power of $\pi$ in the $n$-loop renormalization-group function $F_n$,  through  lower-loop rational coefficients in terms without $\pi$. We now introduce the degree-$p$ forms that appear in the relations below. To this end, we define the set  $\mathscr{C}([n],p)$ of all compositions of an integer $n$ into $p$ parts, $\mathscr{C}([n],p) = \left\{  m_1,\dots, m_p : \sum_{i=1}^p m_i = n, m_i \geqslant 1  \right\}$ (i.e.,\ repetitions are allowed and the order of the summands matters). Unless stated otherwise, all summations below run over all $p$-tuples $\bm{m}(p)= m_1,\dots,m_p$ belonging to $\mathscr{C}([n],p)$. For products of coefficients, we use the shorthand notation
\begin{align}
\beta^{\text{R}}_{\bm{m}(r)} = \prod_{i=1}^{r} \beta^{\text{R}}_{m_{i}} \quad \text{ for } r \geqslant 1 
\qquad\text{ and } \qquad
\beta^{\text{R}}_{\bm{m}(r)} = 1 \quad \text{ for } r \leqslant 0 ,
\end{align}
where $F_n^{\text{R}}$ stands for the rational part of $F_n$.

The degree-$p$ forms entering the relations are denoted as 
\begin{align}
R^{F,n}_{(r;t \vert s)}(\zeta_{\boldsymbol{\Lambda}_1}, \dots, \zeta_{\boldsymbol{\Lambda}_t} \vert \zeta_{\boldsymbol{\Sigma}_1}, \dots, \zeta_{\boldsymbol{\Sigma}_s})
\end{align} 
with $p = r+t+s$ and $F = \beta$ or  $\gamma$. Every term of the form is a product of $r$ rational parts of AD or $\beta$-function coefficients and $t+s$ transcendental parts. The $s$ MZVs to the right of the vertical line are related to the circumflexed MZVs~\eqref{eq:KTformula} and \eqref{eq:zeta53}--\eqref{eq:zeta8611}; consequently, the corresponding forms are accompanied by factors of expansion coefficients $\hat{\zeta}_{{\boldsymbol{\Sigma}_i}(n)}^{2a_i;{\boldsymbol{\Sigma}'_i}}$, $i=1,\dots, s$ defined in eq.~\eqref{eq:zetaexpansatz}. If the MZV $\varepsilon$-expansions are known, then a given relation can be constructed combinatorially as a sum of all possible forms,
\begin{align}
F_{n}^{\zeta_{2m} \zeta_{\boldsymbol{\Lambda}}} = \sum Q \left( \prod_{i=1}^s  \hat{\zeta}_{{\boldsymbol{\Sigma}_i}(g_i)}^{{2a_i; \boldsymbol{\Sigma}'_i}} \right) R^{F,n}_{(r;t \vert s)}(\zeta_{\boldsymbol{\Lambda}_1}, \dots, \zeta_{\boldsymbol{\Lambda}_t} \vert \zeta_{\boldsymbol{\Sigma}_1}, \dots, \zeta_{\boldsymbol{\Sigma}_s}),
\end{align} 
so that, in every term above, $p =r +t+s =1+\sum_{i=1}^s g_i $,  $\zeta_{\boldsymbol{\Lambda}} = \zeta_{\boldsymbol{\Lambda}_1;\dots;\boldsymbol{\Lambda}_t;\boldsymbol{\Sigma}'_1;\dots;\boldsymbol{\Sigma}'_s}$, $m = \sum_{i=1}^s a_i$, and $Q = \left( \prod \zeta_{2a_i} \right)/\zeta_{2m}$ is a rational number.

The explicit expressions for the forms required in the following are listed below.
\begin{align}
R^{F,n}_{(r;0 \vert 1)} (\zeta_{\boldsymbol{\lambda}}) = \sum_{\bm{m}(r+1) \in \mathscr{C}([n],r+1)} \beta^{\text{R}}_{\bm{m}(r)} \left[ F_{m_{r+1}}^{\zeta_{\boldsymbol{\lambda}}}
- r \frac{m_{r+1}}{\sum_{i=1}^{r} m_i} \beta_{m_r}^{\zeta_{\boldsymbol{\lambda}}} F^{\text{R}}_{m_{r+1}} \right];
\end{align}

\begin{align}
R^{F,n}_{(r;1 \vert 1)} (\zeta_{\boldsymbol{\lambda}} \vert \zeta_{\boldsymbol{\sigma}})
={}&
(r+1) \sum \beta^{\text{R}}_{\bm{m}(r)} \beta_{m_{r+1}}^{\zeta_{\boldsymbol{\lambda}}} F_{m_{r+2}}^{\zeta_{\boldsymbol{\sigma}}}
\notag\\
&{}-(r+1) \sum \frac{m_{r+2}}{\sum_{i=1}^{r+1} m_i} \beta^{\text{R}}_{\bm{m}(r)} \beta_{m_{r+1}}^{\zeta_{\boldsymbol{\sigma}}} F_{m_{r+2}}^{\zeta_{\boldsymbol{\lambda}}}
\notag\\
&{}-(r+1)r  \sum \frac{m_{r+2}}{\sum_{i=1}^{r+1} m_i} \beta^{\text{R}}_{\bm{m}(r-1)} \beta_{m_r}^{\zeta_{\boldsymbol{\lambda}}} \beta_{m_{r+1}}^{\zeta_{\boldsymbol{\sigma}}} F^{\text{R}}_{m_{r+2}}.
\end{align}

\begin{multline}
R^{F,n}_{(r;2 \vert 1)} (\zeta_{\boldsymbol{\rho}}, \zeta_{\boldsymbol{\lambda}} \vert \zeta_{\boldsymbol{\sigma}})
=
\frac12 (r+1) (r+2) \sum \beta^{\text{R}}_{\bm{m}(r)} \beta_{m_{r+1}}^{\zeta_{\boldsymbol{\rho}}} \beta_{m_{r+2}}^{\zeta_{\boldsymbol{\lambda}}} F_{m_{r+3}}^{\zeta_{\boldsymbol{\sigma}}}
\\
{}-(r+1) (r+2) \sum \frac{m_{r+3}}{\sum_{i=1}^{r+2} m_i} \beta^{\text{R}}_{\bm{m}(r)} \beta_{m_{r+1}}^{\zeta_{\boldsymbol{\rho}}} \beta_{m_{r+2}}^{\zeta_{\boldsymbol{\sigma}}} F_{m_{r+3}}^{\zeta_{\boldsymbol{\lambda}}}
\\
{}-\frac12 r (r+1) (r+2)  \sum \frac{m_{r+3}}{\sum_{i=1}^{r+2} m_i} \beta^{\text{R}}_{\bm{m}(r-1)} \beta_{m_r}^{\zeta_{\boldsymbol{\rho}}} \beta_{m_{r+1}}^{\zeta_{\boldsymbol{\lambda}}} \beta_{m_{r+2}}^{\zeta_{\boldsymbol{\sigma}}} F^{\text{R}}_{m_{r+3}};
\end{multline}

\begin{align}
R^{F,n}_{(1;0 \vert 2)} (\zeta_{\boldsymbol{\lambda}},\zeta_{\boldsymbol{\sigma}})
={}&
 \sum \frac{m_1-m_2-m_3}{m_1} \beta_{m_1}^{\zeta_{\boldsymbol{\lambda}}} F_{m_2}^{\zeta_{\boldsymbol{\sigma}}} \beta^{\text{R}}_{m_3}
\notag\\
&{}+ \sum \frac{m_3 \left[ m_1^2+m_2^2 + m_3 (m_1+m_2) \right] }{2 m_1 m_2 (m_1+m_2)} \beta_{m_1}^{\zeta_{\boldsymbol{\lambda}}} \beta_{m_2}^{\zeta_{\boldsymbol{\sigma}}} F^{\text{R}}_{m_3};
\end{align}

\begin{multline}
R^{F,n}_{(0;1\vert 2)} (\zeta_{\boldsymbol{\rho}} \vert \zeta_{\boldsymbol{\lambda}}, \zeta_{\boldsymbol{\sigma}})
=\sum \frac{m_1-m_2-m_3}{m_1} \beta_{m_1}^{\zeta_{\boldsymbol{\lambda}}} F_{m_2}^{\zeta_{\boldsymbol{\sigma}}} \beta_{m_3}^{\zeta_{\boldsymbol{\rho}}}
\\
+ \sum \frac{m_3 \left[ m_3 (m_1+m_2) + m_1^2+m_2^2 \right] }{2 m_1 m_2 (m_1+m_2)} \beta_{m_1}^{\zeta_{\boldsymbol{\lambda}}} \beta_{m_2}^{\zeta_{\boldsymbol{\sigma}}} F_{m_3}^{\zeta_{\boldsymbol{\rho}}};
\end{multline}

\begin{multline}
R^{F,n}_{(3;0 \vert 2)} (\zeta_{\boldsymbol{\lambda}},\zeta_{\boldsymbol{\sigma}})
\\
{}=
 \sum \left[ 4 - \frac12 m_{12345} \left( \frac{1}{m_1} + \frac{1}{m_{134}} + \frac{1}{m_{135}} + \frac{1}{m_{145}} \right)  \right] \beta_{m_1}^{\zeta_{\boldsymbol{\lambda}}} F_{m_2}^{\zeta_{\boldsymbol{\sigma}}} \beta^{\text{R}}_{m_3} \beta^{\text{R}}_{m_4} \beta^{\text{R}}_{m_5}
\\
{}- \sum \frac34 m_3 \left[ \frac{8}{m_{1245}} -  \left( \frac{m_{12345}}{m_2 m_{145}} + \frac{m_{12345}}{m_1 m_{245}} \right)  \right] \beta_{m_1}^{\zeta_{\boldsymbol{\lambda}}} \beta_{m_2}^{\zeta_{\boldsymbol{\sigma}}} F^{\text{R}}_{m_3}  \beta^{\text{R}}_{m_4} \beta^{\text{R}}_{m_5},
\end{multline}
where $m_{i_1\dots i_K} = \sum_{k=1}^K m_{i_k}$. 

The last two relations for $F^{\zeta_3 \zeta_{10}}_n$ and $F^{\zeta_5 \zeta_{10}}_n$ in eqs.~\eqref{eq:FZ3Z10} and \eqref{eq:FZ5Z10}, respectively, also contain the degree-5 forms $R^{F,n}_{(2;1\vert 2)} (\zeta_{\boldsymbol{\rho}} \vert \zeta_{\boldsymbol{\lambda}}, \zeta_{\boldsymbol{\sigma}})$, for which we were unable to find a closed-form representation. It appears that $n=15$ is not large enough to determine the corresponding rational function of the tuple $\mathbf{m}(5)$ unambiguously (at least when considering only the 48 sequences of relations without unfixed parameters).

The explicit values of the coefficients $\hat{\zeta}^{2m;\boldsymbol{\Lambda}}_{\bm{\sigma}(n)}$ are defined by eqs.~\eqref{eq:KTformula} and \eqref{eq:zeta53}--\eqref{eq:zeta8611}. The 33 relations that do not contain free parameters and are linked to the circumflexed MZVs~\eqref{eq:KTformula} and  \eqref{eq:zeta53}--\eqref{eq:zeta8611}  can now be written as follows: 
\begin{align}\label{eq:FZ4}
F_{n}^{\zeta_4} = \hat\zeta^{4}_{3(1)} R^{F,n}_{(1;0 \vert 1)} (\zeta_3), 
\qquad n \geqslant 4;
\end{align}

\begin{align}
F_{n}^{\zeta_6} = 
\hat\zeta^{6}_{5(1)} R^{F,n}_{(1;0 \vert 1)} (\zeta_5)
+ \hat\zeta^{6}_{3(3)} R^{F,n}_{(3;0 \vert 1)} (\zeta_3), 
\qquad n \geqslant 5;
\end{align}

\begin{multline}
F_{n}^{\zeta_8} = 
\hat\zeta^{8}_{5,3(0)} F_{n}^{\zeta_{5,3}}
+ \hat\zeta^{8}_{7(1)} R^{F,n}_{(1;0 \vert 1)} (\zeta_7)
+ Z_{3;3}^8 R^{F,n}_{(2;0 \vert 1)} (\zeta_3^2)
\\
+ Z_{3;3}^8 R^{F,n}_{(1;0 \vert 2)} (\zeta_3, \zeta_3)
+ \hat\zeta^{8}_{5(3)} R^{F,n}_{(3;0 \vert 1)} (\zeta_5)
+\hat\zeta^{8}_{3(5)} R^{F,n}_{(5;0 \vert 1)} (\zeta_3), 
\qquad n \geqslant 6,
\\
\qquad Z_{3;3}^8 =\frac{ \left( \zeta_4 \hat\zeta^{4}_{3(1)} \right)^2 }{ \zeta_8 };
\end{multline}

\begin{align}
F_{n}^{\zeta_3 \zeta_4} =
2 \hat\zeta^{4}_{3(1)} R^{F,n}_{(1;0 \vert 1)} (\zeta_3^2)
+ \hat\zeta^{4}_{3(1)} R^{F,n}_{(0; 1 \vert 1)} (\zeta_3 \vert \zeta_3), 
\qquad n \geqslant 5;
\end{align}

\begin{multline}
F_{n}^{\zeta_4 \zeta_\lambda} =
\hat\zeta^{4}_{3(1)} R^{F,n}_{(1;0 \vert 1)} (\zeta_3 \zeta_\lambda)
+ \hat\zeta^{4;7}_{\lambda,3(1)} R^{F,n}_{(1;0 \vert 1)} (\zeta_{\lambda,3})
+ \hat\zeta^{4}_{3(1)} R^{F,n}_{(0; 1 \vert 1)} (\zeta_\lambda \vert \zeta_3), 
\\
\lambda=5,\;7,
\qquad n \geqslant (\lambda+7)/2;
\end{multline}

\begin{multline}
F_{n}^{\zeta_4 \zeta_9} =
 \hat\zeta^{4;9}_{5,5,3(0)} F_{n}^{\zeta_{5,5,3}}
+ \hat\zeta^{4;9}_{7,3,3(0)} F_{n}^{\zeta_{7,3,3}}
+ \hat\zeta^{4;9}_{6,4,1,1(1)} R^{F,n}_{(1;0 \vert 1)} (\zeta_{6,4,1,1})
\\
+ \hat\zeta^{4}_{3(1)} R^{F,n}_{(1;0 \vert 1)} (\zeta_3 \zeta_9)
+ \hat\zeta^{4;9}_{9,3(1)} R^{F,n}_{(1;0 \vert 1)} (\zeta_{9,3})
+ \hat\zeta^{4}_{3(1)} R^{F,n}_{(0; 1 \vert 1)} (\zeta_9 \vert \zeta_3), 
\qquad n \geqslant 8;
\end{multline}

\begin{multline}
F_{n}^{\zeta_4 \zeta_{11}} =
 \hat\zeta^{4;11}_{7,3,5(0)} F_{n}^{\zeta_{7,3,5}}
+ \hat\zeta^{4;11}_{9,3,3(0)} F_{n}^{\zeta_{9,3,3}}
+ \hat\zeta^{4;11}_{6,4,3,1,1(0)} F_{n}^{\zeta_{6,4,3,1,1}}
+ \hat\zeta^{4}_{3(1)} R^{F,n}_{(1;0 \vert 1)} (\zeta_3 \zeta_{11})
\\
+ \hat\zeta^{4;11}_{5,3,3,3(1)} R^{F,n}_{(1;0 \vert 1)} (\zeta_{5,3,3,3})
+ \hat\zeta^{4;11}_{11,3(1)} R^{F,n}_{(1;0 \vert 1)} (\zeta_{11,3})
+ \hat\zeta^{4;11}_{9,5(1)} R^{F,n}_{(1;0 \vert 1)} (\zeta_{9,5})
\\
+ \hat\zeta^{4}_{3(1)} R^{F,n}_{(0; 1 \vert 1)} (\zeta_{11} \vert \zeta_3), 
\qquad n \geqslant 9;
\end{multline}

\begin{multline}
F_{n}^{\zeta_4 \zeta_{5,3}} =
\hat\zeta^{4}_{3(1)} R^{F,n}_{(1; 0 \vert 1)} (\zeta_3 \zeta_{5,3})
+ \hat\zeta^{4;\lambda,3}_{\lambda,3,3(0)} R^{F,n}_{(1; 0 \vert 1)} (\zeta_{5,3,3})
+ \hat\zeta^{4}_{3(1)} R^{F,n}_{(0; 1 \vert 1)} (\zeta_{5,3} \vert \zeta_3)
\\
+ \hat\zeta^{4;5,3}_{6,4,1,1(0)} F_n^{\zeta_{6,4,1,1}}, 
\qquad n \geqslant 8;
\end{multline}

\begin{multline}
F_{n}^{\zeta_4 \zeta_{7,3}} =
\hat\zeta^{4}_{3(1)} R^{F,n}_{(1; 0 \vert 1)} (\zeta_3 \zeta_{7,3})
+ \hat\zeta^{4;\lambda,3}_{\lambda,3,3(0)} R^{F,n}_{(1; 0 \vert 1)} (\zeta_{7,3,3})
+ \hat\zeta^{4}_{3(1)} R^{F,n}_{(0; 1 \vert 1)} (\zeta_{7,3} \vert \zeta_3), 
\qquad n \geqslant 9;
\end{multline}

\begin{multline}
F_{n}^{\zeta_4 \zeta_3 \zeta_{5,3}} =
\hat\zeta^{4}_{3(1)} R^{F,n}_{(0 ; 1 \vert 1)} (\zeta_3 \vert \zeta_3 \zeta_{5,3})
+ \hat\zeta^{4;5,3}_{5,3,3(1)} R^{F,n}_{(1 ; 0 \vert 1)} (\zeta_3 \zeta_{5,3,3})
+ \hat\zeta^{4;5,3}_{5,3,3(1)} R^{F,n}_{(0 ; 1 \vert 1)} (\zeta_3 \vert \zeta_{5,3,3})
\\
+ \hat\zeta^{4}_{3(1)} R^{F,n}_{(0 ; 1 \vert 1)} (\zeta_3 \zeta_{5,3} \vert \zeta_3)
+ 2 \hat\zeta^{4}_{3(1)} R^{F,n}_{(0 ; 1 \vert 1)} (\zeta_{5,3} \vert \zeta_3^2)
+ 2 \hat\zeta^{4}_{3(1)} R^{F,n}_{(1 ; 0 \vert 1)} (\zeta_3^2 \zeta_{5,3})
\\
+ \hat\zeta^{4;5,3}_{6,4,1,1(0)} F_n^{\zeta_{6,4,3,1,1}}
+ \hat\zeta^{4;5,3}_{6,4,1,1(0)} F_n^{\zeta_3 \zeta_{6,4,1,1}},
\qquad n \geqslant 9;
\end{multline}

\begin{multline}
F_{n}^{\zeta_4 \zeta_{5,3,3}} =
\hat\zeta^{4;5,3,3}_{5,3,3,3(1)} R^{F,n}_{(1; 0 \vert 1)} (\zeta_{5,3,3,3})
+ \hat\zeta^{4}_{3(1)} R^{F,n}_{(1; 0 \vert 1)} (\zeta_3 \zeta_{5,3,3})
+ \hat\zeta^{4}_{3(1)} R^{F,n}_{(0; 1 \vert 1)} (\zeta_{5,3,3} \vert \zeta_3)
\\
+ \hat\zeta^{4;5,3,3}_{6,4,3,1,1(0)} F_n^{\zeta_{6,4,3,1,1}}, 
\qquad n \geqslant 9;
\end{multline}

\begin{multline}
F_{n}^{\zeta_4 \zeta_3^2 \zeta_{5}} =
\hat\zeta^{4;5}_{5,3(1)} R^{F,n}_{(0 ; 1 \vert 1)} (\zeta_3 \vert \zeta_3 \zeta_{5,3})
+\hat\zeta^{4}_{3(1)} R^{F,n}_{(0 ; 1 \vert 1)} (\zeta_3^2 \vert \zeta_3 \zeta_5)
+\hat\zeta^{4;5}_{5,3(1)} R^{F,n}_{(0 ; 1 \vert 1)} (\zeta_3^2 \vert \zeta_{5,3})
\\
+ 3 \hat\zeta^{4}_{3(1)} R^{F,n}_{(0 ; 1 \vert 1)} (\zeta_5 \vert \zeta_3^3)
+2 \hat\zeta^{4}_{3(1)} R^{F,n}_{(0 ; 1 \vert 1)} (\zeta_3 \zeta_5 \vert \zeta_3^2)
+ 2 \hat\zeta^{4}_{3(1)} R^{F,n}_{(0 ; 1 \vert 1)} (\zeta_3 \vert \zeta_3^2 \zeta_5)
\\
+ \hat\zeta^{4}_{3(1)} R^{F,n}_{(0 ; 1 \vert 1)} (\zeta_3^2 \zeta_5 \vert \zeta_3)
+ 3 \hat\zeta^{4}_{3(1)} R^{F,n}_{(1 ; 0 \vert 1)} (\zeta_3^3 \zeta_5)
+ \hat\zeta^{4;5}_{5,3(1)} R^{F,n}_{(1 ; 0 \vert 1)} (\zeta_3^2 \zeta_{5,3})
\\
 + \hat\zeta^{3;3;4;5}_{6,4,3,1,1(0)} F_n^{\zeta_{6,4,3,1,1}}
 + \hat\zeta^{3;4;5}_{6,4,1,1(0)} F_n^{\zeta_3 \zeta_{6,4,1,1}},
\qquad n \geqslant 9;
\end{multline}

\begin{multline}
F_{n}^{\zeta_3 \zeta_4 \zeta_5} =
2  \hat\zeta^{4}_{3(1)} R^{F,n}_{(1; 0 \vert 1)} (\zeta_3^2 \zeta_5) 
 + \hat\zeta^{4;5}_{5,3(1)} R^{F,n}_{(1; 0 \vert 1)} (\zeta_{5,3} \zeta_3) 
 + \hat\zeta^{4;5}_{5,3(1)} R^{F,n}_{(0; 1 \vert 1)} (\zeta_3 \vert \zeta_{5,3}) 
 \\{}
 + 2 \hat\zeta^{4}_{3(1)} R^{F,n}_{(0; 1 \vert 1)} (\zeta_{5} \vert \zeta_3^2) 
 + \hat\zeta^{4}_{3(1)} R^{F,n}_{(0; 1 \vert 1)} (\zeta_3 \vert \zeta_3 \zeta_5) 
 + \hat\zeta^{4}_{3(1)} R^{F,n}_{(0; 1 \vert 1)} (\zeta_3 \zeta_5 \vert \zeta_3)
 \\
 + \hat\zeta^{3;4;5}_{6,4,1,1(0)} F_n^{\zeta_{6,4,1,1}},
\qquad n \geqslant 8;
\end{multline}

\begin{multline}
F_{n}^{\zeta_3 \zeta_4 \zeta_7} =
2  \hat\zeta^{4}_{3(1)} R^{F,n}_{(1; 0 \vert 1)} (\zeta_3^2 \zeta_7) 
 + \hat\zeta^{4;5}_{\lambda,3(1)} R^{F,n}_{(1; 0 \vert 1)} (\zeta_{7,3} \zeta_3) 
 + \hat\zeta^{4;5}_{\lambda,3(1)} R^{F,n}_{(0; 1 \vert 1)} (\zeta_3 \vert \zeta_{7,3}) 
 \\{}
 + 2 \hat\zeta^{4}_{3(1)} R^{F,n}_{(0; 1 \vert 1)} (\zeta_{7} \vert \zeta_3^2) 
 + \hat\zeta^{4}_{3(1)} R^{F,n}_{(0; 1 \vert 1)} (\zeta_3 \vert \zeta_3 \zeta_7) 
 + \hat\zeta^{4}_{3(1)} R^{F,n}_{(0; 1 \vert 1)} (\zeta_3 \zeta_7 \vert \zeta_3)
 \\
 + \hat\zeta^{4;7}_{5,3,3(0)} F_n^{\zeta_3 \zeta_{5,3,3}},
\qquad n \geqslant 9;
\end{multline}

\begin{align}
F_{n}^{\zeta_3^2 \zeta_4} =
3 \hat\zeta^{4}_{3(1)} R^{F,n}_{(1; 0 \vert 1)} (\zeta_3^{3}) 
+ \sum_{m=1}^{2} m \hat\zeta^{4}_{3(1)} R^{F,n}_{(0; 1 \vert 1)} (\zeta_3^{3-m} \vert \zeta_3^m),
\qquad n \geqslant 7;
\end{align}

\begin{multline}
F_{n}^{\zeta_3^3 \zeta_4} =
4 \hat\zeta^{4}_{3(1)} R^{F,n}_{(1; 0 \vert 1)} (\zeta_3^{4}) 
+ \hat\zeta^{3;3;3;4}_{6,4,1,1(1)} R^{F,n}_{(1; 0 \vert 1)} (\zeta_{6,4,4,1}) 
+ \sum_{m=1}^{3} m \hat\zeta^{4}_{3(1)} R^{F,n}_{(0; 1 \vert 1)} (\zeta_3^{4-m} \vert \zeta_3^m),
\\
\quad n \geqslant 8;
\end{multline}

\begin{multline}
F_{n}^{\zeta_4 \zeta_5^2} =
\hat\zeta^{4;5;5}_{7,3,3(1)} R^{F,n}_{(1; 0 \vert 1)} (\zeta_{7,3,3}) 
+ \hat\zeta^{4;5;5}_{5,5,3(1)} R^{F,n}_{(1; 0 \vert 1)} (\zeta_{5,5,3}) 
+ \hat\zeta^{4}_{3(1)} R^{F,n}_{(1; 0 \vert 1)} (\zeta_3 \zeta_5^2) 
\\{}
+ \hat\zeta^{4;5}_{5,3(1)} R^{F,n}_{(1; 0 \vert 1)} (\zeta_5 \zeta_{5,3}) 
+ \hat\zeta^{4;5}_{5,3(1)} R^{F,n}_{(0; 1 \vert 1)} (\zeta_5 \vert \zeta_{5,3}) 
+ \hat\zeta^{4}_{3(1)} R^{F,n}_{(0; 1 \vert 1)} (\zeta_5^2 \vert \zeta_3) 
\\{}
+ \hat\zeta^{4}_{3(1)} R^{F,n}_{(0; 1 \vert 1)} (\zeta_5 \vert \zeta_3 \zeta_5),
\qquad n \geqslant 9;
\end{multline}

\begin{multline}
F_{n}^{\zeta_3 \zeta_6} =
\hat\zeta^{6}_{5(1)} R^{F,n}_{(1; 0 \vert 1)} (\zeta_3 \zeta_5) 
+ \hat\zeta^{6}_{5(1)} R^{F,n}_{(0; 1 \vert 1)} (\zeta_3 \vert \zeta_5) 
+ 2 \hat\zeta^{6}_{3(3)} R^{F,n}_{(3; 0 \vert 1)} (\zeta_3^2) 
+ \hat\zeta^{6}_{3(3)} R^{F,n}_{(2; 1 \vert 1)} (\zeta_3 \vert \zeta_3),
\\
\qquad n \geqslant 7;
\end{multline}

\begin{multline}
F_{n}^{\zeta_5 \zeta_6} =
\hat\zeta^{5;6}_{5,3,3(0)} F_{n}^{\zeta_{5,3,3}}
+ 2 \hat\zeta^{6}_{5(1)} R^{F,n}_{(1; 0 \vert 1)} (\zeta_5^2) 
+ \hat\zeta^{5;6}_{7,3(1)} R^{F,n}_{(1; 0 \vert 1)} (\zeta_{7,3}) 
+ \hat\zeta^{6}_{5(1)} R^{F,n}_{(0; 1 \vert 1)} (\zeta_5 \vert \zeta_5) 
\\
+ \hat\zeta^{5;6}_{5,3(3)} R^{F,n}_{(3; 0 \vert 1)} (\zeta_{5,3}) 
+ \hat\zeta^{6}_{3(3)} R^{F,n}_{(3; 0 \vert 1)} (\zeta_3 \zeta_5) 
+ \hat\zeta^{6}_{3(3)} R^{F,n}_{(2; 1 \vert 1)} (\zeta_5 \vert \zeta_3),
\qquad n \geqslant 7;
\end{multline}

\begin{multline}
F_{n}^{\zeta_6 \zeta_7} =
\hat\zeta^{6;7}_{7,3,3(0)} F_{n}^{\zeta_{7,3,3}}
+ \hat\zeta^{6}_{5(1)} R^{F,n}_{(1; 0 \vert 1)} (\zeta_5 \zeta_7) 
+ \hat\zeta^{6;7}_{9,3(1)} R^{F,n}_{(1; 0 \vert 1)} (\zeta_{9,3}) 
+ \hat\zeta^{6;7}_{6,4,1,1(1)} R^{F,n}_{(1; 0 \vert 1)} (\zeta_{6,4,1,1}) 
\\
+ \hat\zeta^{6}_{5(1)} R^{F,n}_{(0; 1 \vert 1)} (\zeta_7 \vert \zeta_5) 
+ \hat\zeta^{6;7}_{7,3(3)} R^{F,n}_{(3; 0 \vert 1)} (\zeta_{7,3}) 
+ \hat\zeta^{6}_{3(3)} R^{F,n}_{(3; 0 \vert 1)} (\zeta_3 \zeta_7) 
+ \hat\zeta^{6}_{3(3)} R^{F,n}_{(2; 1 \vert 1)} (\zeta_7 \vert \zeta_3),
\\
\qquad n \geqslant 8;
\end{multline}

\begin{multline}
F_{n}^{\zeta_6 \zeta_9} =
\hat\zeta^{6;9}_{7,3,5(0)} F_{n}^{\zeta_{7,3,5}}
+ \hat\zeta^{6;9}_{9,3,3(0)} F_{n}^{\zeta_{9,3,3}}
+ \hat\zeta^{6;9}_{6,4,3,1,1(0)} F_{n}^{\zeta_{6,4,3,1,1}}
+ \hat\zeta^{6}_{5(1)} R^{F,n}_{(0 ; 1 \vert 1)} (\zeta_9 \vert \zeta_5)
\\
+ Z^{6;9}_{5,3,3;3} R^{F,n}_{(0 ; 1 \vert 1)} (\zeta_{5,3,3} \vert \zeta_3)
+ \hat\zeta^{6}_{5(1)} R^{F,n}_{(1 ; 0 \vert 1)} (\zeta_5 \zeta_9)
+ \hat\zeta^{6;9}_{9,5(1)} R^{F,n}_{(1 ; 0 \vert 1)} (\zeta_{9,5})
\\
+ \hat\zeta^{6;9}_{11,3(1)} R^{F,n}_{(1 ; 0 \vert 1)} (\zeta_{11,3})
+ \hat\zeta^{6;9}_{5,3,3,3(1)} R^{F,n}_{(1 ; 0 \vert 1)} (\zeta_{5,3,3,3})
+ Z^{6;9}_{5,3,3;3} R^{F,n}_{(1 ; 0 \vert 1)} (\zeta_3 \zeta_{5,3,3})
\\
+ \hat\zeta^{6}_{3(3)} R^{F,n}_{(2 ; 1 \vert 1)} (\zeta_9 \vert \zeta_3)
+ \hat\zeta^{6}_{3(3)} R^{F,n}_{(3 ; 0 \vert 1)} (\zeta_3 \zeta_9)
+ \hat\zeta^{6;9}_{9,3(3)} R^{F,n}_{(3 ; 0 \vert 1)} (\zeta_{9,3})
\\
+ \hat\zeta^{6;9}_{6,4,1,1(3)} R^{F,n}_{(3 ; 0 \vert 1)} (\zeta_{6,4,1,1}),
\qquad n \geqslant 9;
\\
\qquad
Z^{6;9}_{5,3,3;3} = \frac{\zeta_2 \hat\zeta^{2;9}_{5,3,3(0)} \zeta_4 \hat\zeta^{4}_{3(1)} }{\zeta_{10}};
\end{multline}

\begin{multline}
F_{n}^{\zeta_6 \zeta_{5,3}} =
\hat\zeta^{6}_{5(1)} R^{F,n}_{(1; 0 \vert 1)} (\zeta_5 \zeta_{5,3}) 
+ \hat\zeta^{6}_{5(1)} R^{F,n}_{(0; 1 \vert 1)} (\zeta_{5,3} \vert \zeta_5) 
+ \hat\zeta^{6;5,3}_{5,3,3(3)} R^{F,n}_{(3; 0 \vert 1)} (\zeta_{5,3,3}) 
\\
+ \hat\zeta^{6}_{3(3)} R^{F,n}_{(3; 0 \vert 1)} (\zeta_3 \zeta_{5,3}) 
+ \hat\zeta^{6}_{3(3)} R^{F,n}_{(2; 1 \vert 1)} (\zeta_{5,3} \vert \zeta_3)
 + \hat\zeta^{6;5,3}_{5,3,3,3(0)} F_n^{\zeta_3 \zeta_{5,3,3,3}},
\qquad n \geqslant 8;
\end{multline}

\begin{multline}
F_{n}^{\zeta_3 \zeta_5 \zeta_6} =
\hat\zeta^{5;6}_{7,3(1)} R^{F,n}_{(1; 0 \vert 1)} (\zeta_3 \zeta_{7,3}) 
+ 2 \hat\zeta^{6}_{5(1)} R^{F,n}_{(1; 0 \vert 1)} (\zeta_3 \zeta_{5}^2) 
+ \hat\zeta^{6}_{5(1)} R^{F,n}_{(0; 1 \vert 1)} (\zeta_3 \zeta_{5} \vert \zeta_5) 
\\
+ \hat\zeta^{6}_{5(1)} R^{F,n}_{(0; 1 \vert 1)} (\zeta_{5} \vert \zeta_3 \zeta_5) 
+ 2 \hat\zeta^{6}_{5(1)} R^{F,n}_{(0; 1 \vert 1)} (\zeta_3 \vert \zeta_5^2) 
+ \hat\zeta^{5;6}_{7,3(1)} R^{F,n}_{(0; 1 \vert 1)} (\zeta_3 \vert \zeta_{7,3}) 
\\
+ \hat\zeta^{5;6}_{5,3(3)} R^{F,n}_{(3; 0 \vert 1)} (\zeta_3 \zeta_{5,3}) 
+ 2 \hat\zeta^{6}_{3(3)} R^{F,n}_{(3; 0 \vert 1)} (\zeta_3^2 \zeta_{5}) 
+ \hat\zeta^{6}_{3(3)} R^{F,n}_{(2; 1 \vert 1)} (\zeta_3 \vert \zeta_3 \zeta_{5}) 
\\
+ \hat\zeta^{6}_{3(3)} R^{F,n}_{(2; 1 \vert 1)} (\zeta_3 \zeta_{5} \vert \zeta_3) 
+ \hat\zeta^{5;6}_{5,3(3)} R^{F,n}_{(2; 1 \vert 1)} (\zeta_3 \vert \zeta_{5,3}) 
+ 2 \hat\zeta^{6}_{3(3)} R^{F,n}_{(2; 1 \vert 1)} (\zeta_5 \vert \zeta_{3}^2)
\\
+ \hat\zeta^{6}_{3(3)} R^{F,n}_{(1; 2 \vert 1)} (\zeta_3, \zeta_5 \vert \zeta_3)
+ \hat\zeta^{6}_{3(3)} R^{F,n}_{(1; 2 \vert 1)} (\zeta_5, \zeta_3 \vert \zeta_3)
 + \hat\zeta^{3;5;6}_{5,3,3(0)} F_n^{\zeta_3 \zeta_{5,3,3}}  ,
\qquad n \geqslant 9;
\end{multline}

\begin{multline}
F_{n}^{\zeta_3^2 \zeta_6} =
\hat\zeta^{6}_{5(1)} R^{F,n}_{(1; 0 \vert 1)} (\zeta_3^2 \zeta_{5}) 
+ \hat\zeta^{6}_{5(1)} R^{F,n}_{(0; 1 \vert 1)} (\zeta_3^2 \vert \zeta_5) 
+ \hat\zeta^{6}_{5(1)} R^{F,n}_{(0; 1 \vert 1)} (\zeta_{3} \vert \zeta_3 \zeta_5) 
\\
+ 3 \hat\zeta^{6}_{3(3)} R^{F,n}_{(3; 0 \vert 1)} (\zeta_3^3) 
+ 2 \hat\zeta^{6}_{3(3)} R^{F,n}_{(2; 1 \vert 1)} (\zeta_3 \vert \zeta_{3}^2)
+ \hat\zeta^{6}_{3(3)} R^{F,n}_{(2; 1 \vert 1)} (\zeta_3^2 \vert \zeta_{3})
\\
+ \hat\zeta^{6}_{3(3)} R^{F,n}_{(1; 2 \vert 1)} (\zeta_3, \zeta_3 \vert \zeta_3) 
+ \hat\zeta^{3;3;6}_{6,4,1,1(0)} F_n^{\zeta_{6,4,1,1}} ,
\qquad n \geqslant 8;
\end{multline}

\begin{multline}
F_{n}^{\zeta_6 \zeta_3^3} =
\hat\zeta^{6}_{5(1)} R^{F,n}_{(0 ; 1 \vert 1)} (\zeta_3 \vert \zeta_3^2 \zeta_5)
+ \hat\zeta^{6}_{5(1)} R^{F,n}_{(0 ; 1 \vert 1)} (\zeta_3^2 \vert \zeta_3 \zeta_5)
+ \hat\zeta^{6}_{5(1)} R^{F,n}_{(0 ; 1 \vert 1)} (\zeta_3^3 \vert \zeta_5)
\\
+ \hat\zeta^{6}_{3(3)} R^{F,n}_{(0 ; 3 \vert 1)} (\zeta_3,\zeta_3,\zeta_3 \vert \zeta_3)
+ \hat\zeta^{6}_{5(1)} R^{F,n}_{(1 ; 0 \vert 1)} (\zeta_3^3 \zeta_5)
+ 2 \hat\zeta^{6}_{3(3)} R^{F,n}_{(1 ; 2 \vert 1)} (\zeta_3,\zeta_3 \vert \zeta_3^2)
\\
+ 2 \hat\zeta^{6}_{3(3)} R^{F,n}_{(1 ; 2 \vert 1)} (\zeta_3^2,\zeta_3 \vert \zeta_3)
+ 2 \hat\zeta^{6}_{3(3)} R^{F,n}_{(2 ; 1 \vert 1)} (\zeta_3^2 \vert \zeta_3^2)
+ 3 \hat\zeta^{6}_{3(3)} R^{F,n}_{(2 ; 1 \vert 1)} (\zeta_3 \vert \zeta_3^3)
\\
+ \hat\zeta^{6}_{3(3)} R^{F,n}_{(2 ; 1 \vert 1)} (\zeta_3^3 \vert \zeta_3)
+ 4 \hat\zeta^{6}_{3(3)} R^{F,n}_{(3 ; 0 \vert 1)} (\zeta_3^4)
+ \hat\zeta^{3;3;3}_{6,4,1,1(3)} R^{F,n}_{(3 ; 0 \vert 1)} (\zeta_{6,4,1,1})
\\
 + \hat\zeta^{3;3;3}_{6,4,3,1,1(0)} F_n^{\zeta_{6,4,3,1,1}}
 + \hat\zeta^{3;3}_{6,4,1,1(0)} F_n^{\zeta_3 \zeta_{6,4,1,1}},
\qquad n \geqslant 9;
\end{multline}

\begin{multline}
F_{n}^{\zeta_3 \zeta_8} =
\hat\zeta^{8}_{5,3(0)} F_{n}^{\zeta_3 \zeta_{5,3}}
+ \hat\zeta^{8}_{7(1)} R^{F,n}_{(1; 0 \vert 1)} (\zeta_3 \zeta_7) 
+ \hat\zeta^{8}_{7(1)} R^{F,n}_{(0; 1 \vert 1)} (\zeta_3 \vert \zeta_7) 
+ Z_{3;3}^8 R^{F,n}_{(0; 1 \vert 2)} (\zeta_3 \vert \zeta_3, \zeta_3)
\\
+ Z_{3;3}^8 R^{F,n}_{(1; 1 \vert 1)} (\zeta_3 \vert \zeta_3^2)
+ 2 Z_{3;3}^8 R^{F,n}_{(1; 0 \vert 2)} (\zeta_3, \zeta_3^2)
+ 2 Z_{3;3}^8 R^{F,n}_{(1; 0 \vert 2)} (\zeta_3^2, \zeta_3)
+ 3 Z_{3;3}^8 R^{F,n}_{(2; 0 \vert 1)} (\zeta_3^3)
\\
+ \hat\zeta^{8}_{5(3)} R^{F,n}_{(3; 0 \vert 1)} (\zeta_3 \zeta_5)
+ \hat\zeta^{8}_{5(3)} R^{F,n}_{(2; 1 \vert 1)} (\zeta_3 \vert \zeta_5)
+ \hat\zeta^{8}_{3(5)} R^{F,n}_{(4; 1 \vert 1)} (\zeta_3 \vert \zeta_3) 
+ 2 \hat\zeta^{8}_{3(5)} R^{F,n}_{(5; 0 \vert 1)} (\zeta_3^2),
\\
 \qquad n \geqslant 8,
\qquad Z_{3;3}^8 =\frac{ \left( \zeta_4 \hat\zeta^{4}_{3(1)} \right)^2 }{ \zeta_8 };
\end{multline}

\begin{multline}
F_{n}^{\zeta_5 \zeta_8} =
\hat\zeta^{8}_{5,3(0)} F_{n}^{\zeta_5 \zeta_{5,3}}
+ \hat\zeta^{5;8}_{5,5,3(0)} F_{n}^{\zeta_{5,5,3}}
+ \hat\zeta^{5;8}_{7,3,3(0)} F_{n}^{\zeta_{7,3,3}}
+ \hat\zeta^{8}_{7(1)} R^{F,n}_{(1; 0 \vert 1)} (\zeta_5 \zeta_7) 
\\
+ \hat\zeta^{5;8}_{9,3(1)} R^{F,n}_{(1; 0 \vert 1)} (\zeta_{9,3}) 
+ \hat\zeta^{8}_{7(1)} R^{F,n}_{(0; 1 \vert 1)} (\zeta_5 \vert \zeta_7) 
+ Z_{3;3}^8 R^{F,n}_{(2; 0 \vert 1)} (\zeta_3^2 \zeta_5)
\\
+ Z_{3;5,3}^{5;8} R^{F,n}_{(2; 0 \vert 1)} (\zeta_3 \zeta_{5,3})
+ \hat\zeta^{5;8}_{5,3,3(2)} R^{F,n}_{(2; 0 \vert 1)} (\zeta_{5,3,3})
+ Z_{3;3}^8 R^{F,n}_{(1; 1 \vert 1)} (\zeta_5, \zeta_3^2)
\\
+ Z_{3;5,3}^{5;8} R^{F,n}_{(1;0 \vert 2)} (\zeta_{5,3}, \zeta_3)
+ Z_{3;5,3}^{5;8} R^{F,n}_{(1;0 \vert 2)} (\zeta_3, \zeta_{5,3})
+ Z_{3;3}^8 R^{F,n}_{(1;0 \vert 2)} (\zeta_3 \zeta_5, \zeta_3)
\\
+ Z_{3;3}^8 R^{F,n}_{(1;0 \vert 2)} (\zeta_3, \zeta_3 \zeta_5)
+ Z_{3;3}^8 R^{F,n}_{(0;1 \vert 2)} (\zeta_5 \vert \zeta_3, \zeta_3)
+ \hat\zeta^{8}_{5(3)} R^{F,n}_{(2;1 \vert 1)} (\zeta_5 \vert \zeta_5)
\\
+ 2 \hat\zeta^{8}_{5(3)} R^{F,n}_{(3;0 \vert 1)} (\zeta_5^2)
+ \hat\zeta^{5;8}_{7,3(3)} R^{F,n}_{(3;0 \vert 1)} (\zeta_{7,3})
+ \hat\zeta^{8}_{3(5)} R^{F,n}_{(4;1 \vert 1)} (\zeta_5 \vert \zeta_3)
\\
+ \hat\zeta^{8}_{3(5)} R^{F,n}_{(5;0 \vert 1)} (\zeta_3 \zeta_5)
+ \hat\zeta^{5;8}_{5,3(5)} R^{F,n}_{(5;0 \vert 1)} (\zeta_{5,3})
+ \hat\zeta^{5;8}_{6,4,1,1(1)} R^{F,n}_{(1;0 \vert 1)} (\zeta_{6,4,1,1}),
\\
\qquad n \geqslant 8, 
 \qquad Z_{3;5,3}^{5;8} = \frac{\zeta_4 \hat\zeta^{4}_{3(1)} \zeta_4 \hat\zeta^{4;5}_{5,3(1)}}{\zeta_8};
\end{multline}

\begin{multline}
F_{n}^{\zeta_8\zeta_7} =
\hat\zeta^{8}_{5,3(0)} F_{n}^{\zeta_7 \zeta_{5,3}}
+ \hat\zeta^{8;7}_{7,3,5(0)} F_{n}^{\zeta_{7,3,5}}
+ \hat\zeta^{8;7}_{9,3,3(0)} F_{n}^{\zeta_{9,3,3}}
+ \hat\zeta^{8;7}_{6,4,3,1,1(0)} F_{n}^{\zeta_{6,4,3,1,1}}
\\
+ \hat\zeta^{8}_{7(1)} R^{F,n}_{(0 ; 1 \vert 1)} (\zeta_7 \vert \zeta_7)
+ Z^{8;7}_{3;5,3,3} R^{F,n}_{(1 ; 0 \vert 1)} (\zeta_3 \zeta_{5,3,3})
+ \hat\zeta^{8;7}_{5,3,3,3(1)} R^{F,n}_{(1 ; 0 \vert 1)} (\zeta_{5,3,3,3})
\\
+ Z^{8;7}_{3;5,3,3} R^{F,n}_{(0 ; 1 \vert 1)} ( \zeta_{5,3,3} \vert \zeta_3)
+ 2 \hat\zeta^{8}_{7(1)} R^{F,n}_{(1 ; 0 \vert 1)} (\zeta_7^2)
+ \hat\zeta^{8;7}_{9,5(1)} R^{F,n}_{(1 ; 0 \vert 1)} (\zeta_{9,5})
\\
+ \hat\zeta^{8;7}_{11,3(1)} R^{F,n}_{(1 ; 0 \vert 1)} (\zeta_{11,3})
+ Z_{3;3}^8 R^{F,n}_{(0 ; 1 \vert 2)} (\zeta_7 \vert \zeta_3, \zeta_3)
+ Z^{8}_{3;3} R^{F,n}_{(1 ; 0 \vert 2)} (\zeta_3,\zeta_3 \zeta_7)
\\ 
+ Z^{8}_{3;3} R^{F,n}_{(1 ; 0 \vert 2)} (\zeta_3 \zeta_7,\zeta_3)
+ Z^{8;7}_{3;7,3} R^{F,n}_{(1 ; 0 \vert 2)} (\zeta_3,\zeta_{7,3})
+ Z^{8;7}_{3;7,3} R^{F,n}_{(1 ; 0 \vert 2)} (\zeta_{7,3},\zeta_3)
\\
+ Z^{8}_{3;3} R^{F,n}_{(1 ; 1 \vert 1)} (\zeta_7 \vert \zeta_3^2)
+ Z^{8}_{3;3} R^{F,n}_{(2 ; 0 \vert 1)} (\zeta_3^2 \zeta_7)
+ Z^{8;7}_{3;7,3} R^{F,n}_{(2 ; 0 \vert 1)} (\zeta_3 \zeta_{7,3})
\\
+ \hat\zeta^{8;7}_{7,3,3(2)} R^{F,n}_{(2 ; 0 \vert 1)} (\zeta_{7,3,3})
+ \hat\zeta^{8}_{5(3)} R^{F,n}_{(2 ; 1 \vert 1)} (\zeta_7 \vert \zeta_5)
+ \hat\zeta^{8}_{5(3)} R^{F,n}_{(3 ; 0 \vert 1)} (\zeta_5 \zeta_7)
\\
+ \hat\zeta^{8;7}_{9,3(3)} R^{F,n}_{(3 ; 0 \vert 1)} (\zeta_{9,3})
+ \hat\zeta^{8;7}_{6,4,1,1(3)} R^{F,n}_{(3 ; 0 \vert 1)} (\zeta_{6,4,1,1})
+ \hat\zeta^{8}_{3(5)} R^{F,n}_{(4 ; 1 \vert 1)} (\zeta_7 \vert \zeta_3)
\\
+ \hat\zeta^{8}_{3(5)} R^{F,n}_{(5 ; 0 \vert 1)} (\zeta_3 \zeta_7)
+ \hat\zeta^{8;7}_{7,3(5)} R^{F,n}_{(5 ; 0 \vert 1)} (\zeta_{7,3}),
\qquad n \geqslant 9;  
\\
\qquad Z^{8;7}_{3;5,3,3} = \frac{\zeta_4^2}{\zeta_8} \hat\zeta^{4;7}_{5,3,3(0)} \hat\zeta^{4}_{3(1)},
\qquad Z^{8;7}_{3;7,3} = \frac{\zeta_4^2}{\zeta_8} \hat\zeta^{7}_{7,3(1)} \hat\zeta^{4}_{3(1)};
\end{multline}

\begin{multline}
F_{n}^{\zeta_3^2 \zeta_8} =
\hat\zeta^{8}_{5,3(0)} F_{n}^{\zeta_3^2 \zeta_{5,3}}
+ \hat\zeta^{8}_{7(1)} R^{F,n}_{(1;0\vert 1)} (\zeta_3^2 \zeta_7) 
+ \hat\zeta^{8}_{7(1)} R^{F,n}_{(0;1\vert 1)} (\zeta_3 \vert \zeta_3 \zeta_7) 
+ \hat\zeta^{8}_{7(1)} R^{F,n}_{(0;1\vert 1)} (\zeta_3^2 \vert \zeta_7) 
\\
+ \hat\zeta^{8}_{5(3)} R^{F,n}_{(3;0\vert 1)} (\zeta_3^2 \zeta_5) 
+ \hat\zeta^{8}_{5(3)} R^{F,n}_{(2;1\vert 1)} (\zeta_3 \vert \zeta_3 \zeta_5) 
+ \hat\zeta^{8}_{5(3)} R^{F,n}_{(1;2\vert 1)} (\zeta_3, \zeta_3 \vert \zeta_5) 
\\
+ \hat\zeta^{8}_{5(3)} R^{F,n}_{(2;1\vert 1)} (\zeta_3^2 \vert \zeta_5) 
+ 3 Z_{3;3}^8 R^{F,n}_{(1;1\vert 1)} (\zeta_3 \vert \zeta_3^3) 
+ 3 Z_{3;3}^8 R^{F,n}_{(1;0\vert 2)} (\zeta_3, \zeta_3^3)
\\
+ 3 Z_{3;3}^8 R^{F,n}_{(1;0\vert 2)} (\zeta_3^3, \zeta_3) 
+ 6 Z_{3;3}^8 R^{F,n}_{(2;0\vert 1)} (\zeta_3^4) 
+ Z_{3;3}^8 R^{F,n}_{(0;2\vert 1)} (\zeta_3, \zeta_3 \vert \zeta_3^2) 
\\
+ Z_{3;3}^8 R^{F,n}_{(0;1\vert 2)} (\zeta_3^2 \vert \zeta_3, \zeta_3) 
+ 2 Z_{3;3}^8 R^{F,n}_{(0;1\vert 2)} (\zeta_3 \vert \zeta_3, \zeta_3^2) 
+ 2 Z_{3;3}^8 R^{F,n}_{(0;1\vert 2)} (\zeta_3 \vert \zeta_3^2, \zeta_3) 
\\
+ Z_{3;3}^8 R^{F,n}_{(1;1\vert 1)} (\zeta_3^2 \vert \zeta_3^2) 
+ 4 Z_{3;3}^8 R^{F,n}_{(1;0\vert 2)} (\zeta_3^2, \zeta_3^2) 
+ 3 \hat\zeta^{8}_{3(5)} R^{F,n}_{(5;0\vert 1)} (\zeta_3^3) 
\\
+ 2 \hat\zeta^{8}_{3(5)} R^{F,n}_{(4;1\vert 1)} (\zeta_3 \vert \zeta_3^2)
+ \hat\zeta^{8}_{3(5)} R^{F,n}_{(3;2\vert 1)} (\zeta_3, \zeta_3 \vert \zeta_3) 
+ \hat\zeta^{8}_{3(5)} R^{F,n}_{(4;1\vert 1)} (\zeta_3^2 \vert \zeta_3)
\\
+ \hat\zeta^{3;3;8}_{6,4,1,1(2)} R^{F,n}_{(2;0\vert 1)} (\zeta_{6,4,1,1}) ,
 \qquad n \geqslant 9;
\end{multline}

\begin{multline}\label{eq:FZ10}
F_{n}^{\zeta_{10}} =
\hat\zeta^{10}_{7,3(0)} F_{n}^{\zeta_{7,3}}
+ \hat\zeta^{10}_{9(1)} R^{F,n}_{(1;0\vert 1)} (\zeta_9) 
+ \hat\zeta^{10}_{5,3(2)} R^{F,n}_{(2;0\vert 1)} (\zeta_{5,3}) 
+ Z^{10}_{3;5} R^{F,n}_{(2;0\vert 1)} (\zeta_3 \zeta_5) 
\\
+ \hat\zeta^{10}_{7(3)} R^{F,n}_{(3;0\vert 1)} (\zeta_7) 
+ 2 Z^{10}_{3;3} R^{F,n}_{(4;0\vert 1)} (\zeta_3^2) 
+ \hat\zeta^{10}_{5(5)} R^{F,n}_{(5;0\vert 1)} (\zeta_5) 
+ \hat\zeta^{10}_{3(7)} R^{F,n}_{(7;0\vert 1)} (\zeta_3) 
\\
+ Z^{10}_{3;5} R^{F,n}_{(1;0\vert 2)} (\zeta_3, \zeta_5) 
+ Z^{10}_{3;5} R^{F,n}_{(1;0\vert 2)} (\zeta_5, \zeta_3) 
+ 2 Z^{10}_{3;3} R^{F,n}_{(3;0\vert 2)} (\zeta_3, \zeta_3),
 \qquad n \geqslant 7,
\\
 \qquad Z_{3;5}^{10} = -Z_{3;3}^{10} = \frac{\zeta_4 \hat\zeta^{4}_{3(1)} \zeta_6 \hat\zeta^{6}_{5(1)} }{\zeta_{10}} = - \frac{ \zeta_4 \hat\zeta^{4}_{3(1)} \zeta_6 \hat\zeta^{6}_{3(3)} }{ \zeta_{10} };
\end{multline}

\begin{multline}\label{eq:FZ3Z10}
F_{n}^{\zeta_3 \zeta_{10}} =
\hat\zeta^{10}_{7,3(0)} F_{n}^{\zeta_3 \zeta_{7,3}}
+ \hat\zeta^{10}_{9(1)} R^{F,n}_{(1;0\vert 1)} (\zeta_3 \zeta_9) 
+ \hat\zeta^{3;10}_{6,4,1,1(1)} R^{F,n}_{(1;0\vert 1)} (\zeta_{6,4,1,1}) 
\\
+ \hat\zeta^{10}_{9(1)} R^{F,n}_{(0;1\vert 1)} (\zeta_3 \vert \zeta_9) 
+ 2 Z^{10}_{3;5} R^{F,n}_{(2;0\vert 1)} (\zeta_3^2 \zeta_5) 
+ \hat\zeta^{10}_{5,3(2)}  R^{F,n}_{(2;0\vert 1)} (\zeta_3 \zeta_{5,3}) 
\\
+ \hat\zeta^{10}_{5,3(2)}  R^{F,n}_{(1;1\vert 1)} (\zeta_3 \vert \zeta_{5,3}) 
+ Z^{10}_{3;5}  R^{F,n}_{(1;1\vert 1)} (\zeta_3 \vert \zeta_3 \zeta_5) 
+ Z^{10}_{3;5}  R^{F,n}_{(1;0\vert 2)} (\zeta_3, \zeta_3 \zeta_5) 
\\
+ Z^{10}_{3;5}  R^{F,n}_{(1;0\vert 2)} (\zeta_3 \zeta_5, \zeta_3) 
+ 2 Z^{10}_{3;5}  R^{F,n}_{(1;0\vert 2)} (\zeta_3^2, \zeta_5) 
+ 2 Z^{10}_{3;5}  R^{F,n}_{(1;0\vert 2)} (\zeta_5, \zeta_3^2) 
\\
+ Z^{10}_{3;5}  R^{F,n}_{(0;1\vert 2)} (\zeta_3 \vert \zeta_3, \zeta_5) 
+ Z^{10}_{3;5}  R^{F,n}_{(0;1\vert 2)} (\zeta_3 \vert \zeta_5, \zeta_3) 
+ 6 Z^{10}_{3;3}  R^{F,n}_{(4;0\vert 1)} (\zeta_3^3) 
\\
+ Z^{10}_{7(3)}  R^{F,n}_{(3;0\vert 1)} (\zeta_3 \zeta_7) 
+ Z^{10}_{7(3)}  R^{F,n}_{(2;1\vert 1)} (\zeta_3 \vert \zeta_7) 
+ 2 Z^{10}_{3;3}  R^{F,n}_{(2;1\vert 2)} (\zeta_3 \vert \zeta_3, \zeta_3) 
\\
+ 2 Z^{10}_{3;3}  R^{F,n}_{(3;1\vert 1)} (\zeta_3 \vert \zeta_3^2) 
+ 4 Z^{10}_{3;3}  R^{F,n}_{(3;0\vert 2)} (\zeta_3, \zeta_3^2) 
+ 4 Z^{10}_{3;3}  R^{F,n}_{(3;0\vert 2)} (\zeta_3^2, \zeta_3) 
\\
+ \hat\zeta^{10}_{5(5)}  R^{F,n}_{(4;1\vert 1)} (\zeta_3 \vert \zeta_5) 
+ \hat\zeta^{10}_{5(5)}  R^{F,n}_{(5;0\vert 1)} (\zeta_3 \zeta_5) 
+ 2 \hat\zeta^{10}_{3(7)} R^{F,n}_{(7;0\vert 1)} (\zeta_3^2),
 \qquad n \geqslant 9;
\end{multline}

\begin{multline}\label{eq:FZ5Z10}
F_{n}^{\zeta_{10}\zeta_5} =
   \hat\zeta^{10}_{7,3(0)} F_{n}^{\zeta_5 \zeta_{7,3}}
+ \hat\zeta^{10;5}_{7,3,5(0)} F_{n}^{\zeta_{7,3,5}}
+ \hat\zeta^{10;5}_{9,3,3(0)} F_{n}^{\zeta_{9,3,3}}
+ \hat\zeta^{10;5}_{6,4,3,1,1(0)} F_{n}^{\zeta_{6,4,3,1,1}}
\\
+ \hat\zeta^{10}_{9(1)} R^{F,n}_{(0 ; 1 \vert 1)} (\zeta_5 \vert \zeta_9)
+ Z^{10;5}_{5,3,3;3} R^{F,n}_{(0 ; 1 \vert 1)} (\zeta_{5,3,3} \vert \zeta_3)
+ Z^{10;5}_{5,3,3;3} R^{F,n}_{(1 ; 0 \vert 1)} (\zeta_3 \zeta_{5,3,3})
\\
+ \hat\zeta^{10;5}_{5,3,3,3(1)} R^{F,n}_{(1 ; 0 \vert 1)} (\zeta_{5,3,3,3})
+ \hat\zeta^{10;5}_{6,4,1,1(3)} R^{F,n}_{(3 ; 0 \vert 1)} (\zeta_{6,4,1,1})
+ \hat\zeta^{10;5}_{5,3,3(4)} R^{F,n}_{(4 ; 0 \vert 1)} (\zeta_{5,3,3})
\\
+ Z_{3;5}^{10} R^{F,n}_{(0 ; 1 \vert 2)} (\zeta_5 \vert \zeta_3,\zeta_5)
+ Z_{3;5}^{10} R^{F,n}_{(0 ; 1 \vert 2)} (\zeta_5 \vert \zeta_5,\zeta_3)
+ \hat\zeta^{10}_{9(1)} R^{F,n}_{(1 ; 0 \vert 1)} (\zeta_5 \zeta_9)
\\
+ \hat\zeta^{10;5}_{11,3(1)} R^{F,n}_{(1 ; 0 \vert 1)} (\zeta_{11,3})
+ Z_{3;5}^{10} R^{F,n}_{(1 ; 0 \vert 2)} (\zeta_5,\zeta_3 \zeta_5)
+ Z_{3;5}^{10} R^{F,n}_{(1 ; 0 \vert 2)} (\zeta_3 \zeta_5,\zeta_5)
\\
+ Z_{3;5}^{10} R^{F,n}_{(1 ; 1 \vert 1)} (\zeta_5 \vert \zeta_3 \zeta_5)
+ 2 Z_{3;5}^{10} R^{F,n}_{(1 ; 0 \vert 2)} (\zeta_3,\zeta_5^2)
+ 2 Z_{3;5}^{10} R^{F,n}_{(1 ; 0 \vert 2)} (\zeta_5^2,\zeta_3)
\\
+ Z^{10;5}_{5,3;5} R^{F,n}_{(1 ; 0 \vert 2)} (\zeta_5,\zeta_{5,3})
+ Z^{10;5}_{5,3;5} R^{F,n}_{(1 ; 0 \vert 2)} (\zeta_{5,3},\zeta_5)
+ Z^{10;5}_{7,3;3} R^{F,n}_{(1 ; 0 \vert 2)} (\zeta_3,\zeta_{7,3})
\\
+ Z^{10;5}_{7,3;3} R^{F,n}_{(1 ; 0 \vert 2)} (\zeta_{7,3},\zeta_3)
+ \hat\zeta^{10}_{5,3(2)} R^{F,n}_{(1 ; 1 \vert 1)} (\zeta_5 \vert \zeta_{5,3})
+ 2 Z_{3;5}^{10} R^{F,n}_{(2 ; 0 \vert 1)} (\zeta_3 \zeta_5^2)
\\
+ \hat\zeta^{10}_{5,3(2)} R^{F,n}_{(2 ; 0 \vert 1)} (\zeta_5 \zeta_{5,3})
+ Z^{10;5}_{7,3;3} R^{F,n}_{(2 ; 0 \vert 1)} (\zeta_3 \zeta_{7,3})
+ \hat\zeta^{10;5}_{5,3,5(2)} R^{F,n}_{(2 ; 0 \vert 1)} (\zeta_{5,3,5})
\\
+ \hat\zeta^{10;5}_{7,3,3(2)} R^{F,n}_{(2 ; 0 \vert 1)} (\zeta_{7,3,3})
+ \hat\zeta^{10}_{7(3)} R^{F,n}_{(2 ; 1 \vert 1)} (\zeta_5 \vert \zeta_7)
+ 2 Z^{10}_{3;3} R^{F,n}_{(2 ; 1 \vert 2)} (\zeta_5 \vert \zeta_3,\zeta_3)
\\
+ \hat\zeta^{10}_{7(3)} R^{F,n}_{(3 ; 0 \vert 1)} (\zeta_5 \zeta_7)
+ \hat\zeta^{10;5}_{9,3(3)} R^{F,n}_{(3 ; 0 \vert 1)} (\zeta_{9,3})
+ 2 Z^{10}_{3;3} R^{F,n}_{(3 ; 0 \vert 2)} (\zeta_3,\zeta_3 \zeta_5)
\\
+ 2 Z^{10}_{3;3} R^{F,n}_{(3 ; 0 \vert 2)} (\zeta_3 \zeta_5,\zeta_3)
+ Z^{10;5}_{5,3;3} R^{F,n}_{(3 ; 0 \vert 2)} (\zeta_3,\zeta_{5,3})
+ Z^{10;5}_{5,3;3} R^{F,n}_{(3 ; 0 \vert 2)} (\zeta_{5,3},\zeta_3)
\\
+ 2 \hat\zeta^{10}_{3(7)} R^{F,n}_{(3 ; 1 \vert 1)} (\zeta_5 \vert \zeta_3^2)
+ Z^{10}_{3;3} R^{F,n}_{(4 ; 0 \vert 1)} (\zeta_3^2 \zeta_5)
+ Z^{10;5}_{5,3;3} R^{F,n}_{(4 ; 0 \vert 1)} (\zeta_3 \zeta_{5,3})
\\
+ \hat\zeta^{10}_{5(5)} R^{F,n}_{(4 ; 1 \vert 1)} (\zeta_5 \vert \zeta_5)
+ 2 \hat\zeta^{10}_{5(5)} R^{F,n}_{(5 ; 0 \vert 1)} (\zeta_5^2)
+ \hat\zeta^{10;5}_{7,3(5)} R^{F,n}_{(5 ; 0 \vert 1)} (\zeta_{7,3})
\\
+ \hat\zeta^{10}_{3(7)} R^{F,n}_{(6 ; 1 \vert 1)} (\zeta_5 \vert \zeta_3)
+ \hat\zeta^{10}_{3(7)} R^{F,n}_{(7 ; 0 \vert 1)} (\zeta_3 \zeta_5)
+ \hat\zeta^{10;5}_{5;3(7)} R^{F,n}_{(7 ; 0 \vert 1)} (\zeta_{5,3}),
\qquad n \geqslant 9;
\\
Z^{10;5}_{5,3;3} = \frac{\zeta_4 \hat\zeta^{4;5}_{5,3(1)} \zeta_6 \hat\zeta^{6}_{3(3)} }{\zeta_{10}},
\qquad
Z^{10;5}_{5,3;5} = \frac{\zeta_4 \hat\zeta^{4;5}_{5,3(1)} \zeta_6 \hat\zeta^{6}_{5(1)} }{\zeta_{10}},
\qquad
Z^{10;5}_{7,3;3} = \frac{\zeta_6 \hat\zeta^{6;5}_{7,3(1)} \zeta_4 \hat\zeta^{4}_{3(1)} }{\zeta_{10}},
\\
Z^{10;5}_{5,3,3;3} = \frac{\zeta_6 \hat\zeta^{6;5}_{5,3,3(0)} \zeta_4 \hat\zeta^{4}_{3(1)} }{\zeta_{10}}.
\end{multline}
The identities \eqref{eq:FZ4}--\eqref{eq:FZ5Z10}, together with the trivial relations \eqref{eq:Fnz2z9}--\eqref{eq:Fnz2z13}, were explicitly verified up to $n=15$ for both anomalous-dimension and $\beta$-function coefficients.

%\bibliographystyle{JHEP}
%\bibliography{nopie}

\providecommand{\href}[2]{#2}\begingroup\raggedright\endgroup

\end{document}